\documentclass[11pt,a4paper]{article}
\pdfoutput=1
\usepackage[applemac]{inputenc}
\usepackage{amsmath, amsthm}
\usepackage{jheppub}
\usepackage[center]{caption}
\usepackage{subcaption}
\usepackage{multirow}
\usepackage{booktabs}
\usepackage{amsmath}
\usepackage{amsfonts}
\usepackage{amssymb}
\usepackage{graphicx}
\usepackage{indentfirst}
\usepackage{slashed}
\usepackage{diagbox}
\usepackage{mathrsfs}
\usepackage{bm}
\usepackage{feynmp}

\makeatletter
\newcommand\makebig[2]{%
  \@xp\newcommand\@xp*\csname#1\endcsname{\bBigg@{#2}}%
  \@xp\newcommand\@xp*\csname#1l\endcsname{\@xp\mathopen\csname#1\endcsname}%
  \@xp\newcommand\@xp*\csname#1r\endcsname{\@xp\mathclose\csname#1\endcsname}%
}
\makeatother

\makebig{biggg} {3.0}
\makebig{Biggg} {3.5}
\makebig{bigggg}{4.0}
\makebig{Bigggg}{4.5}

\newcommand{\refE}[1]{eq.~(\ref{#1})}
\newcommand{\refF}[1]{fig.~\ref{#1}}

\newcommand{\refS}[1]{Section~\ref{#1}}

\newcommand{\abs}[1]{\left\lvert#1\right\rvert}

\newcommand{\Li}[0]{\text{Li}}

\newcommand{\Gram}{\text{Gram}}
\newcommand{\plainJ}{J}
\newcommand{\plainK}{K}
\newcommand{\tildeJ}{\widetilde J}

\newcommand{\hypgeo}[4]{\,_2F_1\left(#1,#2;#3;#4\right)}

\newcommand{\jtilde}{\widetilde J}
\newcommand{\jhat}{\widehat{J}}

\newcommand{\zbar}{\bar{z}}
\newcommand{\zz}{z}

\newcommand{\cutRes}{\mathcal{C}}

\newcommand{\wbar}{\bar w}

\newcommand{\bea}{\begin{eqnarray}}
\newcommand{\eea}{\end{eqnarray}}
\newcommand{\bean}{\begin{eqnarray*}}
\newcommand{\eean}{\end{eqnarray*}}

\def\abs#1{\left| #1\right|}

\def\det{\mathop{\rm det}}
\def\Li{{\rm Li}}

\def\eref#1{(\ref{#1})}

\def\eps{\epsilon}

\newcommand{\xb}{\bar{x}}

\def\ord{{\cal O}}
\def\cS{{\cal S}}
\def\cP{{\cal P}}

\def\Label#1{\label{#1}
  \smash{\hbox to0pt{\raise1ex\hbox{\tiny[#1]}\hss}}}

\def\beq{\begin{equation}}
\def\eeq{\end{equation}}
\def\bsp#1\esp{\begin{split}#1\end{split}}

\newcommand{\cA}{\mathcal{A}}
\newcommand{\cC}{\mathcal{C}}
\newcommand{\cH}{\mathcal{H}}

\def \sh{{\,\amalg\hskip -3.6pt\amalg\,}}

\renewcommand{\ln}{\log}

\theoremstyle{definition}

\preprint{CERN-TH-2017-092
\rightline{CP3-17-11}
\rightline{Edinburgh 2017/09}
\rightline{FR-PHENO-2017-010}
\rightline{TCDMATH-17-09}
}

\title{Diagrammatic Hopf algebra of cut Feynman integrals: the one-loop case}

\author[a]{Samuel Abreu,}
\author[b,c,d,e]{Ruth Britto,}
\author[f,g]{Claude Duhr,}
\author[h]{and Einan Gardi}

\affiliation[a]{Physikalisches Institut, Albert-Ludwigs-Universit\"at Freiburg, D-79104 Freiburg, Germany}
\affiliation[b]{School of Mathematics, Trinity College, Dublin 2, Ireland}
\affiliation[c]{Hamilton Mathematics Institute, Trinity College, Dublin 2, Ireland}
\affiliation[d]{Institut de Physique Th{\'e}orique, Universit\'e Paris Saclay, 
CEA, CNRS, F-91191 Gif-sur-Yvette cedex, France}
\affiliation[e]{Kavli Institute for Theoretical Physics, University of California, Santa Barbara, CA 93106, USA}
\affiliation[f]{Theoretical Physics Department, CERN, Geneva, Switzerland}
\affiliation[g]{Center for Cosmology, Particle Physics and Phenomenology (CP3), Universit\'e Catholique de Louvain, 1348 Louvain-La-Neuve, Belgium}
\affiliation[h]{Higgs Centre for Theoretical Physics, 
School of Physics and Astronomy, \\
The University of Edinburgh, Edinburgh EH9 3FD, Scotland, UK}

\emailAdd{abreu.samuel@physik.uni-freiburg.de}
\emailAdd{ruth.britto@tcd.ie}
\emailAdd{claude.duhr@cern.ch}
\emailAdd{einan.gardi@ed.ac.uk}

\abstract{
We construct a diagrammatic coaction acting on one-loop Feynman graphs and their cuts. The graphs are naturally identified with the corresponding (cut) Feynman integrals in dimensional regularization, whose coefficients of the Laurent expansion in the dimensional regulator are multiple polylogarithms (MPLs). Our main result is the conjecture that this diagrammatic coaction reproduces the combinatorics of the coaction on MPLs order by order in the Laurent expansion. We show that our conjecture holds in a broad range of nontrivial one-loop integrals. We then explore its consequences for the study of discontinuities of Feynman integrals, and  the differential equations that they satisfy. In particular, using the diagrammatic coaction along with information from cuts, we explicitly derive differential equations for any one-loop Feynman integral. We also explain how to construct the symbol of any one-loop Feynman integral recursively. Finally, we show that our diagrammatic coaction follows, in the special case of one-loop integrals, from a more general coaction proposed recently, which is constructed by pairing master integrands with corresponding master contours.
}


\ifx \tadInsert \undefined
	\def\tadInsert [#1]{
		\raisebox{-4mm}{\includegraphics[keepaspectratio=true, width=.8cm]{./diagrams/#1}}
	}
\fi

\ifx \tadInsertLow \undefined
	\def\tadInsertLow [#1]{
		\raisebox{-4.1mm}{\includegraphics[keepaspectratio=true, width=.81cm]{./diagrams/#1}}
	}
\fi

\ifx \bubInsertHigh \undefined
	\def\bubInsertHigh [#1]{
		\raisebox{-2.2mm}{\includegraphics[keepaspectratio=true, width=2cm]{./diagrams/#1}}
	}
\fi

\ifx \bubInsert \undefined
	\def\bubInsert [#1]{
		\raisebox{-3.6mm}{\includegraphics[keepaspectratio=true, width=2cm]{./diagrams/#1}}
	}
\fi

\ifx \bubInsertLow \undefined
	\def\bubInsertLow [#1]{
		\raisebox{-4.2mm}{\includegraphics[keepaspectratio=true, width=2cm]{./diagrams/#1}}
	}
\fi

\ifx \triInsert \undefined
	\def\triInsert [#1]{
		\raisebox{-6.3mm}{\includegraphics[keepaspectratio=true, width=2cm]{./diagrams/#1}}
	}
\fi

\ifx \triInsertLow \undefined
	\def\triInsertLow [#1]{
		\raisebox{-6.9mm}{\includegraphics[keepaspectratio=true, width=2cm]{./diagrams/#1}}
	}
\fi

\ifx \boxInsert \undefined
	\def\boxInsert [#1]{
		\raisebox{-4.9mm}{\includegraphics[keepaspectratio=true, width=2.3cm]{./diagrams/#1}}
	}
\fi

\ifx \boxInsertLow \undefined
	\def\boxInsertLow [#1]{
		\raisebox{-7.45mm}{\includegraphics[keepaspectratio=true, width=2.3cm]{./diagrams/#1}}
	}
\fi

\begin{document}
\maketitle





\section{Introduction}
\label{sec:introduction}

Feynman integrals are central objects in the perturbative approach to quantum field theory. In particular, they are one of the main ingredients in the calculation of scattering amplitudes, which are the quantities that one needs to compute to make precise theoretical predictions for high-energy collider experiments such as the Large Hadron Collider at CERN. One of the main challenges in the computation of scattering amplitudes is the evaluation of multi-loop and multi-leg Feynman integrals. A good understanding of the analytic structure of Feynman integrals is therefore fundamental in finding more efficient ways of computing them. 

In recent years, the realization that a large class of Feynman integrals can be written in terms of so-called multiple polylogarithms (MPLs) \cite{GoncharovMixedTate,Goncharov:2005sla} has led to major advances in precision calculations. Indeed, understanding the mathematics and algebraic structure of MPLs has led both to new efficient techniques to evaluate Feynman integrals \cite{Brown:2008um,Anastasiou:2013srw,Henn:2013pwa,Ablinger:2014yaa,Bogner:2014mha,Panzer:2015ida,Bogner:2015nda} and to tools to handle the complicated analytic expressions inherent to these computations~\cite{Goncharov:2010jf}. It is known, however, that starting from two loops there are Feynman integrals that cannot be written in terms of MPLs only, and generalizations to elliptic curves appear \cite{Caffo:1998du,Adams:2013kgc,Bloch:2013tra,Bloch:2014qca,Adams:2014vja,Adams:2015gva,Adams:2015ydq,Bloch:2016izu,Adams:2016xah,Remiddi:2016gno,Primo:2016ebd,Primo:2017ipr}. Given the important role played by understanding the mathematics of MPLs in precision computations in recent years, extending the techniques developed for MPLs to more general classes of functions is a pressing question.

A particularly important tool when working with MPLs is their coaction \cite{GoncharovMixedTate,Brown:2011ik,Duhr:2012fh}, a mathematical operation that exposes properties of MPLs through a decomposition into simpler functions. Coactions are closely related to Hopf algebras, and it has been speculated for a long time that there should be a natural Hopf algebra defined directly in terms of Feynman graphs~\cite{Connes:1998qv,Connes:1999yr,Kreimer:1997dp,Kreimer:2009iy,Kreimer:2009jt} such that the coaction on MPLs is reproduced if the graphs are replaced by their respective analytic expressions~\cite{Bloch:2005bh}.
In ref.~\cite{Brown:motivicperiods} a coaction was defined that acts naturally on Feynman integrals~\cite{Panzer:2016snt,Brown:2015fyf}, though it is not obvious how to interpret the objects appearing inside the coaction directly in terms of physical quantities such as Feynman graphs or integrals.

In parallel, it was observed in ref.~\cite{Abreu:2014cla,Abreu:2015zaa} that certain entries in the coaction of some one- and two-loop integrals can be expressed in terms of so-called \emph{cut} Feynman integrals, i.e., variants of Feynman integrals where some of the propagators are on their mass-shell. The empirical results of ref.~\cite{Abreu:2014cla,Abreu:2015zaa} gave first indications that any interpretation of a coaction on Feynman integrals in terms of physical quantities should not only involve Feynman graphs or integrals, but also their cut analogues.
This statement was made more precise in ref.~\cite{Abreu:2017enx}, where a coaction on certain classes of integrals was proposed that interprets the objects appearing inside the coaction in terms of so-called `master integrals' and `master contours'. It was argued that, when acting on one-loop integrals in dimensional regularization, this coaction can be cast in a form that only involves Feynman graphs and their cuts. As such, this diagrammatic coaction provides the first concrete realization of a coaction on Feynman graphs that reduces to the coaction on MPLs when graphs are replaced by their analytic expressions.

The main goal of this paper is to expand on the findings of ref.~\cite{Abreu:2017enx}.
By analyzing concrete examples of one-loop integrals, we give compelling evidence that one can define a coaction on Feynman graphs that decomposes a graph into `simpler' ones. There are two natural ways to construct simpler graphs: by contracting some of its edges, which corresponds to a Feynman integral with fewer propagators, or by `cutting' some of its edges, which corresponds to a Feynman integral with a modified integration contour {corresponding to on-shell restrictions for a set of propagators}. The restriction to these two operations is motivated by the fact that they appear in the study of the Landau conditions \cite{Landau:1959fi}, and that cut Feynman integrals encode information about the discontinuities of the uncut integrals~\cite{Cutkosky:1960sp,SMatrix,tHooft:1973pz,PhamBook,PhamInTeplitz,HwaTeplitz,Bloch:2015efx,Abreu:2017ptx}. 
At the end, we show
that our diagrammatic coaction can be seen as a special case of the coaction on integrals proposed in~\cite{Abreu:2017enx}.

After we have established the validity of our conjecture on various nontrivial examples, 
we explore its consequences for the study of the discontinuities and differential equations of Feynman integrals. As an important application of our result, we present an efficient way to derive differential equations for one-loop Feynman integrals. We show that the differential equation satisfied by any one-loop integral is completely fixed by its cuts where at most two of the propagators are not on their mass-shell. We explicitly compute all the relevant cuts of integrals depending on an arbitrary number of scales using the techniques of ref.~\cite{Abreu:2017ptx}. The knowledge of these cuts also allows us to iteratively construct the full alphabet and even the symbol of any one-loop integral. 

The paper is organized as follows. In Section \ref{sec:one-loop-diagrams} we define our basis of one-loop Feynman integrals, we summarize some properties of one-loop cut integrals that will be important in the rest of the paper, and we give a brief review of multiple polylogarithms. In Section \ref{sec:firstExamples} we motivate our main results by analyzing simple examples of coactions on one-loop integrals. In Section \ref{sec:general_formulation} we discuss the general formulation of the diagrammatic coaction, which is the main result of this paper. In Section \ref{sec:dimreg} we discuss the interplay between the diagrammatic coaction and dimensional regularization. In Sections \ref{sec:moreExamples} and \ref{sec:dci} we give support to our conjecture, first by exploring its predictions in a range of one-loop integrals, and second by showing how it is consistent with the extended dual conformal invariance of the finite integrals in our basis. In Sections~\ref{sec:discontinuities} and \ref{sec:diffEqAndAlphabet} we discuss the consequences of the diagrammatic coaction in the study of the discontinuities and differential equations satisfied by one-loop integrals. In Section~\ref{sec:master_formula} we show how we can recover the diagrammatic coaction from the more general conjecture of ref.~\cite{Abreu:2017enx}. Finally, in Section \ref{sec:summary} we draw our conclusions. We include several appendices where we summarize the notation and review some mathematical concepts used throughout the paper, and we present explicit results for the (cut) integrals.


\section{One-loop Feynman integrals and their cuts}
\label{sec:one-loop-diagrams}

\subsection{One-loop Feynman integrals}

In this section we give a short review of one-loop Feynman integrals, which are the main actors in this work.
We focus especially on linear relations among Feynman integrals, and identify a basis of one-loop integrals with nice analytic properties.

In order to write a convenient basis of one-loop Feynman integrals, let us define the scalar  $n$-point Feynman integrals as
\beq\bsp\label{eq:one-loop}
I_n^D&(\{p_i\cdot p_j\};\{m_i^2\};\eps) = e^{\gamma_E\eps}\int \frac{d^Dk}{i\pi^{D/2}}\prod_{j=1}^n\frac{1}{(k-q_j)^2-m_j^2+i0}\,.
\esp\eeq
We work in dimensional regularization in $D=d-2\eps$ dimensions, where $d$ is an even positive integer, $\eps$ is a formal variable, and $\gamma_E=-\Gamma'(1)$ is the Euler-Mascheroni constant. 
The external momenta are labelled by $p_i$ and satisfy momentum conservation, $\sum_{i=1}^np_i=0$. In the integrand, the propagator labelled by $j$ carries momentum $k-q_j$, where each $q_j$ is a  linear combination of the external momenta,  resulting from imposing momentum conservation at each vertex of the diagram corresponding to the integral $I_n^D$:
\begin{equation}
	q_j=\sum_{i=1}^nc_{ji}p_i,\qquad c_{ji}\in\left\{-1,0,1\right\}\,.
\end{equation}
For concreteness, we can define the loop momentum $k$ as the one carried by the propagator labelled by 1, so that $q_1$ is the zero vector.

We do not need to consider numerator factors in the definition of the one-loop integrals, nor the case where the exponents of the propagators are different from unity. Indeed, at one-loop we can always express polynomials in $k^2$ and $k\cdot p_i$ in the numerator in terms of propagators, and we can use integration-by-parts identities (IBPs)~\cite{Tkachov:1981wb,Chetyrkin:1981qh,Laporta:2001dd} to reduce all such one-loop integrals to linear combinations of scalar integrals with unit numerators and powers of the propagators. 
As a consequence, integrals of the type~\eqref{eq:one-loop} form a linear basis (a so-called set of `master integrals') for all one-loop integrals in a given dimension~$D$. However, for our purposes it will be convenient to work with a different one-loop basis where scalar integrals are chosen in different dimensions according to the number of propagators, as we now describe.

It is well known that there are relations between Feynman integrals in dimensions $D$ and $D+2$~\cite{Tarasov:1996br,Bern:1992em,Lee:2009dh}, and conjecturally all linear relations between Feynman integrals arise from either IBP identities or dimensional shift identities. Hence, rather than choosing master integrals for a fixed value of $D$, we can choose different basis integrals to live in different numbers of dimensions. For our purposes, a particularly convenient choice for a one-loop basis is\footnote{We are grateful to V.~Smirnov and R.~Lee for providing us with a rigorous proof of the fact that the $\jtilde_n$ form a basis.}
\beq\label{eq:J_n}
\jtilde_n(\{p_i\cdot p_j\};\{m_i^2\};\eps) \equiv I_n^{D_n}(\{p_i\cdot p_j\};\{m_i^2\};\eps)\,,
\eeq
where
\beq\label{eq:defDn}
D_n=\left\{\begin{array}{ll}
n-2\eps\,, & \textrm{ for } n \textrm{ even}\,,\\
n+1-2\eps\,, & \textrm{ for } n \textrm{ odd}\,.
\end{array}\right.
\eeq
We will usually omit writing  the dependence of a Feynman integral on its arguments and the dimension~$D$. In particular, the dependence on the dimensional regularization parameter $\eps$ will always be implicitly assumed.

This choice of basis is motivated by the expectation that the $\jtilde_n$ have a particularly nice functional form: they are expected to be expressible, up to an overall algebraic prefactor, in terms of multiple polylogarithms of uniform weight. A corollary is that all one-loop integrals can be expressed in terms of polylogarithmic functions. If $\eps$ is formally assigned a weight of $-1$, then all the terms in the Laurent expansion of $\jtilde_n$ have weight $\lceil n/2\rceil$. 
Although we are able to write this basis of one-loop Feynman integrals, very few of the integrals $\tildeJ_n$ are known analytically, and even then usually only through $\ord(\eps^0)$. 
Specifically, we know all one-loop integrals up to $n = 4$ in $D = 4-2\eps$ up to finite terms (see ref.~\cite{Ellis:2007qk} and references therein), as well as a few special cases for $n=5$ and $n=6$ in $D=6$ dimensions~\cite{Dixon:2011ng,DelDuca:2011ne,DelDuca:2011jm,DelDuca:2011wh,Papadopoulos:2014lla,Spradlin:2011wp,Kozlov:2015kol}.

We stress that the discussion in this paper is not specific to any particular quantum field theory, as we will study Feynman integrals and graphs as interesting objects in their own right. The Feynman integrals relevant for any given theory can be written in terms of our basis $\jtilde_n$ through the relations described above. One might worry about some integrals not being defined for some theories, such as tadpole integrals in massless theories, but these issues are settled in dimensional regularization.

\subsection{One-loop cut integrals}
In order to state the results of this paper, we need to consider \emph{cut (Feynman) integrals}. In the following we give a short review of the results of ref.~\cite{Abreu:2017ptx}, which introduced a version of one-loop cut integrals in dimensional regularization.

It is well known that the solutions to the Landau conditions for one-loop integrals~\cite{Landau:1959fi} can be classified into two categories, called singularities of the first and second type, 
corresponding to kinematic configurations where either the Gram determinant $\Gram_C$ or the modified Cayley determinant $Y_C$ vanishes for a subset $C$ of propagators. These determinants are defined in the usual way:
\beq\bsp\label{eq:Gram_def}
\Gram_C &\,= \det\left((q_i-q_*)\cdot(q_j-q_*)\right)_{i,j\in C\setminus{*}}\,,
\esp\eeq
and
\beq\bsp\label{eq:Cayley_def}
Y_C &\,= \det\left(\frac{1}{2}(-(q_i-q_j)^2+m_i^2+m_j^2)\right)_{i,j\in C}\,,
\esp\eeq
where $C\subseteq [n] \equiv \{1,\ldots,n\}$ denotes a subset of propagators and $*$ denotes any particular element of $C$, for example the one with lowest index.

A singularity of the first type corresponds to the vanishing of the modified Cayley determinant $Y_C$ associated to a subset $C$ of propagators. To such a singularity we associate the cut integral $\cC_C\tildeJ_n$, defined as the integral where the integration contour is deformed such as to encircle the poles of the propagators in $C$, effectively putting those propagators on shell. Evaluating the integral in terms of residues, we find the following expression for cut integrals associated to singularities of the first type,
\beq\bsp\label{eq:cutdef}
\mathcal{C}_C&\jtilde_n = \frac{(2\pi i)^{\lfloor n_C/2\rfloor}e^{\gamma_E\eps}}{2^{n_C}\sqrt{\mu^{n_C}Y_C}}\left(\mu\frac{Y_C}{\Gram_C}\right)^{(D_n-n_C)/2}\!\!\!\!
\int\frac{d\Omega_{D_n-n_C}}{i\pi^{D_n/2}}\left[\prod_{j \notin C}\frac{1}{(k-q_j)^2-m_j^2}\right]_{C}\!\!\!\!\!\!\!\mod i\pi,
\esp\eeq
where $n_C=|C|$ is the number of cut propagators, $\mu=+1\, (-1)$ in Euclidean (Minkowskian) kinematics, and $[ \cdot]_{C}$ indicates that the function inside the square brackets is evaluated on the zero locus of the inverse cut propagators. We stress that our cut integrals are defined only modulo $i\pi$, i.e., up to branch cuts.\footnote{A more precise definition is necessary to put our discussion on more solid mathematical footing, but this goes beyond the scope of this paper. 
As we highlight in the conclusion in Section \ref{sec:summary}, this is one issue on which we hope for input from the mathematics community.} As we will see shortly when we introduce the coaction, this  is not a restriction for the results of this paper. Because we have not specified an orientation for the contour,  the overall sign in \refE{eq:cutdef}  is a priori undefined. Relative signs will be fixed by requiring certain relations between cuts, which we discuss below.

A singularity of the second type is attributed to a pinch singularity at large loop momentum and corresponds to the vanishing of one of the Gram determinants $\Gram_C$~\cite{SMatrix,SecondType1,SecondType2}. To this singularity we associate the cut integral $\cC_{\infty C}\tildeJ_n$, defined as the integral over a contour which not only encircles the poles of the propagators in $C$, but also winds around the branch point at infinite loop momentum.\footnote{In dimensional regularization, the singularity at infinity loop momentum is a branch point rather than a pole. See ref.~\cite{Abreu:2017ptx} for more details.} In ref.~\cite{PhamInTeplitz,Abreu:2017ptx} it was shown that a cut integral associated to a singularity of the second type can be written as a linear combination of cut integrals associated to singularities of the first type, as follows.
\begin{itemize}
\item For $n_C$ even,
\beq\label{eq:homo_even}
\cC_{\infty C}\tildeJ_n = \sum_{i\in [n]\setminus C} \cC_{Ci}\tildeJ_n + \sum_{\substack{i,j\in [n]\setminus C\\ i<j}}\cC_{C ij}\tildeJ_n\mod i\pi\,.
\eeq
\item For $n_C$ odd,
\beq\label{eq:homo_odd}
\cC_{\infty C}\tildeJ_n = -2\cC_{C}\tildeJ_n-\sum_{i\in [n]\setminus C} \cC_{Ci}\tildeJ_n \mod i\pi\,.
\eeq
\end{itemize}
In the case where $C=\emptyset$, it is possible to compute explicitly the cut integral associated to the Landau singularity of the second type, $\cC_{\infty}\tildeJ_n = -\eps\,\tildeJ_n$. This implies a remarkable relation between the single and double cuts of an integral that will play an important role throughout this paper,
\beq\label{eq:PCI}
\sum_{i\in [n]} \cC_{i}\tildeJ_n + \sum_{\substack{i,j\in [n]\\ i<j}}\cC_{ij}\tildeJ_n = -\eps\,\tildeJ_n\mod i\pi\,.
\eeq

Equations~(\ref{eq:homo_even}) through~(\ref{eq:PCI}) make it clear that cut integrals are not all linearly independent. In ref.~\cite{Abreu:2017ptx} it was argued that there are two types of linear relations among one-loop cut integrals:
\begin{enumerate}
\item Linear relations among cut integrals with different propagators but the same set of cut propagators. Relations of this type are consequences of integration-by-parts (IBP) and dimensional shift identities for uncut integrals, with the restriction that an integral vanishes if a cut propagator is raised to a nonnegative power. This is similar to the reverse-unitarity principle of ref.~\cite{Anastasiou:2002yz,Anastasiou:2003yy,Anastasiou:2002wq,Anastasiou:2002qz,Anastasiou:2003ds}.
\item Linear relations among cut integrals with the same propagators but a different set of cut propagators. These are a consequence of relations among the generators of the homology groups associated to one-loop integrals. Equations~\eqref{eq:homo_even} and~\eqref{eq:homo_odd} are examples of such relations.
\end{enumerate}
The fact that one-loop cut integrals satisfy IBP and dimensional shift identities implies that the basis $\tildeJ_n$ of one-loop integrals defined in eq.~\eqref{eq:J_n} can immediately be lifted to a basis $\cC_C\tildeJ_n$ of one-loop cut integrals, while eq.~\eqref{eq:homo_even} and~\eqref{eq:homo_odd} imply that the basis may be chosen to contain  only cut integrals associated to Landau singularities of the first type. The basis  $\cC_C\tildeJ_n$ solves the same linear system of differential equations as their uncut analogues \cite{Abreu:2017ptx,Frellesvig:2017aai,Zeng:2017ipr,Bosma:2017ens}, and so one-loop cut integrals can always be expressed in terms of multiple polylogarithms. This agrees with the fact that cut integrals compute discontinuities of Feynman integrals~\cite{Cutkosky:1960sp,tHooft:1973pz,Veltman:1994wz}.
Let us also mention that for $n_C\le 3$, it is possible to state some sufficient conditions for a one-loop cut integral to vanish~\cite{Abreu:2017ptx}:
\begin{enumerate}
\item If $n_C=1$, the cut integral vanishes if the cut propagator is massless.
\item If $n_C=2$, the cut integral vanishes if the total momentum flowing through the cut propagators is lightlike.
\item If $n_C=3$, the cut integral vanishes if any pair of the cut propagators isolate a three-point vertex where three massless lines (external and internal) meet.
\end{enumerate}

In the previous section, we mentioned that the basis integrals $\tildeJ_n$ are polylogarithmic functions of uniform weight, up to an overall algebraic prefactor. This prefactor is related to the maximal cut of the integrals in integer dimensions. We define 
\beq\label{eq:LS_J_n}
j_n \equiv \lim_{\eps\to 0}\mathcal{C}_{[n]}\tildeJ_n= \left\{\begin{array}{ll}
\displaystyle 2^{1-n/2}\,i^{n/2}\,Y_{[n]}^{-1/2}\,, & \textrm{ for } n \textrm{ even}\,,\\
\displaystyle 2^{(1-n)/2}\,i^{(n-1)/2}\Gram_{[n]}^{-1/2}\,, & \textrm{ for } n \textrm{ odd}\,.
\end{array}\right.
\eeq
The reason for the normalization in eq.~\eqref{eq:LS_J_n} will become clear in the next section.
It will often be convenient to normalize the basis integrals~\eqref{eq:J_n} to their maximal cut, resulting in integrals $J_n$ defined as 
\beq
{J}_n = \jtilde_n\,/\,j_n\,.
\eeq
The quantities $J_n$ still form a basis of all one-loop Feynman integrals, but unlike the $\jtilde_n$, they are \emph{pure} functions~\cite{ArkaniHamed:2010gh}. Consequently, the basis $J_n$ satisfies a particularly nice linear system of first-order differential equations~\cite{Kotikov:1990kg,Kotikov:1991pm,Kotikov:1991hm,Gehrmann:1999as,Spradlin:2011wp,Henn:2013pwa}.

\subsection{Review of multiple polylogarithms}
\label{subsec:MPLs}
Since we have argued that all one-loop Feynman integrals, as well as their cuts, can be expressed in terms of multiple polylogarithms (MPLs), we now give a short review of the mathematics of MPLs and their algebraic properties.

Multiple polylogarithms are defined by the iterated integral~\cite{Lappo:1927,Goncharov:1998kja,GoncharovMixedTate}
 \beq\label{eq:Mult_PolyLog_def}
 G(a_1,\ldots,a_n;z)=\,\int_0^z\,\frac{d t}{t-a_1}\,G(a_2,\ldots,a_n;t)\,,\\
\eeq
with algebraic $a_i$ and $z$. In the special case where all the $a_i$'s are zero, we define
\beq
G(\underbrace{0,\ldots,0}_{n\textrm{ times}};z) = \frac{1}{n!}\,\ln^n z\,.
\eeq
The number $n$ of integrations in eq.~\eqref{eq:Mult_PolyLog_def}, or equivalently the number of $a_i$'s, is called the \emph{weight} of the multiple polylogarithm.
In the following we denote by $\cA$ the $\mathbb{Q}$-vector space spanned by all multiple polylogarithms.
In addition, $\cA$ can be turned into an algebra. Indeed, iterated integrals form a \emph{shuffle algebra},
\beq
G(\vec a_1;z)\,G(\vec a_2;z) = \sum_{\vec a\,\in\,\vec a_1\sh \vec a_2}G(\vec a;z)\,,
\eeq
where $\vec a_1\sh \vec a_2$ denotes the set of all shuffles of $\vec a_1$ and $\vec a_2$, i.e., the set of all permutations of their union that preserve the relative orderings inside $\vec a_1$ and $\vec a_2$. It is obvious that the shuffle product preserves the weight, and hence the product of two multiple polylogarithms of weights $n_1$ and $n_2$ is a linear combination of multiple polylogarithms of weight $n_1+n_2$.

Multiple polylogarithms can be endowed with more algebraic structures (see e.g.~\cite{Goncharov:2005sla,GoncharovMixedTate}). If we look at the quotient space $\cH = \cA/(i\pi\,\cA)$ (the algebra $\cA$ modulo $i\pi$), then $\cH$ is conjectured to form a Hopf algebra. In particular, $\cH$ can be equipped with a coassociative coproduct $\Delta_{\rm MPL}$ that respects the multiplication and the weight. 
While $\cH$ is a Hopf algebra, we are practically interested in the full algebra $\cA$ where we have kept all factors of $i\pi$. We can reintroduce $i\pi$ into the construction by considering the trivial comodule $\cA=\mathbb{Q}[i\pi]\otimes\cH$. The coproduct is then lifted to a coaction\footnote{By abuse of notation, we denote both the coproduct on $\cH$ and the coaction on $\cA$ by $\Delta_{\rm MPL}$.} $\Delta_{\rm MPL}:\cA\to \cA\otimes \cH$ which coacts on $i\pi$ according to~\cite{Brown:2011ik,Duhr:2012fh}
\beq\label{eq:Delta_ipi}
\Delta_{\rm MPL}(i\pi)=i\pi\otimes1\,.
\eeq 
Moreover, 
$\mathcal{A}$ and $\mathcal{H}$ are in fact conjectured to be graded algebras, because the product of two MPLs of weights $w_1$ and $w_2$ naturally has  weight $w_1+w_2$, and the action of the coaction $\Delta_{\rm MPL}$ respects the weight. This conjecture is supported by all known relations among MPLs and multiple $\zeta$-values. 
It allows us to define $\Delta_{w_1,w_2}$ as the operator that selects the terms in the coaction of weight $w_1$ in the first entry and $w_2$ in the second entry, and we refer to this sum of terms as the $(w_1,w_2)$ component of the coaction.

Next, let us discuss how differentiation and taking discontinuities interact with the coaction. One can show that derivatives only act on the last entry of the coaction while discontinuities only act on the first entry,
\begin{align}
\label{eq:disc_coproduct}
\Delta_{\rm MPL}\,\textrm{Disc} & = (\textrm{Disc}\otimes\textrm{id})\,\Delta_{\rm MPL}\,,\\
\label{eq:der_coproduct}
\Delta_{\rm MPL}\,\frac{\partial}{\partial z} & = \left(\textrm{id}\otimes\frac{\partial}{\partial z}\right)\,\Delta_{\rm MPL}\,.
\end{align}

\begin{figure}[!t]
\begin{center}
\includegraphics[scale=1.0]{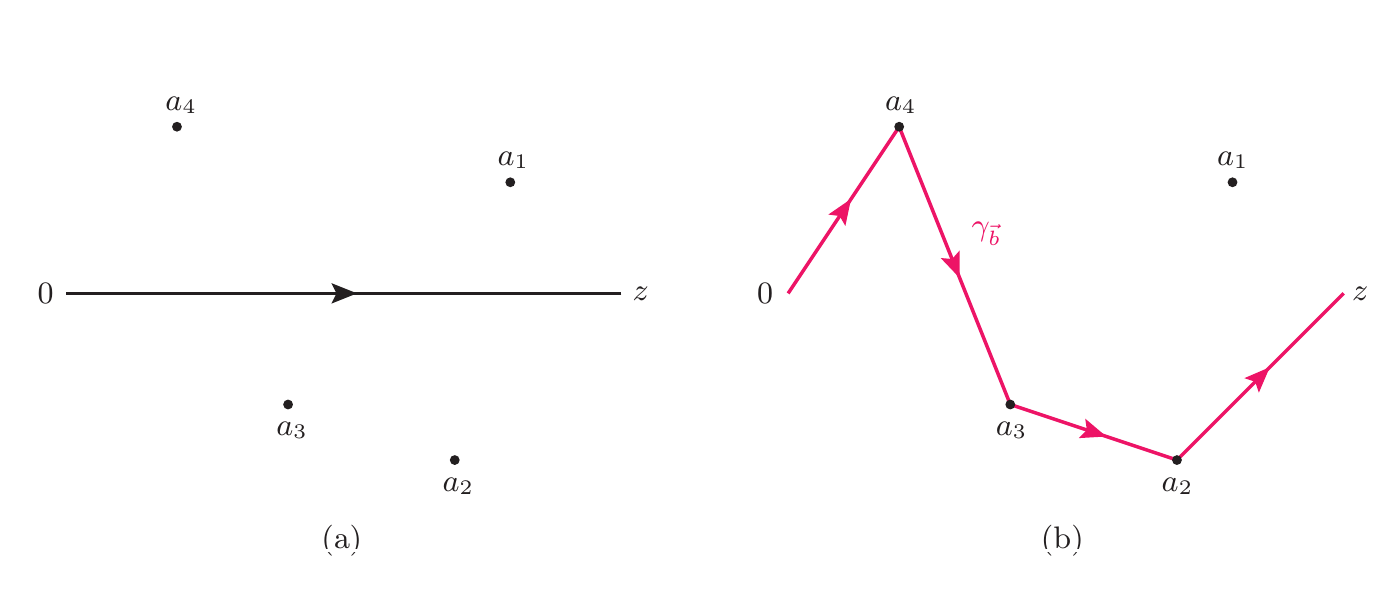}
\end{center}
\caption{\label{fig:paths}(a) The integration path for $G(a_1,a_2,a_3,a_4;z)$. (b) The integration path for $G_{\vec b}(a_1,a_2,a_3,a_4;z)$ for $\vec b=(a_2,a_3,a_4)$. The path $\gamma_{\vec b}$ encircles each of the singularities $a_i\in \vec b$ counter-clockwise, in the order in which they appear in $\vec b$.}
\end{figure}

The coaction on $G(\vec a;z)$ can be written in the following suggestive way, at least in the generic case, i.e., when the arguments take generic values,
\beq\label{eq:Delta_MPL}
\Delta_{\rm MPL}(G(\vec a;z)) = \sum_{\vec b\subseteq \vec a} G(\vec b;z)\otimes G_{\vec b}(\vec a;z)\,,
\eeq
where the sum runs over all order-preserving subsets $\vec b$ of $\vec a$, including the empty set. $G_{\vec b}(\vec a;z)$ denotes the iterated integral with the same integrand as $G(\vec a;z)$, but integrated over the contour $\gamma_{\vec b}$ that encircles the singularities at the points $z=a_i$, $a_i\in \vec b$, in the order in which the elements appear in $\vec b$. This is equivalent to taking the residues at these points, and we divide by $2\pi i$ per residue (see fig.~\ref{fig:paths}). We see that the coaction on MPLs has a very simple combinatorial interpretation: the different terms in this sum correspond to the tensor product of the MPL with a restricted set of poles and the integral over the differential form obtained by taking the residues at those poles.

The operation of taking residues has a direct analogue in terms of Feynman integrals. The residues at the propagators of a Feynman integral are naturally identified with the cuts of the integral, where some of the propagators have been put on shell. Since all one-loop Feynman integrals are expressible in terms of MPLs order by order in the dimensional regulator, it is natural to ask whether the coaction on one-loop Feynman integrals admits a similarly simple combinatorial description. In the rest of this paper we give evidence that this is indeed the case, and we conjecture a formula for the coaction on one-loop Feynman integrals which is purely diagrammatic in nature and very reminiscent of eq.~\eqref{eq:Delta_MPL}. Before stating this conjecture in Section~\ref{sec:general_formulation}, we introduce and motivate our construction in the next section using some simple examples of one-loop integrals.

%


\section{First examples of the diagrammatic representation of the coaction}
\label{sec:firstExamples}

In this section, we present some simple examples  as  motivation for the main conjecture. We investigate some one-loop Feynman integrals with up to three propagators, and we show that, empirically, we can rearrange the terms in the coaction in such a way that it can be written entirely in terms of Feynman graphs and their cuts. These examples illustrate all the main features of the general conjecture.

{
Although $\Delta_{\rm MPL}$ was defined in the previous section as acting on MPLs, throughout this section and the rest of the paper we will write expressions where  $\Delta_{\rm MPL}$ acts on one-loop Feynman integrals.  In dimensional regularization, these integrals are not MPLs, but the coefficients in their Laurent expansion in $\eps$ are. The action of  $\Delta_{\rm MPL}$ should thus be understood order by order in $\eps$.
}

\subsection{The tadpole integral}

We start with the tadpole integral in $D=2-2\epsilon$ dimensions, given by
\begin{equation}
	\jtilde_1\left(m^2\right)=e^{\gamma_E\eps}\int\frac{d^Dk}{i\pi^{D/2}}\,\frac{1}{k^2-m^2+i0} = -\frac{e^{\gamma_E\epsilon}\Gamma(1+\epsilon)\left(m^2\right)^{-\epsilon}}{\epsilon}\,.
\end{equation}
There is only one cut integral associated to its Landau singularities of the first type, because there is only one propagator that can be cut. This cut integral is given by (see \refE{eq:cut_tadpole}) 
\begin{equation}
	\cutRes_e \jtilde_1\left(m^2\right)=\frac{e^{\gamma_E\epsilon}\left(-m^2\right)^{-\epsilon}}{\Gamma(
	1-\epsilon)}\,,
\end{equation}
where the subscript $e$ labels the propagator. Normalizing with \refE{eq:LS_J_n} where now $j_1(m^2) = 1$, we have
\beq
J_1 \left(m^2\right)=-\frac{e^{\gamma_E\epsilon}\Gamma(1+\epsilon)\left(m^2\right)^{-
	\epsilon}}{\epsilon} {\rm~~and~~} \cC_eJ_1 \left(m^2\right)=\frac{e^{\gamma_E\epsilon}\left(-m^2\right)^{-\epsilon}}{\Gamma(
	1-\epsilon)}\,.
\eeq
It will be convenient to introduce a diagrammatic representation for the integrals $J_n$ and their cuts. We represent $J_n$ by its underlying Feynman graph, and we label the edges by the propagators (the label may be omitted if no confusion arises). Thick lines in a graph represent off-shell external lines or massive propagators. Cut edges in a cut graph $\cC_CJ_n$ are represented by a (red) dashed line through the propagator. For example, in the  case of the tadpole, we have
\begin{equation}
	\tadInsert[tadEdge]\empty \equiv
	J_1 \left(m^2\right) {\rm~~and~~}\tadInsert[tadEdgeCutM] \equiv\cC_eJ_1 \left(m^2\right)\,.
\end{equation}

Next, let us analyze the image of the tadpole under the coaction $\Delta_{\rm MPL}$. Using the identities
 \beq\bsp\label{eq:relationsForTadpole}
 \Delta_{\rm MPL}\left[\ln^n z\right]&\,=\sum_{k=0}^{n}\binom{n}{k}\ln^kz\otimes\ln^{n-k}z\,,\\
 e^{\gamma_E\eps}\Gamma(1+\eps) &\,= \exp\left[\sum_{k=2}^\infty\frac{(-\eps)^k}{k}\,\zeta_k\right]\,,\\
 \Gamma(1-\eps)\Gamma(1+\eps) &\,= \frac{\pi\eps}{\sin(\pi\eps)} = 1+\frac{(\pi\eps)^2}{6}+\mathcal{O}\left((\pi\eps)^4 \right)=1\mod i\pi\,,
 \esp\eeq
 one can easily show that
 \begin{align}
 \begin{split}
 \Delta_{\rm MPL}\left[(m^2)^{-\eps}\right] &= (m^2)^{-\eps}\otimes (m^2)^{-\eps}\,,\\ \Delta_{\rm MPL} \left[e^{\gamma_E\eps}\Gamma(1+\eps)\right]  &= \left(e^{\gamma_E\eps}\Gamma(1+\eps)\right)\otimes \left(e^{\gamma_E\eps}\Gamma(1+\eps)\right)\\
 &= \left(e^{\gamma_E\eps}\Gamma(1+\eps)\right)\otimes \frac{e^{\gamma_E\eps}}{\Gamma(1-\eps)}
 \,,
 \end{split}
 \end{align}
 where these identities hold order by order in the $\eps$ expansion. We have used the last relation of eq.~\eqref{eq:relationsForTadpole} in the rightmost coaction component since it is defined mod $i\pi$, as indicated in  eq.~\eqref{eq:Delta_ipi} and the preceding discussion. Hence,\footnote{The fact that $\cC_eJ_1$ contains a power of $-m^2$ while $J_1$ contains a power of $m^2$ is because their physical interpretations are most natural in different kinematic regions, which plays a role in the discussion of discontinuities below. It has no effect on the coaction formula, since $(m^2)^{-\eps}\otimes (-m^2)^{-\eps}=(m^2)^{-\eps}\otimes (m^2)^{-\eps}$.} we can write
\begin{equation}
	\Delta_{\rm MPL}\left[J_1\left(m^2\right)\right]=
	J_1\left(m^2\right)\otimes \cutRes_eJ_1\left(m^2\right)\,,
\end{equation}
or in terms of graphs,
\begin{equation}\label{graphCoprod_tad}
	\Delta_{\rm MPL}\left[\tadInsert[tadEdge]\empty \right]=\tadInsert[tadEdge]\otimes\tadInsert[tadEdgeCutM]\,.
\end{equation}
While the derivation of eq.~\eqref{graphCoprod_tad} is easy, this result is instructive in illustrating some general features. First, it is remarkable that the coaction on $J_1\left(m^2\right)$ can be written in a compact form, completely and to all orders in epsilon, in terms of (cut) Feynman graphs. Indeed, we have to interpret $J_1\left(m^2\right)$ as an infinite Laurent series in dimensional regularization, and the coaction in the left-hand side acts order by order in the expansion in $\eps$.
 The right-hand side of eq.~\eqref{graphCoprod_tad}, however, is a tensor product of two Laurent series, and it is a priori not at all clear that the two sides should agree. 
We note that the second entry on the right involves only information about the underlying cut graph.
Also, we observe that  the  second entry contains the single cut of the tadpole, while the first entry contains the tadpole integral itself. This is consistent with the fact that the single-cut integral corresponds to a discontinuity in the mass of the cut propagator~\cite{Abreu:2015zaa},
\beq
\textrm{disc}_{m^2}J_1(m^2) = \theta(-m^2)\, \cC_eJ_1(m^2)\mod i\pi\,,
\eeq
where the discontinuity operator $\textrm{disc}_x$ is defined by
\beq
\textrm{disc}_xf(x) = \frac{1}{2\pi i}\,\lim_{\eta\to 0}\left[f(x+i\eta)-f(x-i\eta)\right]\,.
\eeq
In particular, it is then easy to show that eq.~\eqref{graphCoprod_tad} is consistent with eq.~\eqref{eq:disc_coproduct}:
\beq\bsp
\Delta_{\rm MPL}\left[\textrm{disc}_{m^2}J_1(m^2)\right] &\,= \theta(-m^2)\, \Delta\cC_e\left[J_1(m^2)\right]\\
&\,= \theta(-m^2)\, \cC_eJ_1(m^2)\otimes \cC_eJ_1(m^2)\\
&\, = \left[\textrm{disc}_{m^2}J_1(m^2)\right]\otimes \cC_eJ_1(m^2)\,.
\esp\eeq
Similarly, we can check that eq.~\eqref{graphCoprod_tad} is consistent with eq.~\eqref{eq:der_coproduct} under differentiation. Indeed, we have
\beq
m^2\frac{\partial}{\partial m^2}J_1(m^2) = -\eps\,J_1(m^2) {\rm~~and~~}
m^2\frac{\partial}{\partial m^2}\cC_e\left[J_1(m^2)\right] = -\eps\,\cC_eJ_1(m^2)\,,
\eeq
and so we find
\beq\bsp
\Delta_{\rm MPL}\left[m^2\frac{\partial}{\partial m^2}J_1(m^2)\right] &\,= -\eps\, \Delta_{\rm MPL} \left[J_1(m^2)\right]\\
&\,= -\eps\, J_1(m^2)\otimes \cC_eJ_1(m^2)\\
&\, = J_1(m^2)\otimes \left[m^2\frac{\partial}{\partial m^2}\cC_eJ_1(m^2)\right]\,.
\esp\eeq

This simple example demonstrates some important features common to all the cases studied in this paper: the coaction on a one-loop integral can be written such that all the first entries are Feynman integrals, while the second entries are cuts of the original integral.

\subsection{The massless bubble integral}
As a second example, let us consider the bubble integral $\jtilde_2\left(p^2\right)$ with two massless propagators in $D=
2-2\epsilon$ dimensions. The bubble with massive propagators will be considered in \refS{sec:bubmassprop}. This integral admits a simple analytic form,
\begin{equation}
	\jtilde_2\left(p^2\right)=-\frac{2c_\Gamma}{\epsilon}\left(-p^2\right)^{-1-\epsilon},
\end{equation}
where we have defined $c_\Gamma$ as the usual ratio of $\Gamma$ functions,
\begin{equation}
	c_\Gamma=\frac{e^{\gamma_E\epsilon}\Gamma^2(1-\epsilon)\Gamma(1+\epsilon)}{\Gamma(1-2\epsilon)}\,.
\end{equation}
Both single-propagator cuts of this integral vanish because the propagators are massless~\cite{Abreu:2017ptx}, and only its two-propagator cut is nonzero,
\begin{equation}
	\cutRes_{e_1 e_2}\jtilde_2\left(p^2\right)=-2\frac{e^{\gamma_E\epsilon}\Gamma(1-\epsilon)}{\Gamma(1-2\epsilon)}
	\left(p^2\right)^{-1-\epsilon}\,.
\end{equation}
The integral, and its cut normalized to its maximal cut $j_2$ as defined in \refE{eq:LS_J_n}, are 
\begin{equation}\bsp\label{eq:bubFG}
	\bubInsertHigh[bub0m]&\,=
	J_2\left(p^2\right)=
	-\frac{c_\Gamma}{\epsilon}\left(-p^2\right)^{-\epsilon}\,,\\
	\bubInsert[bub0mCut]&\,=
	\cC_{e_1e_2}J_2\left(p^2\right)=\frac{e^{\gamma_E\epsilon}\Gamma(1-\epsilon)}{\Gamma(1-2\epsilon)}
	\left(p^2\right)^{-\epsilon}\,.	
\esp\end{equation}
Using the same reasoning as for the tadpole, it is easy to check that we can rearrange all the terms in the coaction on $J_2(p^2)$ into the form
\begin{equation}
	\Delta_{\rm MPL}\left[J_2\left(p^2\right)\right]=J_2\left(p^2\right)\otimes
	\cutRes_{e_1 e_2} J_2\left(p^2\right)\,,
\end{equation}
or, in terms of graphs,
\begin{equation}\label{diagConjBub}
	\Delta_{\rm MPL}\left[\bubInsertHigh[bub0m]\right]
	=\bubInsertHigh[bub0m]
	\otimes\bubInsert[bub0mCut]\,.
\end{equation}
It is easy to check that eq.~\eqref{diagConjBub} is consistent with the derivatives and the discontinuities of the $J_2(p^2)$ upon noting that $J_2(p^2)$ has a single branch cut for $p^2>0$, and the discontinuity across the cut is~\cite{Landau:1959fi,Cutkosky:1960sp,tHooft:1973pz}
\beq
\textrm{disc}_{p^2}J_2(p^2) = -\theta(p^2)\,\cC_{e_1 e_2}J_2(p^2)\,.
\eeq

\subsection{The triangle integral with three massive external legs}
\label{sec:tp1p2p3}
Since the previous examples were one-scale integrals, their functional dependence on the scale was very simple. We now analyze the scalar triangle with three external scales, but no internal propagator masses, computed in $D=4-2\epsilon$
as per \refE{eq:defDn}.
The external masses are denoted $p_i^2$, with $i\in\{1,2,3\}$, and the integral evaluates to
\begin{equation}
\label{J3tild_LO}
	\jtilde_3\left(p_1^2,p_2^2,p_3^2\right)=
	\frac{(-p_1^2)^{-1-\epsilon}}{(\zz-\zbar)}\left[\mathcal{T}(\zz,\zbar)
	+\mathcal{O}(\epsilon)\right]\,.
\end{equation}
The function $\mathcal{T}(\zz,\zbar)$ is given by
\beq
\mathcal{T}(\zz,\zbar) = -2\Li_2(\zz)+2\Li_2(\zbar)-\ln(\zz\zbar)
	\ln\left(\frac{1-\zz}{1-\zbar}\right)\,.
\eeq
Higher-order terms in the $\eps$ expansion in eq.~\eqref{J3tild_LO} are also known~\cite{Birthwright:2004kk,Chavez:2012kn}, but for simplicity we focus on the leading term in the expansion.
The arguments of $\mathcal{T}(\zz,\zbar)$ are the two dimensionless variables
$\zz$ and $\zbar$ defined by
\beq
\label{eq:ZZbarDef}
\zz\,\zbar = \frac{p_2^2}{p_1^2} {\rm~~and~~} (1-\zz)\,(1-\zbar) = \frac{p_3^2}{p_1^2}\,.
\eeq
It is easy to write down the coaction on $\mathcal{T}(\zz,\zbar)$. We find
\begin{align}
\label{eq:trianglecoproduct}
	\Delta_{\rm MPL}\left[\mathcal{T}(\zz,\zbar)\right]
	=\,&\mathcal{T}(\zz,\zbar)\otimes1 + 1\otimes\mathcal{T}(\zz,\zbar) +
	\ln(\zz\bar\zz)\otimes \ln\frac{1-\bar\zz}{1-\zz}\\
\nonumber	\,&+ \ln[(1-\zz)(1-\bar\zz)] \otimes \ln\frac{\zz}{\bar\zz}\nonumber \\
\nonumber	=\,& \mathcal{T}(\zz,\zbar)\otimes1 + 1\otimes\mathcal{T}(\zz,\zbar) +\ln\left(-p_2^2\right)\otimes
	\ln\frac{1-\bar\zz}{1-\zz} + \ln\left(-p_3^2\right)\otimes \ln\frac{\zz} {\bar\zz}\\
\nonumber	\,& + \ln (-p_1^2) \otimes  \ln\frac{\zbar(1-\zz)}{\zz(1-\zbar)}\,,
\end{align}
where the second line displays the \emph{first-entry condition}~\cite{Gaiotto:2011dt,Abreu:2014cla}, which states that in the case of massless propagators it is possible to arrange the terms in the coaction in such a way that all left-most entries of weight one in the coaction are logarithms of Mandelstam invariants.

We want to see that the coaction in eq.~\eqref{eq:trianglecoproduct} can be written entirely in terms of Feynman integrals and their cuts. In the following, we concentrate on the leading term in the $\eps$ expansion, but we have checked explicitly that the conclusion holds at least through weight $4$. In order to understand the analysis, it is convenient to group the terms in \refE{eq:trianglecoproduct} into three categories: the term $\mathcal{T}(\zz,\zbar)\otimes1$, the term
$1\otimes\mathcal{T}(\zz,\zbar)$, and the rest.  We call the first two of these categories \emph{trivial}.

We start by interpreting the term $\mathcal{T}(\zz,\zbar)\otimes1$ in the coaction.
Working with the normalized integral $J_3$, we have 
\begin{equation}\label{eq:tp1p2p3FG}
	\triInsert[tp1p2p3LabelEdges]=
	J_3=\mathcal{T}(\zz,\zbar)+\mathcal{O}(\epsilon)\, ,
\end{equation}
and
\begin{equation}\label{eq:tp1p2p3Cut123FG}
	\triInsert[tp1p2p3Cut123LabelEdges]=\mathcal{C}_{e_1e_2e_3}J_3=1+\mathcal{O}(\epsilon)\,,
\end{equation}
where the integers labeling the external edges refer to the external momentum flowing through that edge.
If we assume that there exists a representation of the coaction on this triangle similar to the cases of the tadpole and bubble, then we would expect that the term $\mathcal{T}(\zz,\zbar)\otimes1$ in the coaction of eq.~(\ref{eq:trianglecoproduct}) should be reproduced using the three-propagator cut of eq.~(\ref{eq:tp1p2p3FG}) in the second entry. Indeed we see that
\begin{equation}
	\triInsert[tp1p2p3LabelEdges]\otimes\triInsert[tp1p2p3Cut123LabelEdges]
	=\mathcal{T}(\zz,\zbar)\otimes1+\mathcal{O}(\epsilon)\,.
\end{equation}

Next, let us analyze the terms in eq.~\eqref{eq:trianglecoproduct} that have a logarithm in both entries, 
\beq\label{eq:tri_1_1}
\Delta_{1,1}\left[\mathcal{T}(\zz,\zbar)\right] = \ln (-p_1^2) \otimes  \ln\frac{\zbar(1-\zz)}{\zz(1-\zbar)} +
\ln\left(-p_2^2\right)\otimes
	\ln\frac{1-\bar\zz}{1-\zz} + \ln\left(-p_3^2\right)\otimes \ln\frac{\zz} {\bar\zz}\,.
\eeq
Following the same logic as for the tadpole and bubble integrals, we would expect that the logarithms in the second entry are related to the discontinuities of the triangle in one of the external scales, i.e., they should correspond to the two-propagator cuts of the triangle. Indeed, computing the two-propagator cuts one obtains: 
\begin{align}\bsp\label{eq:tp1p2p3CutsFG}
	\triInsert[tp1p2p3CutP1LabelEdges]=
	\cutRes_{e_1e_2}J_3&=
	\ln\frac{\zbar(1-\zz)}{\zz(1-\zbar)}+\mathcal{O}(\epsilon)\, ,\\
	\triInsert[tp1p2p3CutP2LabelEdges]=
	\cutRes_{e_2e_3}J_3&=
	\ln\frac{1-\zbar}{1-\zz}+\mathcal{O}(\epsilon)\, ,\\
	\triInsert[tp1p2p3CutP3LabelEdges]=
	\cutRes_{e_1e_3}J_3&= \ln\frac{\zz}{\zbar}+\mathcal{O}(\epsilon)\, .
\esp\end{align}
We expect that the logarithms in the left-hand side of eq.~\eqref{eq:tri_1_1} arise from Feynman integrals that have a discontinuity precisely when the logarithms develop an imaginary part. The natural choice for such a Feynman integral is a bubble integral, 
\begin{equation}\label{eq:bubtp1p2p3FG}
	\bubInsertHigh[bub0m1]
	=J_2\left(p_1^2\right)=-\frac{1}{\epsilon}+\ln(-p_1^2)+\mathcal{O}(\epsilon)\,.
\end{equation}
Ignoring for the moment the fact that the bubble is divergent, we see that we can indeed write
\begin{align}\bsp
	\Delta_{1,1}\left[\mathcal{T}(\zz,\zbar)\right]
	=&\bubInsert[bub0m1LabelEdges]\big\vert_{\epsilon^0}\otimes
	\triInsert[tp1p2p3CutP1LabelEdges]\Bigg\vert_{\epsilon^{0}}
	+\bubInsert[bub0m2LabelEdges]\big\vert_{\epsilon^0}\otimes
	\triInsert[tp1p2p3CutP2LabelEdges]\Bigg\vert_{\epsilon^{0}}\\
	&+\bubInsert[bub0m3Edges]\big\vert_{\epsilon^0}\otimes
	\triInsert[tp1p2p3CutP3LabelEdges]\Bigg\vert_{\epsilon^{0}}\,,
\esp\end{align}
where $\left.X\right|_{\eps^k}$ denotes the coefficient of $\eps^k$ in the Laurent expansion of $X$.

Finally, let us turn to the 
 term $1\otimes\mathcal{T}(\zz,\zbar)$ in eq.~\eqref{eq:trianglecoproduct}. 
To see how this term arises, we rely on \refE{eq:PCI}, which relates the result of a Feynman integral to a specific sum of its cuts: 
\begin{equation}\label{eq:poleCancelationTp1p2p3}
	\triInsert[tp1p2p3CutP1LabelEdges]\Bigg\vert_{\epsilon^{n}}
	+\triInsert[tp1p2p3CutP2LabelEdges]\Bigg\vert_{\epsilon^{n}}
	+\triInsert[tp1p2p3CutP3LabelEdges]\Bigg\vert_{\epsilon^{n}}
	=-\triInsert[tp1p2p3LabelEdges]\Bigg\vert_{\epsilon^{n-1}}\mod i\pi\,.
\end{equation}
The term $1\otimes\mathcal{T}(\zz,\zbar)$ is then reproduced from eq.~\eqref{eq:poleCancelationTp1p2p3} with $n=1$,
\begin{align}\bsp
	1\otimes\mathcal{T}(\zz,\zbar)=&
	\bubInsert[bub0m1LabelEdges]\big\vert_{\epsilon^{-1}}\otimes\triInsert[tp1p2p3CutP1LabelEdges]\Bigg\vert_{\epsilon^{1}}
	+\bubInsert[bub0m2LabelEdges]\big\vert_{\epsilon^{-1}}\otimes\triInsert[tp1p2p3CutP2LabelEdges]\Bigg\vert_{\epsilon^{1}}\\
	&+\bubInsert[bub0m3Edges]\big\vert_{\epsilon^{-1}}\otimes\triInsert[tp1p2p3CutP3LabelEdges]\Bigg\vert_{\epsilon^{1}}.
\esp\end{align}
Note that this relation shows how the poles introduced by the bubble integrals cancel, but also that the pole in eq.~\eqref{eq:bubtp1p2p3FG} is actually essential to reproduce the correct term $1\otimes\mathcal{T}(\zz,\zbar)$. 

Putting everything together, we have shown that, at least through $\ord(\eps^0)$, we can write
\beq\bsp
	\Delta_{\rm MPL}&\left[J_3\left(p_1^2,p_2^2,p_3^2\right)\right]\,=J_2\left(p_1^2\right)\otimes
	\cutRes_{e_1e_2} J_3\left(p_1^2,p_2^2,p_3^2\right)\\
	&\qquad+J_2\left(p_2^2\right)\otimes
	\cutRes_{e_2e_3} J_3\left(p_1^2,p_2^2,p_3^2\right)+J_2\left(p_3^2\right)\otimes \cutRes_{e_1e_3} J_3\left(p_1^2,p_2^2,p_3^2\right)
	\\
	&\qquad+J_3\left(p_1^2,p_2^2,p_3^2\right)\otimes \cutRes_{e_1e_2e_3}J_3\left(p_1^2,p_2^2,p_3^2\right)\,,
\esp\eeq
or in terms of graphs,
\begin{align}\bsp\label{eq:diagCoprodThreeMass}
	\Delta_{\rm MPL}&\left[\triInsert[tp1p2p3LabelEdges]\right]=
	\bubInsert[bub0m1LabelEdges]\otimes\triInsert[tp1p2p3CutP1LabelEdges]
	+\bubInsert[bub0m2LabelEdges]\otimes\triInsert[tp1p2p3CutP2LabelEdges]\\
	&\,+\bubInsert[bub0m3Edges]\otimes\triInsert[tp1p2p3CutP3LabelEdges]
	+
	\triInsert[tp1p2p3LabelEdges]\otimes\triInsert[tp1p2p3Cut123LabelEdges].
\esp\end{align}
This equation has the same structure as the equivalent ones for the tadpole and bubble integrals.
We sum over all possible ways to select a subset of the propagators. The first factor in the coaction is then the Feynman integral with this subset of propagators, while the second entry corresponds to the cut of the original integral, where precisely the set of propagators that appear in the first factor are cut. We can also check that eq.~\eqref{eq:diagCoprodThreeMass} agrees with eqs.~\eqref{eq:disc_coproduct} and~\eqref{eq:der_coproduct}. Finally, we repeat that, although we have only discussed eq.~\eqref{eq:diagCoprodThreeMass} through  finite terms in the $\epsilon$ expansion, we have verified that it continues to hold at higher orders  (up to weight 4, i.e., order $\eps^2$), and we conjecture that it holds to all orders. Higher orders in $\epsilon$ for the cut integrals can be obtained by expanding the all-order expressions given in Appendix~\ref{sec:tp1p2p3Res}, and for the uncut integrals from refs.~\cite{Birthwright:2004kk,Chavez:2012kn}.

From the three cases that we have seen so far,
 a pattern emerges in the coaction on one-loop integrals: in all cases the left factor is constructed
by pinching all uncut propagators of the right factor. (The absence of single-propagator cuts 
in the last two examples
follows from the fact that both the single cut of a massless propagator and
the corresponding pinched graph,
the massless tadpole, are zero in dimensional regularization.)
Despite its validity in the above examples, it turns out that this simple rule is not correct for general
one-loop graphs. In the next section we show an example where it fails, and we explain how the
rule for the diagrammatic coaction should be extended.

\subsection{The bubble integral with massive propagators}
\label{sec:bubmassprop}

Let us consider the bubble integral with two propagators with masses $m_1^2$ and $m_2^2$.
This integral is finite in $D=2-2\epsilon$
dimensions. It is convenient to introduce
 variables $w$ and $\wbar$ satisfying the relations 
\begin{equation}\label{eq:defWWbar}
	w \,\wbar=\frac{m_1^2}{p^2}\,,\qquad (1-w)(1-\wbar)=\frac{m_2^2}{p^2}\,.
\end{equation}
The bubble with massive propagators is finite in two dimensions, and the coaction on the leading term in the $\eps$ expansion is
(see eqs.~\eqref{eq:bubM1M2} and \eqref{bubM1M2LS})
\begin{equation}\label{eq:Delta_jt2}
	\Delta_{\rm MPL}\left[J_2\right]=
	\frac{1}{2}\left(\ln\frac{w(1-\wbar)}{\wbar(1-w)}\otimes 1
	+1\otimes \ln\frac{w(1-\wbar)}{\wbar(1-w)}\right) + \ord(\eps)\,.
\end{equation}
If we apply the naive diagrammatic rule for the coaction stated at the end of the previous section, then the coaction on the bubble with massive propagators
should be given by (see eqs.~\eqref{bubM1M2CutM1}, \eqref{bubM1M2CutM2} and \eqref{bubM1M2CutP})
\begin{align}\bsp\label{eq:wrongCoprod}
	&\bubInsertLow[bub2m12Edges]\otimes \bubInsert[bub2m12CutPEdges]
	+\tadInsert[tad1]\otimes\bubInsert[bub2m12Cut1Edges]
	+\tadInsert[tad2]\otimes\bubInsert[bub2m12Cut2Edges]=\\
	&=\frac{1\otimes 1}{\epsilon}+\mathcal{O}(\epsilon^0)\,,
\esp\end{align}
and we see that this combination exhibits a pole in $\eps$, unlike the expresssion in  eq.~\eqref{eq:Delta_jt2}.

We claim that the correct rule to obtain the coaction on a one-loop
Feynman integral is stated as follows:
the second entries run over cut integrals with the same propagators as the original integral, cutting all possible nonempty subsets of (originally uncut) propagators. 
We distinguish the cases with even and odd numbers of cut propagators.
Then,\footnote{\label{foot:Formulation}There is an alternative way to state this rule, where instead of adding graphs with pinched edges in the first factor, we add graphs with additional cut edges in the second entry. We return to this point in section~\ref{sec:general_formulation}.}
\begin{itemize}
	\item if the number of cut edges is odd, then the first entry is the
	graph obtained by pinching the uncut edges;
	\item if the number of cut edges is even, then the first entry is the
	graph obtained by pinching the uncut edges,
	plus one-half times the sum of all graphs obtained by pinching an extra edge.
\end{itemize}
Applying this rule to the bubble integral with massive propagators, we find
\begin{align}\bsp\label{diagConjBubM1M2}
	\Delta_{\rm MPL}\left[\bubInsertLow[bub2m12Edges]\right]
	=&\left(\bubInsertLow[bub2m12Edges]+\frac{1}{2}\tadInsert[tad1]
	+\frac{1}{2}\tadInsert[tad2]\right)\otimes\bubInsert[bub2m12CutPEdges]\\
	&+\tadInsert[tad1]\otimes\bubInsert[bub2m12Cut1Edges]
	+\tadInsert[tad2]\otimes\bubInsert[bub2m12Cut2Edges].
\esp\end{align}
We have checked that eq.~\eqref{diagConjBubM1M2} correctly reproduces the coaction on $J_2$ up to $\ord(\eps^3)$, and we thus conjecture that it holds to all orders.
This is a nontrivial relation,
given that the~$\epsilon$ expansions of the two-mass bubble and its cuts contain nontrivial
polylogarithms in $w$ and~$\bar{w}$. Moreover, the left-hand side of eq.~\eqref{diagConjBubM1M2} is finite, while the right-hand side involves divergent tadpole integrals.
Just as in the case of the triangle integral, the poles cancel due to relations
among the different cuts of the two-mass bubble --- see eqs.~\eqref{eq:homo_even} and \eqref{eq:homo_odd}.
Stated more precisely, the poles introduced by the tadpoles cancel because the single and double cuts of the bubble are related by \refE{eq:PCI},
\begin{equation}
	 \cutRes_{e_1}J_2+\cutRes_{e_2}J_2+\cutRes_{e_1 e_2}J_2= -\eps\,J_2 \mod\,i\pi\,.
\end{equation}

Let us conclude this example by looking at the limit where the propagators become massless. Since massless tadpoles vanish in dimensional regularization, we see that the diagrammatic relation of eq.~\eqref{diagConjBubM1M2} reduces to eq.~\eqref{diagConjBub} in the limit where both propagators are massless. Note that this becomes a nontrivial statement about integrals, because the $\eps$-expansion does not commute with the massless limit, and the limit $m_1^2,m_2^2\to0$ can only be taken before expansion in $\eps$. Similarly, we obtain from eq.~\eqref{diagConjBubM1M2} a prediction for the coaction on the bubble integral with one massive and one massless propagator. If, say, $m_2^2\to0$, then we drop all massless tadpoles in eq.~\eqref{diagConjBubM1M2}, and we obtain
\beq\bsp\label{diagConjBubM}
	\Delta_{\textrm{MPL}}\left[\bubInsert[bub1mEdges] \right]=&\,
	\left(\bubInsert[bub1mEdges]+\frac{1}{2}\tadInsert[tad1]\right)\otimes
	\bubInsert[bub1mCutPEdges]\\
	&\,+\tadInsert[tad1]\otimes\bubInsert[bub1mCut1Edges].
\esp\eeq
{This relation was checked explicitly order by order in $\epsilon$ up to weight four, and we conjecture it to be valid to all orders in $\eps$ (see Appendix~\ref{sec:resBubOneMass} for explicit expressions for the contributing integrals).}

In the examples above, we have observed that the coaction on MPLs can be interpreted as acting on complete Feynman integrals in their $\eps$ expansion (conjecturally to all orders), yielding tensor products of (cut) Feynman integrals. In the following section, we will see that our graphical rules satisfy the axioms of a coaction, and we state our main conjecture of the equivalence of the two coactions.


\section{A coaction on cut graphs}
\label{sec:general_formulation}

In this section, we introduce a family of simple combinatorial operations on graphs and establish that they are coactions. One element of the family will turn out to reproduce the diagrammatic rules of the previous section. We are then able to formulate the conjecture that the diagrammatic coaction reproduces $\Delta_{\rm MPL}$ on the expansion of one-loop scalar Feynman integrals.  

 We define  objects called \emph{cut graphs}, which are pairs $(G,C)$ where $G$ is a Feynman graph and $C$ is a subset of edges of $G$. Any edge in $C$ is called \emph{cut}, and all other edges are \emph{uncut}. Note that every Feynman graph $G$ is naturally also a cut graph $(G,\emptyset)$. 
The construction  proceeds in two steps.  First, we show that there is a natural way to equip the algebra of cut graphs with a {coaction}. Second, we introduce a deformation of the coaction to account for the terms proportional to $1/2$ that appear in the coaction of the bubble integral in Section~\ref{sec:bubmassprop}. {We refer the reader to Appendix \ref{app:notation} for a summary of the notation used in this paper.}

\subsection{The incidence coaction on cut graphs}

We denote the free algebra generated by all cut graphs by ${\bf A}$.
There is a simple and natural way to define a coaction on ${\bf A}$, namely
\beq\label{eq:incidence_coaction}
\Delta_{\textrm{Inc}}(G,C) = \sum_{\substack{C\subseteq X\subseteq E_G\\ X\neq \emptyset}}(G_X,C)\otimes  (G,X)\,,
\eeq
where $E_G$ denotes the set of edges of $G$ (that are not external edges), and $G_X$ denotes the graph obtained from $G$ by contracting all edges that do not belong to $X$. 

Equation~\eqref{eq:incidence_coaction} has a very simple combinatorial interpretation: $\Delta_{\textrm{Inc}}(G,C)$ is obtained by summing over all possible terms whose second entries are the same graph, but with additional cut edges. The corresponding first entries are obtained by contracting all edges that are not cut in the second entry. 
{Alternatively, we can say that the first entries are contractions of all possible subsets of uncut edges, and the corresponding second entries have cuts of all the edges left uncontracted in the first entry.}

We point out that eq.~\eqref{eq:incidence_coaction} does not define a coproduct, but rather a coaction, on ${\bf A}$. 
The definition of a coaction requires the existence of a Hopf algebra (cf. the relationship between $\cA$ and $\cH$ in Section~\ref{subsec:MPLs} --- see Appendix~\ref{app:hopf} for more details). In particular, we show that this coaction is closely related to \emph{incidence Hopf algebras}. In order to keep the present discussion as light as possible, we only introduce the main definitions that we need to state our conjectural relationship between $\Delta_{\textrm{Inc}}$ and $\Delta_{\textrm{MPL}}$, and we refer to Appendix~\ref{app:hopf} for a more rigorous exposition.

Let us look at some simple examples of how $\Delta_{\textrm{Inc}}$ acts on cut graphs.
As in the graphs drawn in the previous section, edges are labelled and cut edges are depicted with red dashed lines. 
For a bubble graph, we have
\begin{align}\bsp
    \Delta_{\text{Inc}}\left[\bubInsertLow[bub2m12Edges]\right]
    =&\bubInsertLow[bub2m12Edges]\otimes\bubInsert[bub2m12CutPEdges]\\
    &+\tadInsert[tad1]\otimes\bubInsert[bub2m12Cut1Edges]+\tadInsert[tad2]\otimes\bubInsert[bub2m12Cut2Edges]\,.
\esp\end{align}
For an example of applying~\eqref{eq:incidence_coaction} to a graph with cut propagators, consider the graph with three edges, of which one is cut. Here we have
\begin{align}\bsp
    \Delta_{\text{Inc}}\left[\triInsertLow[t123Cut1Edges]\right]&=
    \triInsert[t123Cut1Edges]\otimes\triInsertLow[t123Cut123Edges]+
    \bubInsert[bub2m12Cut1Edges]\otimes\triInsertLow[t123Cut12Edges]\\
    &+\bubInsert[bub2m13Cut1Edges]\otimes\triInsertLow[t123Cut13Edges]
    +\tadInsert[tad1CutM]\otimes\triInsertLow[t123Cut1Edges]\,.
\esp\end{align}
As a final simple example, it is obvious that the coaction on a \emph{maximally cut graph}, i.e., a cut graph of the form $(G,E_G)$, is
\beq
\Delta_{\textrm{Inc}}(G,E_G) = (G,E_G)\otimes (G,E_G)\,.
\eeq

\subsection{The deformed incidence coaction}

We now present
 a deformation of the coaction in eq.~\eqref{eq:incidence_coaction} that captures the extended rules for the coaction on one-loop Feynman integrals of Section~\ref{sec:bubmassprop}.

We start by defining an algebra morphism (i.e., a map that preserves the algebra structure) $\varphi_a: {\bf A} \to {\bf A}$, via
\beq\bsp\label{eq:phidef}
\varphi_a(G,C) &\,= (G,C) + a_{E_G}\,\sum_{e\in E_G/C}(G/e,C)\,,
\esp\eeq
where $C\subseteq E_G$ and $a_{E_G}$ is defined through
\beq\bsp\label{eq:aXdef}
a_X &\,= \left\{\begin{array}{ll}\displaystyle
a\,, & \textrm{ if } n_X \textrm{ is even},\\
0 \,, & \textrm{ if } n_X \textrm{ is odd}\,,
\end{array}\right. \quad\text{with}\quad a\in \mathbb{Q}\,,
\esp\eeq
where $X$ is a set of edges, and $n_X=|X|$.
Here, $G/e$ denotes the graph obtained from $G$ by contracting the edge $e$.
This map is obviously invertible for any value of $a$, and it therefore induces a new coaction on ${\bf A}$ by
\beq\bsp
\label{eq:phipsiaction}
\Delta_a &\,\equiv (\varphi_a\otimes \varphi_a)\Delta_{\textrm{Inc}}\varphi_{a}^{-1}
= (\varphi_a\otimes\textrm{id})\Delta_{\textrm{Inc}} \\
&\,=\sum_{\substack{C\subseteq X\subseteq E_G,\\X\neq \emptyset}} \left((G_X,C) +a_X \sum_{e\in X\setminus C} (G_{X\setminus e},C) \right) \otimes (G,X)\,.
\esp\eeq
Note that the deformed coaction reduces to the incidence coaction for $a=0$, $\Delta_0=\Delta_{\textrm{Inc}}$. 
As an example, let us consider the action of $\Delta_a$ on the uncut bubble graph,
\begin{align}\bsp\label{bubble_coaction_example}
	\Delta_{a}\left[\bubInsertLow[bub2m12Edges]\right]
	=&\left(\bubInsertLow[bub2m12Edges]+a\,\tadInsert[tad1]
	+a\,\tadInsert[tad2]\right)\otimes\bubInsert[bub2m12CutPEdges]\\
	&+\tadInsert[tad1]\otimes\bubInsert[bub2m12Cut1Edges]
	+\tadInsert[tad2]\otimes\bubInsert[bub2m12Cut2Edges].
\esp\end{align}
Comparing the previous equation to eq.~\eqref{diagConjBubM1M2}, we see that the two coactions agree for $a=1/2$. In the next section we will argue that this is a general feature, and that there is a connection between the coaction $\Delta_{1/2}$ on {(cut)} one-loop graphs and the coaction $\Delta_{\textrm{MPL}}$ on {(cut)} one-loop Feynman integrals.

While  the deformation parameter $a$ appears only in the first factor in eq.~\eqref{eq:phipsiaction}, it is easy to show that  one can write $\Delta_a$ equivalently with the deformation parameter appearing only in the second factor,
\beq\label{eq:psi_coaction}
\Delta_a = (\textrm{id}\otimes \psi_a)\Delta_{\textrm{Inc}}\,,
\eeq
where $\psi_a$ is defined by
\begin{equation}
    \psi_a(G,C) \,= (G,C) + (a-{a}_C)\,\sum_{e\in E_G/C}(G,C\cup e)\,,
\end{equation}
with $a_C$ defined in \refE{eq:aXdef}.
For example, it is easy to check that we can rearrange terms and write eq.~\eqref{bubble_coaction_example} in the equivalent form
\begin{align}
\nonumber
	\Delta_{a}\left[\bubInsertLow[bub2m12Edges]\right]
	=&\bubInsertLow[bub2m12Edges]\otimes\bubInsert[bub2m12CutPEdges]+\tadInsert[tad1]\otimes\left(\bubInsert[bub2m12Cut1Edges]+a\,\bubInsert[bub2m12CutPEdges]\right)\\
&	+\tadInsert[tad2]\otimes\left(\bubInsert[bub2m12Cut2Edges]+a\,\bubInsert[bub2m12CutPEdges]\right)\,.
\end{align}

\subsection{The main conjecture}\label{sec:diagCoprodDef}
We can now formulate our main conjecture: the coaction $\Delta_{\textrm{MPL}}$ on one-loop Feynman integrals admits a purely diagrammatic description. To be more precise, 
we will write the formula in terms of the basis integrals.
 If $(G,C)$ is a one-loop cut graph, we denote by $\mathcal{I}$ the linear map that associates to $(G,C)$ its cut integral $\mathcal{I}(G,C)\equiv\cC_C {J}_G$ in $D_n$ dimensions, as defined in \refE{eq:defDn}. Here $J_G$ denotes the one-loop Feynman integral 
 associated to the graph $G$, and   normalized to the leading order of its maximal cut as given in
\refE{eq:LS_J_n}.
Our conjecture can then be written in the explicit form
\beq\label{eq:graph_conjecture}
\boxed{
\Delta_{\textrm{MPL}} \, \mathcal{I}(G,C) = (\mathcal{I}\otimes\mathcal{I}) \, \Delta_{1/2}(G,C)\,.
}
\eeq
Equation~\eqref{eq:graph_conjecture} implies that the combinatorics of the coaction $\Delta_{\textrm{MPL}}$ of one-loop integrals is entirely captured by the coaction $\Delta_{1/2}$ on cut graphs. It is remarkable that such a formula exists at all: the left-hand side of eq.~\eqref{eq:graph_conjecture} involves the coaction on MPLs of refs.~\cite{Goncharov:2005sla,GoncharovMixedTate}, order by order in dimensional regularization, while the right-hand side only involves the coaction on cut graphs, which reproduces the left-hand side if each factor in the coaction is separately expanded in $\eps$. It is noteworthy that eq.~\eqref{eq:graph_conjecture} makes no  distinction between massive and massless integrals, while in general the $\eps$-expansion does not commute with the massless limit, where new infrared singularities might develop.

Using eq.~\eqref{eq:phipsiaction} we can write our conjecture for the basis integrals and their cuts more explicitly as
\beq\bsp
\label{eq:Coaction_cut_integrals_phi}
\Delta_{\rm MPL} \left({\cal C}_C {J}_G\right)
&\,=\sum_{\substack{C\subseteq X\subseteq E_G,\\X\neq \emptyset}}
\left( {\cal C}_C {J}_{G_X} + a_{X} \sum_{e\in X\setminus C} {\cal C}_C {J}_{{G_X}\setminus e}\right) \otimes  {\cal C}_X {J}_G\,,
\esp\eeq
with $a_X=1/2$ if $n_X$ is even and $a_X=0$ otherwise.
The case of uncut integrals is recovered for $C=\emptyset$.
It is useful to introduce a shorthand notation for the combination of integrals appearing in the left component of the coaction,
\beq
\label{hatG_J}
{\widehat J}_{G} \equiv J_{\varphi_{1/2}(G)} = {J}_G + a_{E_G} \sum_{e\in E_G} {J}_{G\setminus e}\, , \quad a=1/2\,,
\eeq
with the straightforward generalization for cut integrals (cut edges are not pinched),
\beq
\label{Cut_hatG_J}
{\cal C}_C{\widehat J}_{G} \equiv   {\cal C}_C {J}_{G} + a_{E_G} \sum_{e\in E_G\setminus C} {\cal C}_C {J}_{{G}\setminus e}\,.
\eeq
With this definition our conjecture takes the simple form:
\beq
\boxed{
\label{eq:Coaction_cut_integrals}
\Delta_{\rm MPL}\left({\cal C}_C {J}_G\right)=
\sum_{\substack{C\subseteq X\subseteq E_G,\\X\neq \emptyset}} {\cal C}_C \widehat{J}_{G_X}\otimes {\cal C}_X {J}_G\, .}
\eeq

Equation~\eqref{eq:psi_coaction} implies that we can cast the conjecture~\eqref{eq:Coaction_cut_integrals} (or (\ref{eq:Coaction_cut_integrals_phi})) in the equivalent form 
\beq\label{eq:Coaction_cut_integrals_psi}
\Delta_{\rm MPL}\left({\cal C}_C {J}_G\right)=
\sum_{\substack{C\subseteq X\subseteq E_G,\\X\neq \emptyset}} {\cal C}_C {J}_{G_X}\otimes \left({\cal C}_X {J}_G+ (a-{a}_C)\,\sum_{e\in E_G\setminus C}{\cal C}_{Xe} {J}_G\right)\, ,
\eeq
with $a=1/2$.

Our conjecture~\eqref{eq:graph_conjecture}, or equivalently eq.~\eqref{eq:Coaction_cut_integrals} {or}~\eqref{eq:Coaction_cut_integrals_psi}, is a special case of a more general coaction presented in ref.~\cite{Abreu:2017enx}, which is formulated in terms of master integrands and master contours. Before making this connection more precise in Section~\ref{sec:master_formula}, we will give ample evidence that eq.~\eqref{eq:graph_conjecture} indeed holds for one-loop graphs, and then explore its implications for the study of discontinuities and differential equations.


\section{Dimensional regularization}
\label{sec:dimreg}

We start with a general discussion of the structure of the conjecture \eqref{eq:graph_conjecture} in dimensional regularization. Our first goal is to be able to identify which diagrams contribute to each coaction component {in the grading by weight}.  Then, we discuss how the trivial coaction components are correctly reproduced.
We focus on the coaction on an uncut Feynman integral, $C=\emptyset$ in \refE{eq:Coaction_cut_integrals}. The generalization to cut integrals is straightforward. 
{We refer the reader to Appendix \ref{app:notation} for a summary of the notation used in this paper.}

\subsection{The diagrammatic coaction in dimensional regularization }
\label{sec:coproductCheck}

$\Delta_{\text{MPL}}$ acts on the coefficients of the Laurent expansion in $\epsilon$ of (cut) Feynman integrals.
It is convenient to introduce the following notation for  terms in the Laurent expansion:
\begin{equation}
\label{J_G_expansions}
	J_G=\sum_{j}\epsilon^j\,J^{(j)}_G\,,\qquad\quad
\widehat{J}_{G_X}=\sum_{j}\epsilon^j\,\widehat{J}_{G_X}^{(j)}
\,,\qquad\quad
	\cutRes_XJ_G=\sum_{j}\epsilon^j \,\cutRes_X^{(j)}J_G\,.
\end{equation}
We would like to sort terms on the two sides of eq.~\eqref{eq:Coaction_cut_integrals} with $C=\emptyset$, order by order, according to their weight.

We denote the weight of a given polylogarithmic function $F$ by $w\left(F\right)$.
We define the weight of a basis function before expansion in $\epsilon$ as the weight of the ${\cal O}(\epsilon^0)$ term\footnote{This operation is well defined because the $J_n$ and their cuts are pure functions.}, namely
\begin{equation}
w_G \equiv \, w\left(J_G\right) \equiv  w\left(J^{(0)}_G\right) = \left\lceil\frac{n_G}{2}\right\rceil\, .
\label{wJ_G}
\end{equation}
Then, the weight of the ${\cal O}(\epsilon^j)$ terms in the expansion of our basis functions is
\begin{equation}
w\left(J^{(j)}_G\right) = w_G + j\,.
\end{equation}
One may verify by explicit calculation  that the $D=2-2\epsilon$ tadpoles and bubbles are functions of weight one, the $D=4-2\epsilon$ triangles and boxes are  functions of weight two, the $D=6-2\epsilon$ pentagons and hexagons are  functions of weight three, and so on (see Appendix~\ref{sec:results} for some examples).
The coaction $\Delta_{\text{MPL}}$ respects the weight grading of the MPLs, and we find
\begin{equation}\label{eq:coprodJN}
	\Delta_{\text{MPL}}\left[ J_G\right]=\sum_{j}\epsilon^j\Delta_{\text{MPL}}\left[ J^{(j)}_G\right]=\sum_{j}\epsilon^j\sum_{k=0}^{w_G+j}\Delta_{k,w_G+j-k}\left[ J^{(j)}_G\right]\,,
\end{equation}
{where $\Delta_{a,b}$ denotes the coaction component where the first and second factors have weights $a$ and $b$ respectively, as defined below \refE{eq:Delta_ipi}.}

On the right-hand side of \eqref{eq:Coaction_cut_integrals}, for $C=\emptyset$, the left and right entries of each term have the following weights, respectively:
\begin{equation}\label{eq:weightCuts}
w_X \equiv 
\, w\left(\widehat{J}_{G_X}\right)= \left\lceil\frac{n_X}{2}\right\rceil
\,,\qquad\quad
w\left(\cutRes_XJ_G\right)
=w_G-w_X\,.
\end{equation}
After sorting the terms of \refE{eq:Coaction_cut_integrals} by powers of $\epsilon$ and by weight,
one obtains the following relation:
\begin{equation}
\label{Weight_contr}
\Delta_{k,w_G+j-k} \left[J^{(j)}_G\right] = \sum_{ X\subseteq E_G,\,X\neq \emptyset}  \widehat{J}_{G_X}^{(k-w_X)} \otimes \cutRes_X^{(w_X+j-k)}J_G\,,
\end{equation}
For finite integrals, $(k-w_X)\geq0$ and $(w_X+j-k)\geq0$. However, it is known that one-loop integrals can be divergent, so we briefly discuss these cases now.

By simple power counting, one finds that $J_1(m^2)$ is ultraviolet-divergent, while all other $J_n$ for $n>1$ are ultraviolet finite. Soft divergences, which arise when all components of the loop momentum become small at the same rate, can also be identified by power counting, and we find that only the $J_n$ with $n\leq 4$ can have soft divergences. We can then use collinear scaling of the loop momentum \cite{Sterman:1995fz} to identify the diagrams with collinear divergences, and again we find that the only $J_n$ that can be divergent are those with $n\leq 4$.
Finally, cuts of finite one-loop integrals are finite \cite{Abreu:2017ptx}.
Thus we can list the divergent integrals explicitly.
For $n$ odd, in which case $\jhat_n=J_n$,
\begin{itemize}
    \item $J_1$ is ultraviolet divergent {and diverges as $\mathcal{O}(\epsilon^{-1})$}. Its cut is finite.
    \item $J_3$ is infrared divergent if and only if $\cutRes_{e_1e_2e_3} J_3=0$.  The single-scale triangle, $J_3(p^2)$, {diverges as $\mathcal{O}(\eps^{-2})$, while all other divergent cases behave like $\mathcal{O}(\eps^{-1})$}. Single or double cuts of divergent triangles are either zero or divergent (see Appendix~\ref{sec:resTriangles}).
    \item For all $n\geq 5$, the $J_n$ and their cuts are finite.
\end{itemize}
For $n$ even, $\jhat_n \neq J_n$, and we find that 
\begin{itemize}
    \item $\jhat_2$  {diverges as $\mathcal{O}(\epsilon^{-1})$}, regardless  of whether any propagator is massive or massless.  The massless bubble $J_2(p^2)$ and one-mass bubble $J_2(p^2;m^2)$ {diverge as $\mathcal{O}(\epsilon^{-1})$} while $J_2(p^2;m_1^2,m_2^2)$ is finite. Cuts of all these integrals are finite.
    \item $\jhat_4$ is always finite, but some $J_4$  {diverge as $\mathcal{O}(\eps^{-2})$ or $\mathcal{O}(\eps^{-1})$}. $J_4$ is infrared divergent if and only if $\cutRes_{e_ie_je_k} J_4=0$ for some choice of $\{e_i,e_j,e_k\}\in E_G$. Single or double cuts of the divergent $J_4$  {diverge at most as $\mathcal{O}(\eps^{-1})$}.
    \item For all $n\geq 6$, the $\jhat_n$, the $J_n$, and their cuts are finite.
\end{itemize}
With this information, we can now determine which (cut) integrals at which order in their Laurent expansion in $\eps$ contribute to \refE{Weight_contr} for each value of $k$.

From \refE{Weight_contr} we see that coaction components with low weight in the left entry (i.e., small $k$) are fully determined by the simplest topologies (i.e., those with small $w_X$). 
The $k=0$ coaction component is determined by tadpoles and bubbles on the left, along with single or double propagator cuts on the right.
With increasing weight of the left entry, one begins to probe more complex topologies (on the left) along with cuts of more propagators (on the right).
Conversely, let us look at components with a function of weight zero in the right entry of the coaction ($k=w_G+j$ in \refE{Weight_contr}) and suppose that we restrict our analysis to finite integrals. These components are  fully determined by the maximal cut for odd $n_G$, and by both the maximal and next-to-maximal cuts for even $n_G$.  Similarly, the components with weight one in the right entry, ($k=w_G+j-1$) are determined by the maximal, next-to-maximal, and next-to-next-to-maximal cut for odd  $n_G$, while for even  $n_G$ we must also include the next-to-next-to-next-to-maximal cut.

\subsection{Divergences and trivial coaction components}\label{sec:cancellationOfPoles}

In this section we address two apparent issues about our conjecture~\eqref{eq:Coaction_cut_integrals} for $C=\emptyset$:
\begin{enumerate}
\item Even for finite integrals ${J}_G$, the right-hand side of eq.~\eqref{eq:Coaction_cut_integrals} involves divergent tadpole and/or bubble integrals.
\item The coaction on a pure function $f$ always takes the form
\beq\label{eq:grouplike}
\Delta_{\rm MPL}(f) = 1\otimes f+f\otimes 1 + \Delta'_{\rm MPL}(f)\,,
\eeq
where $\Delta'_{\rm MPL}(f)$ denotes the reduced coaction,  i.e., the terms in the coaction where each of the two factors has weight at least one.
It is not immediately clear how eq.~\eqref{eq:Coaction_cut_integrals} reproduces the specific terms separated in  eq.~\eqref{eq:grouplike} with weight 0 entries, which we call the \emph{trivial components}.
\end{enumerate}
In the remainder of this section, we show how these two issues are resolved, giving strong support to the validity of the conjecture.

We consider the integrals $\plainJ_G$, with $n_G \geq 3$, in the generic case where all external legs are off-shell, postponing the discussion of divergent integrals  to  Section~\ref{sec:moreExamples}. 
Given a finite integral $\plainJ_G$, it is clear that for eq.~\eqref{eq:Coaction_cut_integrals} to be consistent, all the poles
coming from the tadpole and bubble integrals on the right-hand side must cancel.

We start from eq.~\eqref{eq:Coaction_cut_integrals} and concentrate on the terms with tadpole and bubble integrals in the first entry,
\beq\bsp\label{eq:delta_pole}
&\Delta_{\rm MPL}\left(\plainJ_G\right) = \sum_{e \in E_G}\plainJ_{e}\otimes \cC_e \plainJ_G 
+ \sum_{\{e,e'\} \subset E_G} \widehat{\plainJ}_{{ee'}} \otimes \cC_{ee'} \plainJ_G
 + \cdots\,,
\esp\eeq
where 
we use the shorthand notation for tadpoles and bubbles,
\beq
J_e \equiv J_{G_{\{e\}}}\,,  \quad J_f \equiv J_{G_{\{e'\}}} \quad \textrm{and} \quad
J_{ee'} \equiv J_{G_{\{e,e'\}}}\,.
\eeq
The dots  in  \refE{eq:delta_pole} indicate terms that do not contain tadpoles or bubbles and are hence finite.

Let us analyze the set of terms in eq.~\eqref{eq:delta_pole} which contain the single cuts. If a propagator is massless, then both the  corresponding tadpole and single cut vanish~\cite{Abreu:2017ptx}, and if a propagator is massive, then the tadpole contributes a single pole in $\epsilon$, $\plainJ_e=-\frac{1}{\eps}+\ord(1)$. The contribution from the single cuts can always be written as 
\beq
\label{tad_pole}
-\frac{1}{\eps}\otimes  \sum_{e \in E_G} \cC_e\plainJ_G + \ord(1)\,,
\eeq
where the only nonvanishing contributions in the sum are the ones corresponding to cuts of massive propagators.

We have seen in the previous subsection that $\widehat{\plainJ}_{ee'}$  always has a simple pole in $\epsilon$. Specifically, we have $\widehat{\plainJ}_{ee'} = -\frac{1}{\eps}+\ord(1)$. We then see that the divergent terms on the right-hand side of eq.~(\ref{eq:delta_pole}) always appear in the following combination:
\beq
\label{div_in_delta_pole}
-\frac{1}{\eps}\otimes\left[\sum_{e \in E_G} \cC_e\plainJ_G+\sum_{\{e,e'\} \subset E_G} \cC_{ee'}\plainJ_G\right]+\ord(1)\,.
\eeq
From \refE{eq:PCI} we know that the single and double cuts of every one-loop integral $J_G$ combine to produce $\epsilon$ times the original integral.
The same relation allows us now to see that all the divergences in (\ref{div_in_delta_pole}) cancel, and the $1/\eps$ poles of the tadpole and bubble integrals combine to contribute  a term of the form
\beq\label{eq:1otJ}
1\otimes \plainJ_G \,.
\eeq
to the right-hand side of eq.~(\ref{eq:delta_pole}).
We emphasize that the leading pole terms in uncut bubbles and tadpoles are the only possible source of weight zero terms in the first entry of the coaction for finite integrals (see \refE{Weight_contr}). Note in particular that the $\ord(1)$ terms which we neglected in eqs.~(\ref{tad_pole}) and (\ref{div_in_delta_pole}) cannot contribute to the coaction component  with weight zero in the left entry.

Thus far we have addressed the first issue raised in the beginning of this section (cancellation of poles on the right-hand side of eq.~\eqref{eq:Coaction_cut_integrals})
as well as part of the second, namely we have seen how the first term in (\ref{eq:grouplike}), $1\otimes f$,  is generated.
Let us now discuss how the second term, $f\otimes 1$, is reproduced.

As already discussed at the end of \refS{sec:coproductCheck}, if $n_G$ is odd, then only the maximal cut contributes to weight zero,
\beq
\plainJ_G \otimes \cC_{E_G}\plainJ_G = \plainJ_G \otimes 1+\ord(\eps)\,, \qquad \textrm{for $n_G$ odd.}
\eeq
If $n_G$ is even, however, then the next-to-maximal cuts also contribute at weight zero. In ref.~\cite{Abreu:2017ptx} it was shown that if $n_G$ is even, then the maximal and next-to-maximal cuts are related by
\beq\label{eq:maxNMaxRel}
 \cC_{E_G \setminus e}\plainJ_G = -\frac{1}{2} \cC_{E_G}\plainJ_G + \ord(\eps)\,.
 \eeq
 The full contribution to weight zero in the second factor of the coaction is then
 \beq\bsp
&\left( \plainJ_G+ \frac{1}{2}\sum_{e \in E_G} \plainJ_{E_G \setminus e}\right) \otimes \cC_{E_G}J_G
+\sum_{e \in E_G} \plainJ_{E_G \setminus e} \otimes \cC_{E_G \setminus e} \plainJ_G = \plainJ_G \otimes 1+\ord(\eps)\,,
\esp\eeq
which
 reproduces the second trivial component on the right-hand side of eq.~\eqref{eq:grouplike}.


\section{Diagrammatic coaction on some one-loop Feynman integrals}
\label{sec:moreExamples}

In this section we discuss some examples of our conjecture.
These examples are chosen to highlight some nontrivial features
and also serve as a check of the conjecture itself. In each case where we display the $\Delta_{1/2}$ coaction diagrammatically, we have verified explicitly that the relation in (\ref{eq:graph_conjecture}) holds. Namely, computing the first few orders in the $\epsilon$ expansion of the integral on the left-hand side, and then taking the coaction on MPLs, is found to be equal to the graphical coaction, where each (cut) diagram on the right-hand side is replaced by the corresponding integral, which is then expanded in $\epsilon$. Explicit expressions for the relevant (cut) integrals are compiled in Appendix~\ref{sec:results}, {and we refer the reader to Appendix \ref{app:notation} for a summary of the notation used in this paper.}

While we discuss only a few examples here,
we have checked that our conjecture holds for all diagrams with up to three propagators, for all assignments of internal and external masses, at least up to terms of weight four in the Laurent expansion in dimensional regularization.\footnote{For the three-point functions with three massive external  legs and massive propagators, in view of  the complexity of the functions, only the cuts were checked to this order. For the uncut integrals, only the leading order was checked.} In addition, we have checked that the coaction on the following one-loop box integrals is correctly reproduced at least up to terms of weight four in the $\eps$-expansion:
\begin{itemize}
\item box integrals with massless propagators and up to two external masses.
\item the box integral with massless external legs and one massive propagator.
\item the box integral with massless external legs and two adjacent massive propagators.
\end{itemize}
Moreover, we have checked that the coaction on the next-to-maximal cut of the four-mass box with massless propagators is correctly predicted by our conjecture through  $\mathcal{O}(\epsilon)$.
Finally, we have checked the consistency of the diagrammatic coaction with the differential equation satisfied by the massless pentagon, and with the symbol of the massless hexagon as studied in ref.~\cite{Spradlin:2011wp}.

\subsection{Reducible one-loop integrals}
\label{sec:redExamples}

In Section~\ref{sec:coproductCheck} we have seen that triangles diverge if and only if their maximal cut vanishes. The vanishing of the maximal cut implies that the integral is reducible via IBP identities to a linear combination of integrals with fewer propagators.
Indeed, for the labelling of internal and external scales of \refF{fig:triNotation} in Appendix \ref{sec:results}, we find: 
\begin{align}\bsp\label{eq:redTriangles}
    &J_3(p^2)=\frac{1}{\epsilon} J_2(p^2)\,,\\
    &J_3\left(p_1^2,p_2^2\right)=\frac{1}{\epsilon}\left( J_2(p_1^2)- J_2(p_2^2)\right)\,,\\
    &J_3\left(p_1^2;m_1^2\right)=
   \frac{1}{2\epsilon}\left(2 J_2(p_1^2;m_{1}^2)- J_1(m_{1}^2)\right)\,,\\
   &J_3\left(p_1^2,p_2^2;m_2^2\right)=\frac{1}{\epsilon}\left( J_2(p_1^2;m_2^2)- J_2(p_2^2;m_2^2)\right)\,.
\esp\end{align}
These relations can be derived via IBP relations and commute with the operation of cutting propagators. For example, 
\begin{equation}\bsp
	 \cC_{e_1}J_3\left(p_1^2;m_1^2\right)&\,=
	\frac{1}{2\epsilon}\left(2 \cC_{e_1}J_2(p_1^2;m_{1}^2)- \cC_{e_1}J_1(m_{1}^2)\right)\,,\\
       \cC_{e_1e_2}J_3\left(p_1^2;m_1^2\right)&\,=\frac{1}{\eps}\,\cC_{e_1e_2}J_2\left(p_1^2;m_{1}^2\right)\,.
\esp\end{equation}
As we have already established in \refS{sec:firstExamples} that tadpole and bubble integrals satisfy the diagrammatic coaction, it then follows that the integrals in \refE{eq:redTriangles} also do. For example, we find

\begin{align}\label{eq:diagConjtp1}
	\Delta_{1/2}\left[\triInsertLow[t1LabelEdges]\right]=
	\bubInsert[bub0m1LabelEdges]\otimes\triInsertLow[t1Cut12LabelEdges]\,,
\end{align}
\begin{align}\label{eq:diagConjtp1Cutp1}
	\Delta_{1/2}\left[\triInsertLow[t1Cut12LabelEdges]\right]=
	\bubInsert[bub0m1CutPEdges]\otimes\triInsertLow[t1Cut12LabelEdges]\,,
\end{align}
\begin{align}\bsp
	\Delta_{1/2}\left[\triInsertLow[t12LabelEdges]\right]=
	\bubInsert[bub0m1LabelEdges]\otimes\triInsertLow[t12Cut12LabelEdges]
	+\bubInsert[bub0m2LabelEdges]\otimes\triInsertLow[t12Cut23LabelEdges]\,,
\esp\end{align}
\begin{align}
	\Delta_{1/2}\left[\triInsertLow[t12Cut12LabelEdges]\right]=
	\bubInsert[bub0m1CutPEdges]\otimes\triInsertLow[t12Cut12LabelEdges]\,,
\end{align}
\begin{equation}\label{eq:diagConjtp1m12}
	\Delta_{1/2}\left[\triInsertLow[t1m1LabelEdges]\right]=
	\tadInsert[tad1]\otimes \triInsertLow[t1m1Cut1LabelEdges]+
	\left(\bubInsertLow[bub1m1LabelEdges]+\frac{1}{2}\tadInsert[tad1]\right)\otimes\triInsertLow[t1m1Cut12LabelEdges],
\end{equation}
\begin{equation}\label{eq:diagConjtp1m12Cutm12}
	\Delta_{1/2}\left[\triInsertLow[t1m1Cut1LabelEdges]\right]=
	\tadInsertLow[tad1CutM]\otimes \triInsertLow[t1m1Cut1LabelEdges]+
	\left(\bubInsert[bub1m1Cut1LabelEdges]+\frac{1}{2}\tadInsert[tad1CutM]\right)\otimes\triInsertLow[t1m1Cut12LabelEdges],
\end{equation}
\begin{align}\label{eq:diagConjtp1m12Cutp1}
	\Delta_{1/2}\left[\triInsertLow[t1m1Cut12LabelEdges]\right]=
	\bubInsertLow[bub1m1CutPLabelEdges]\otimes\triInsertLow[t1m1Cut12LabelEdges].
\end{align}
These examples illustrate how our conjecture~\eqref{eq:Coaction_cut_integrals} correctly reproduces the coaction on one-loop integrals that are divergent and reducible.


\subsection{Cancellation of poles}

We now illustrate in a specific example of a finite integral how the cancellation of the poles introduced by divergent bubbles and tadpoles can be explained by the relation \eqref{eq:PCI} between an uncut integral and its one- and two-propagator cuts. Consider the triangle with one massive external leg and a single non-adjacent massive propagator whose integrated expression is given in \refE{eq:tp1m23}.
 Its diagrammatic coaction is given by:
\begin{align}\bsp\label{eq:diagConjtp1m23}
	\Delta_{1/2}\left[\triInsertLow[t1m3LabelEdges]\right]=&
	\tadInsert[tad3]\otimes \triInsertLow[t1m3Cut3LabelEdges]+
	\bubInsert[bub0m1LabelEdges]\otimes\triInsertLow[t1m3Cut12LabelEdges]\\
	&+\triInsertLow[t1m3LabelEdges]\otimes\triInsertLow[t1m3Cut123LabelEdges]\,,
\esp\end{align}

\begin{align}\label{eq:diagConjtp1m23CutM23}
	\Delta_{1/2}\left[\triInsertLow[t1m3Cut3LabelEdges]\right]=
	\tadInsertLow[tad3CutM]\otimes \triInsertLow[t1m3Cut3LabelEdges]
	+\triInsertLow[t1m3Cut3LabelEdges]\otimes\triInsertLow[t1m3Cut123LabelEdges]\,,
\end{align}

\begin{align}\label{eq:diagConjtp1m23CutP1}
	\Delta_{1/2}\left[\triInsertLow[t1m3Cut12LabelEdges]\right]=
	\bubInsert[bub0m1CutPEdges]\otimes \triInsertLow[t1m3Cut12LabelEdges]
	+\triInsertLow[t1m3Cut12LabelEdges]\otimes\triInsertLow[t1m3Cut123LabelEdges]\,,
\end{align}

\begin{align}\label{eq:diagConjtp1m23CutP1M23}
	\Delta_{1/2}\left[\triInsertLow[t1m3Cut123LabelEdges]\right]=
	\triInsertLow[t1m3Cut123LabelEdges]\otimes \triInsertLow[t1m3Cut123LabelEdges]\,.
\end{align}
On the right-hand-side of \refE{eq:diagConjtp1m23} there are two divergent contributions:
\begin{equation*}
	\tadInsert[tad3]\Bigg\vert_{\epsilon^{-1}}\otimes \triInsertLow[t1m3Cut3LabelEdges]\Bigg\vert_{\epsilon^{0}}
	=-1\otimes\triInsertLow[t1m3Cut3LabelEdges]\Bigg\vert_{\epsilon^{0}}\, ,
\end{equation*}
and
\begin{equation*}
	\bubInsert[bub0m1LabelEdges]\Bigg\vert_{\epsilon^{-1}}\otimes\triInsertLow[t1m3Cut12LabelEdges]\Bigg\vert_{\epsilon^{0}}
	=-1\otimes\triInsertLow[t1m3Cut12LabelEdges]\Bigg\vert_{\epsilon^{0}}\, .
\end{equation*}
Given that the left-hand-side of \refE{eq:diagConjtp1m23} is finite, these contributions must cancel.
As discussed in Section \ref{sec:cancellationOfPoles}, this happens because of \refE{eq:PCI}, which for our example reads
\begin{equation}
\triInsertLow[t1m3Cut3LabelEdges] + \triInsertLow[t1m3Cut12LabelEdges] = \ord(\eps)\,.
\end{equation}
The previous equation can be seen to hold in this specific case by inserting the explicit results for the cuts given in eqs.~\eqref{eq:tp1m23Cutm23} and~\eqref{eq:tp1m23CutP1}.

\subsection{Infrared divergent integrals}

In this section we discuss the diagrammatic coaction in the context of an infrared divergent integral, the massless box $J_4(s,t)$ which diverges as $\mathcal{O}(\eps^{-2})$. We have already discussed some simple infrared  divergent integrals in Sections \ref{sec:firstExamples} and \ref{sec:redExamples}, but all those examples were either trivial or reducible to integrals with fewer propagators.
In this section we illustrate two features that do not follow from the general discussion of Section \ref{sec:cancellationOfPoles}: we first show how the poles are correctly reproduced by the diagrammatic coaction, and then how the trivial coaction components arise. 

The diagrammatic coaction for the massless box is:
\begin{align}\bsp\label{eq:diagConjbst}
	\Delta_{1/2}&\left[\boxInsert[bstLabelEdges]\right]=
	\bubInsert[bub0msLabelEdges]\otimes \boxInsertLow[bstCut13LabelEdges]
	+\bubInsert[bub0mtLabelEdges]\otimes \boxInsert[bstCut24LabelEdges]\\
	&\qquad\qquad\quad+\left(\boxInsert[bstLabelEdges]+\triInsert[tsLabelEdges]+\triInsert[ttLabelEdges]\right)\otimes \boxInsertLow[bstCut1234LabelEdges]\,.
\esp\end{align}
Note that all one- and three-propagator cuts of the massless box vanish.
For the double cut we find
\begin{equation}\bsp\label{eq:diagConjbstCutS}
	\Delta_{1/2}\left[\boxInsertLow[bstCut13LabelEdges]\right]=&
	\bubInsert[bub0msCutPLabelEdges]\otimes \boxInsertLow[bstCut13LabelEdges]\\
	+&\left(\boxInsertLow[bstCut13LabelEdges]+\triInsert[tsCut13LabelEdges]\right)\otimes \boxInsertLow[bstCut1234LabelEdges]\,,
\esp\end{equation}
and the quadruple cut gives
\begin{equation}\label{eq:diagConjbstCutST}
	\Delta_{1/2}\left[\boxInsertLow[bstCut1234LabelEdges]\right]=
	\boxInsertLow[bstCut1234LabelEdges]\otimes \boxInsertLow[bstCut1234LabelEdges]\,.
\end{equation}

Let us now discuss how the correct infrared pole structure arises in the right-hand side of eq.~\eqref{eq:diagConjbst}. We have seen in Section~\ref{sec:coproductCheck} that $\jhat_4$ is finite, and so the second line of eq.~\eqref{eq:diagConjbst} is free of poles in $\eps$. Hence, the poles must arise entirely form the terms involving bubbles in the first line. 
The relation between an uncut integral and its one- and two-propagator cuts presented in \refE{eq:PCI} plays a central role, in this case not to cancel poles introduced by the bubbles, but rather to reproduce the correct poles of the uncut integrals. For our example, it reads: 
\begin{equation}
	\boxInsertLow[bstCut13LabelEdges]
	+\boxInsert[bstCut24LabelEdges]=-\epsilon \,\boxInsert[bstLabelEdges]=\mathcal{O}(\eps^{-1})\,.
\end{equation}
The contribution of the poles of the bubbles thus correctly reproduces all coaction terms with weight zero in the first factor, and in particular the double pole of the massless box.

The other feature we wish to illustrate with this example is more subtle: we will explain how the coaction terms of the form $J_4(s,t)\otimes 1$ are reproduced. 
Because the massless box is not reducible to simpler topologies,  the structure of the diagrammatic coaction implies that the maximal cut cannot vanish if we are to correctly reproduce these coaction components. However, it is not just the box that appears with the maximal cut in the diagrammatic coaction, but rather the combination~$\jhat_4(s,t)$, as seen in the second line of \refE{eq:diagConjbst}.
In \refS{sec:cancellationOfPoles}, we argued that for  finite integrals 
the issue is resolved through \refE{eq:maxNMaxRel}, relating maximal cuts to next-to-maximal cuts of diagrams with an even number of propagators. The mechanism must be different here, because all next-to-maximal cuts of $J_4(s,t)$ vanish.

To see how the correct behaviour is restored, note that the triangle diagrams
in $\jhat_4(s,t)$
  correspond to the integrals $J_3(s)$ and $J_3(t)$: these are divergent and thus reducible to bubbles according to \refE{eq:redTriangles}, leading to the relation 
\begin{equation}
	\jhat_4(s,t)=J_4(s,t)+\frac{1}{\epsilon}J_2(s)+\frac{1}{\epsilon}J_2(t)\,.
\end{equation}
We can then rewrite \refE{eq:diagConjbst} as 
\begin{align}
	\Delta_{1/2}&\left[\boxInsert[bstLabelEdges]\right]=
	\bubInsert[bub0msLabelEdges]\otimes \left(\boxInsertLow[bstCut13LabelEdges]+\frac{1}{\epsilon}\boxInsertLow[bstCut1234LabelEdges]\right)\\
	+&\bubInsert[bub0mtLabelEdges]\otimes \left(\boxInsert[bstCut24LabelEdges]+\frac{1}{\epsilon}\boxInsertLow[bstCut1234LabelEdges]\right)
	+\boxInsert[bstLabelEdges]\otimes \boxInsertLow[bstCut1234LabelEdges]\,.\nonumber
\end{align}
The sums in parentheses are finite weight one functions and cannot contribute to coaction components with weight zero in the second entry. It follows that $J_4(s,t)\otimes 1$ is correctly reproduced to all orders in the Laurent expansion by the last term on the right-hand side.

\subsection{Two-mass-easy and two-mass-hard boxes}

In this section we illustrate the difference between the so-called `two-mass-easy' box $J_{4}(s,t,p_1^2,p_3^2)$ and the `two-mass-hard' box $J_4(s,t,p_1^2,p_2^2)$. Both integrals are divergent. The main difference between the two is that in the two-mass-easy case all triple cuts vanish, and in the two-mass-hard case one of them is nonzero.

The diagrammatic coaction on the two-mass-easy box is:
\begin{align}\bsp\label{eq:diagConjbstp1p3}
	&\Delta_{1/2}\left[\boxInsert[bst13LabelEdges]\right]=
	\bubInsert[bub0m1LabelEdges]\otimes \boxInsertLow[bst13Cut12LabelEdges]
	+\bubInsert[bub0m3BoxLabelEdges]\otimes \boxInsert[bst13Cut34LabelEdges]\\
	&+\bubInsert[bub0msLabelEdges]\otimes \boxInsertLow[bst13Cut13LabelEdges]
	+\bubInsert[bub0mtLabelEdges]\otimes \boxInsert[bst13Cut24LabelEdges]
	+\left\{\boxInsert[bst13LabelEdges]\right.\\
	&\left.+\frac{1}{2}\left(\triInsert[ts1LabelEdges]+\triInsert[ts3LabelEdges]
	+\triInsert[tt1LabelEdges]+\triInsert[tt3LabelEdges]\right)\right\}
	\otimes\boxInsertLow[bst13Cut1234LabelEdges]\, .
\esp\end{align}
Cuts of one or three propagators vanish on general grounds.  For cuts of two propagators, we have, for example,
\begin{align}\bsp
	\Delta_{1/2}\left[\boxInsertLow[bst13Cut13LabelEdges]\right]=&
	\bubInsert[bub0msCutPLabelEdges]\otimes \boxInsertLow[bst13Cut13LabelEdges]
	+\left\{\boxInsert[bst13LabelEdges]\right.\\
	&\left.+\frac{1}{2}\left(\triInsert[ts1Cut13LabelEdges]+\triInsert[ts3Cut13LabelEdges]\right)\right\}
	\otimes\boxInsertLow[bst13Cut1234LabelEdges]\,.
\esp\end{align} 
The other double cuts are similar, and for the quadruple cut we find
\begin{align}\bsp
	\Delta_{1/2}\left[\boxInsertLow[bst13Cut1234LabelEdges]\right]=
	\boxInsertLow[bst13Cut1234LabelEdges]
	\otimes\boxInsertLow[bst13Cut1234LabelEdges]\,.
\esp\end{align}

The diagrammatic coaction on the two-mass-hard box is:
\begin{align}\label{eq:diagConjbstp1p2}
	\Delta_{1/2}&\left[\boxInsert[bst12LabelEdges]\right]=
	\bubInsert[bub0m1LabelEdges]\otimes \boxInsertLow[bst12Cut12LabelEdges]
	+\bubInsert[bub0m2LabelEdges]\otimes \boxInsert[bst12Cut23LabelEdges]\nonumber\\
	&+\bubInsert[bub0msLabelEdges]\otimes \boxInsertLow[bst12Cut13LabelEdges]
	+\bubInsert[bub0mtLabelEdges]\otimes \boxInsert[bst12Cut24LabelEdges]\nonumber\\
	&+\triInsert[ts12LabelEdges] \otimes \boxInsertLow[bst12Cut123LabelEdges]
	+\left\{\boxInsert[bst12LabelEdges]+\frac{1}{2}\left(
	\triInsert[ts12LabelEdges]+\triInsert[tsBisLabelEdges] \right.\right. \nonumber\\
	&\left.\left.+\triInsert[tt1LabelEdges]+\triInsert[tt2LabelEdges]\right)\right\}
	\otimes\boxInsertLow[bst12Cut1234LabelEdges]\, .
\end{align}
For its cuts, we have 
\begin{align}\bsp
	\Delta_{1/2}&\left[\boxInsertLow[bst12Cut13LabelEdges]\right]=
	\bubInsert[bub0msCutPLabelEdges]\otimes\boxInsertLow[bst12Cut13LabelEdges]
	+\triInsert[ts12Cut13LabelEdges] \otimes\boxInsertLow[bst12Cut123LabelEdges]\\
	&+\left\{\boxInsertLow[bst12Cut13LabelEdges]+\frac{1}{2}\left(
	\triInsert[ts12Cut13LabelEdges]+\triInsert[tsBisCut13LabelEdges]\right)\right\}
	\otimes\boxInsertLow[bst12Cut1234LabelEdges]\, ,
\esp\end{align}
\begin{align}\bsp
	\Delta_{1/2}&\left[\boxInsert[bst12Cut24LabelEdges]\right]=
	\bubInsert[bub0mtCutPLabelEdges]\otimes\boxInsert[bst12Cut24LabelEdges]\\
	&+\left\{\boxInsert[bst12Cut24LabelEdges]+\frac{1}{2}\left(
	\triInsert[tt1Cut24LabelEdges]+\triInsert[tt2Cut24LabelEdges]\right)\right\}
	\otimes\boxInsertLow[bst12Cut1234LabelEdges]\, .
\esp\end{align} 
All other double cuts are similar. For the triple and quadruple cuts, we find,
\begin{align}\bsp
	\Delta_{1/2}\left[\boxInsertLow[bst12Cut123LabelEdges]\right]&=
	\triInsert[ts12Cut123LabelEdges] \otimes\boxInsertLow[bst12Cut123LabelEdges]\\
	&+\left(\boxInsertLow[bst12Cut123LabelEdges]+\frac{1}{2}
	\triInsert[ts12Cut123LabelEdges]\right)
	\otimes\boxInsertLow[bst12Cut1234LabelEdges]\, ,
\esp\end{align} 
\begin{align}\bsp
	\Delta_{1/2}\left[\boxInsertLow[bst12Cut1234LabelEdges]\right]&=
	\boxInsertLow[bst12Cut1234LabelEdges]
	\otimes\boxInsertLow[bst12Cut1234LabelEdges]\, .
\esp\end{align}

We first comment on the appearance of two-mass triangles in the coaction of the two boxes. While the triangle with two massive external legs is obviously symmetric under exchange of the two channels, it is antisymmetric after normalization by the maximal cut (see eqs. \eqref{eq:tp2p3} and \eqref{eq:lstp2p3} for explicit expressions), and we must specify how we relate the two-mass triangles in eqs.~\eqref{eq:diagConjbstp1p3} and \eqref{eq:diagConjbstp1p2} with the results given in Appendix \ref{sec:tp2p3Res}. This is fixed by requiring that the coactions on two-mass boxes reduce to \refE{eq:diagConjbst} when external legs become massless. The correct interpretation is 
\begin{equation}
 	J_3(s,p_i^2)\equiv \left(p_i^2-s\right)\widetilde J_3\left(s,p_i^2\right)=
 	J_3(s)-J_3(p_i^2)
 	\,,
\end{equation}
and similarly for $ J_3(t,p_i^2)$, taking $s\to t$ in the above expression. If $p_i^2\to0$, we do get $J_3(s,p_i^2)\to J_3(s)$ as needed.

The main difference between these two boxes is  that in one case all next-to-maximal cuts vanish, whereas in the other one of them is nonzero. The main implication of this difference is in the way the trivial coaction components $J_{4}(s,t,p_1^2,p_3^2)\otimes 1$ and $J_{4}(s,t,p_1^2,p_2^2)\otimes 1$ are reproduced. For the two-mass-easy box, where all triple cuts vanish, they are reproduced by exactly the same procedure as for the massless box. For the two-mass hard box, we have a mix of that procedure with the general mechanism described in \refS{sec:cancellationOfPoles}: while the contribution of the three-mass triangle to coaction components with weight zero in the right factor cancels because of the relation between maximal and next-to-maximal cuts given in  \refE{eq:maxNMaxRel}, the contributions to those coaction terms coming from the divergent triangles that appear only in $\jhat_{4}(s,t,p_1^2,p_2^2)$ cancel by the same mechanism as in the massless or two-mass-easy boxes.

\subsection{Box with massive internal propagator}\label{sec:bstm}

As a final example, we present the box with a single massive propagator, which is the first example of a four-point integral for which at least one cut of each type does not vanish. The relation between the  diagrammatic coaction and  the coaction on MPLs thus relies on the interplay of many different uncut and cut integrals.  There are no essential new features in this example compared to the ones discussed above, so we simply write the diagrammatic coaction below for illustration purposes.
\begin{align}
	\Delta_{1/2}&\left[\boxInsert[bstm1LabelEdges]\right]=
	\tadInsert[tad1]\otimes\boxInsertLow[bstm1Cut1LabelEdges]+
	\left(\bubInsert[bub1msLabelEdges]+\frac{1}{2}\tadInsert[tad1]\right)
	\otimes\boxInsertLow[bstm1Cut13LabelEdges]\nonumber\\
	+&\bubInsert[bub0mtLabelEdges]
	\otimes\boxInsert[bstm1Cut24LabelEdges]+\triInsert[ttm1LabelEdges]\otimes\boxInsertLow[bstm1Cut124LabelEdges]
	+\Biggg\{\boxInsert[bstm1LabelEdges] \\
	+& \frac{1}{2}\left(\triInsert[tsm1LabelEdgesBis]+
	\triInsert[tsm1LabelEdges]+\triInsert[ttLabelEdgesBis]+\triInsert[ttm1LabelEdges]\right)\Biggg\}\otimes\boxInsertLow[bstm1Cut1234LabelEdges]\nonumber\, ,
\end{align}
\begin{align}
	\Delta_{1/2}&\left[\boxInsertLow[bstm1Cut1LabelEdges]\right]=
	\tadInsertLow[tad1CutM]\otimes\boxInsertLow[bstm1Cut1LabelEdges]\nonumber\\
	+&
	\left(\bubInsert[bub1msCut1LabelEdges]+\frac{1}{2}\tadInsert[tad1CutM]\right)
	\otimes\boxInsertLow[bstm1Cut13LabelEdges]+\triInsert[ttm1Cut1LabelEdges]\otimes\boxInsertLow[bstm1Cut124LabelEdges]\\
	+&\left\{\boxInsertLow[bstm1Cut1LabelEdges]+\frac{1}{2}\left(
	\triInsert[tsm1Cut1LabelEdgesBis]+\triInsert[tsm1Cut1LabelEdges]+
	\triInsert[ttm1Cut1LabelEdges]\right)\right\}\otimes\boxInsertLow[bstm1Cut1234LabelEdges]\nonumber\, ,
\end{align}
\begin{align}\bsp\label{eq:diagConjbstm1Cut13}
	\Delta_{1/2}&\left[\boxInsertLow[bstm1Cut13LabelEdges]\right]=
	\bubInsert[bub1msCutPLabelEdges]\otimes\boxInsertLow[bstm1Cut13LabelEdges]\\
	+&\left\{\boxInsertLow[bstm1Cut13LabelEdges]+\frac{1}{2}\left(
	\triInsert[tsm1Cut13LabelEdgesBis]+\triInsert[tsm1Cut13LabelEdges]\right)\right\}\otimes\boxInsertLow[bstm1Cut1234LabelEdges]\, ,
\esp\end{align}
\begin{align}\bsp\label{eq:diagConjbstm1Cut24}
	\Delta_{1/2}&\left[\boxInsert[bstm1Cut24LabelEdges]\right]=
	\bubInsert[bub0mtCutPLabelEdges]\otimes\boxInsert[bstm1Cut24LabelEdges]
	+\triInsert[ttm1Cut24LabelEdges]\otimes\boxInsertLow[bstm1Cut124LabelEdges]\\
	+&\left\{\boxInsert[bstm1Cut24LabelEdges]+\frac{1}{2}\left(
	\triInsert[ttm1Cut24LabelEdges]+\triInsert[ttCut24LabelEdgesBis]\right)\right\}\otimes\boxInsertLow[bstm1Cut1234LabelEdges]\, .
\esp\end{align}
All other double cuts are similar. For the triple and quadruple cuts, we find,
\begin{align}\bsp
	\Delta_{1/2}\left[\boxInsertLow[bstm1Cut124LabelEdges]\right]=&
	\triInsert[ttm1Cut124LabelEdges]\otimes\boxInsertLow[bstm1Cut124LabelEdges]\\
	+&\left(\boxInsertLow[bstm1Cut124LabelEdges]+\frac{1}{2}
	\triInsert[ttm1Cut124LabelEdges]\right)\otimes\boxInsertLow[bstm1Cut124LabelEdges]\, ,
\esp\end{align}
\begin{align}\bsp
	\Delta_{1/2}\left[\boxInsertLow[bstm1Cut1234LabelEdges]\right]=
	\boxInsertLow[bstm1Cut1234LabelEdges]\otimes\boxInsertLow[bstm1Cut1234LabelEdges].
\esp\end{align}


\section{The coaction and extended dual conformal invariance}
\label{sec:dci}

It is well known that in the limit where the number of dimensions matches the number of propagators, $D=n$, the integrals develop an enhanced symmetry, known as (extended) dual conformal symmetry~\cite{Drummond:2006rz,Henn:2010ir,Henn:2011xk,Caron-Huot:2014lda,Broadhurst:1993ib}. In the case of massless propagators, the dual conformal symmetry reduces to the conformal symmetry in the dual coordinates $q_i$. Since we consider the number of dimensions to be even, this implies that only integrals with an even number of propagators are dual conformally invariant (DCI) in our setting. It is easy to check that one-loop cut integrals are DCI whenever the corresponding uncut integral is. In this section we show that the coaction on DCI integrals can be expressed entirely in terms of DCI integrals.

Let us define 
\beq
K_G \equiv  \lim_{\eps\to0} J_G\,,
\eeq
whenever the limit exists. In the following we only discuss the case where all propagators are massive. The extension to massless propagators is straightforward. If all propagators are massive, then the only divergent integrals in our basis are tadpoles, and we know from the discussion in Section~\ref{sec:cancellationOfPoles} that the pole cancels from the coaction. If $n_G$ is even, then $K_G$ is DCI. The coaction~\eqref{eq:Coaction_cut_integrals} on $K_G$, however, involves integrals in the first entry with an odd number of propagators, which are not DCI. We now show that all integrals with an odd number of propagators drop out of the coaction.
If we collect all the contributions from integrals with an odd number of propagators (which appear only in the first entry), we find
\beq\bsp\label{eq:DCI_proof_1}
\sum_{\substack{C\subset E_G\\ n_C\textrm{ odd}}} \plainK_{G_C}\otimes \Bigg(
{\cC_{C}  \plainK_{G}  + \frac{1}{2} \sum_{e\in E_G\setminus C}\cC_{Ce}  \plainK_{G} }
\Bigg)\,,
\esp\eeq
with $\cC_{Ce}\equiv \cC_{C\cup e}$. In refs.~\cite{HwaTeplitz,Abreu:2017ptx} it was shown that there are relations among cut integrals in integer dimensions,
\beq\label{eq:DCI_cut_relations_app_anyN}
2\,\cC_C
\plainK_{G} 
+ \sum_{e\in E_G \setminus C}\,\cC_{Ce}
\plainK_{G} = 0\,,\qquad 
\textrm{for $n_G$ even and $n_{C}$ odd.}
\eeq
This relation is a direct consequence of the homological relation~\eqref{eq:homo_odd} and the fact that a DCI integral has no singularity at infinity~\cite{Abreu:2017ptx}.
It implies that the terms inside the brackets in eq.~\eqref{eq:DCI_proof_1} vanish. The coaction on a DCI integral therefore only has integrals with an even number of propagators in the first factor, which are themselves DCI. Since all the cuts of a DCI integral are themselves DCI, the coaction on a DCI integral only involves DCI integrals. Explicitly, we find,
\beq\label{eq:coaction_DCI}
\Delta_{\rm MPL}\left({K}_G\right)=
1\otimes K_G+K_G\otimes1 +\!\!\!\!\!\!\!\!\!\!\!\! \sum_{\substack{X\subset E_G,\\0<n_X<n_G;\, n_X\textrm{ even}}}\!\!\!\!\!\!\!\!\!\! {K}_{G_X}\otimes {\cal C}_X {K}_G\,.
\eeq
We find it remarkable that the coaction on DCI integrals is given by the incidence coproduct. The coaction~\eqref{eq:coaction_DCI} has already appeared in the work of Goncharov~\cite{Goncharov:1996}, where he has considered the coaction (or rather, the coproduct) on certain classes of integrals in projective space with singularities along some quadric. In Appendix~\ref{app:compact} we show that every one-loop integral, cut or uncut, can be cast in precisely such a form, and in the case of an even number of dimensions we reproduce precisely the class of integrals considered by Goncharov. In other words, our conjecture~\eqref{eq:Coaction_cut_integrals} reduces to the coproduct defined by Goncharov in ref.~\cite{Goncharov:1996} in the special case of DCI integrals in an even number of dimensions.

Integrals with an odd number of propagators are not DCI, but  can be obtained from DCI integrals by sending a point  to infinity. {We refer the reader to Appendix~\ref{app:pointToInf} where this is worked out explicitly.} The same conclusion holds for cut integrals with an odd number of propagators, provided that the edge corresponding to the point that is sent to infinity is not cut. Otherwise, we rewrite the cut integral in terms of integrals where this edge is not cut. Consider for example the case where the number of cut propagators is even, and we want to send the cut edge $e_0$ to infinity. We can then use eq.~\eqref{eq:DCI_cut_relations_app_anyN} and express this cut integral as a linear combination where the edge $e_0$ is uncut,
\beq\label{eq:infty_cut}
\cC_{Ce_0}
\plainK_{G} = -2\,\Bigg(\cC_C
\plainK_{G} 
+ \frac{1}{2}\sum_{\substack{e\in E_G \setminus C\\ e\neq e_0}}\,\cC_{Ce}
\plainK_{G}\Bigg)\,.
\eeq
A similar relation can be derived from eq.~\eqref{eq:homo_even} in the case where the number of cut propagators is odd.
The resulting formula for the coaction on integrals with an odd number of propagators in integer dimensions is in perfect agreement with the conjecture~\eqref{eq:Coaction_cut_integrals}. In particular, we see how the terms proportional to factors of 1/2 appear in a natural way in the coaction starting from eq.~\eqref{eq:coaction_DCI} and inserting eq.~\eqref{eq:infty_cut}, which is a nontrivial consistency check of our conjecture.


\section{Discontinuities}\label{sec:discontinuities}

In this section we show the consistency of our conjecture (\ref{eq:graph_conjecture}) with known results on discontinuities of one-loop Feynman integrals, in particular with the so-called first-entry condition.

\subsection{Discontinuities and the diagrammatic coaction}

We have seen in eq.~\eqref{eq:disc_coproduct} that the coaction on the discontinuity of a function can be obtained by computing the discontinuity of the function appearing in its first entry. Equation~\eqref{eq:disc_coproduct} provides a strong consistency check on our conjectured coaction on Feynman graphs. Indeed, it is well known that cut integrals compute discontinuities of Feynman integrals. The aim of this section is to analyze in detail the interplay between discontinuities of Feynman integrals in our conjectured coaction.

We start by defining more precisely what we mean by discontinuities and how they are related to cut integrals. Since this is in principle well known, we will be brief and refer to the literature (see, e.g., refs.~\cite{Abreu:2014cla,Abreu:2015zaa,Bloch:2015efx,Cutkosky:1960sp,Abreu:2017ptx,PhamInTeplitz,HwaTeplitz,Froissart,PhamBook,PhamCompact,tHooft:1973pz,Veltman:1994wz} and references therein). In a nutshell, singularities or branch points may occur for kinematical configurations for which the Landau conditions are satisfied. At one loop, there are two types of solutions to the Landau conditions:
\begin{itemize}
\item Singularities of the first type: those correspond to pinch singularities where a subset $C$ of propagators go on shell. The kinematic configurations for which this happens form the Landau variety $L_C$, and they are characterized by a vanishing of the modified Cayley determinant, $Y_C=0$.
\item Singularities of the second type: those corresponds to the loop momentum being pinched at infinity in addition to the singularities from a subset $C$ of on-shell propagators. The kinematic configurations for which this happens form the Landau variety $L_{\infty C}$, and they are characterized by a vanishing of the Gram determinant, $\textrm{Gram}_C=0$.
\end{itemize}
The discontinuity of a one-loop integral around the Landau variety $L_{C}$ (where $C$ may or may not contain $\infty$) is defined as the difference before and after analytically continuing the external kinematics along a small positively-oriented circle around $L_{C}$. The discontinuity can be expressed in terms of cut integrals
\beq\label{eq:disc_cut}
\textrm{Disc}_CJ_G = -N_C\,\cC_CJ_G\mod i\pi\,,\quad C\subseteq E_G\cup \{\infty\}\,,
\eeq
where $N_C$ is an integer whose precise value is irrelevant in the following.\footnote{For the precise normalization, we refer to ref.~\cite{Abreu:2017ptx}.} If $C\nsubseteq E_G\cup\{\infty\}$, then $\textrm{Disc}_CJ_G=0$.

Let us now show that our conjecture is consistent with eqs.~\eqref{eq:disc_coproduct} and~\eqref{eq:disc_cut}. We start by analyzing discontinuities around Landau varieties of the first type. If $C\subseteq E_G$, we have
\beq\bsp
\Delta{_{\textrm{MPL}}}(\textrm{Disc}_CJ_G) &\,= -N_C\,\Delta{_{\textrm{MPL}}}(\cC_CJ_G)=-N_C\sum_{C\subseteq X\subseteq E_G}\cC_C\widehat{J}_{G_X}\otimes \cC_XJ_G \\
&\,=\sum_{C\subseteq X\subseteq E_G}\left(\textrm{Disc}_C\widehat{J}_{G_X}\right)\otimes \cC_XJ_G\\
&\,=(\textrm{Disc}_C\otimes \textrm{id})\sum_{C\subseteq X\subseteq E_G}\widehat{J}_{G_X}\otimes \cC_XJ_G\\
&\,=(\textrm{Disc}_C\otimes \textrm{id})\Delta{_{\textrm{MPL}}}(J_G)\,,
\esp\eeq
where in the last step we used the fact that $\textrm{Disc}_CJ_{G_X}=0$ if $X\nsubseteq C$. In the case of singularities of the second type, we can use eqs.~\eqref{eq:homo_even} and~\eqref{eq:homo_odd} and reduce the problem to the one we have just studied. Hence we see that our conjecture is consistent with eqs.~\eqref{eq:disc_coproduct} and~\eqref{eq:disc_cut} for singularities of both the first and second type. This provides a strong consistency check on the conjecture.

\subsection{The first-entry condition}

Closely connected to discontinuities and the coaction on Feynman integrals is the \emph{first-entry condition}. In its original and weakest version, the first entry condition states that the symbol (i.e., the maximal iteration of the coproduct) of a Feynman integral with massless propagators can always be cast in a form such that the first entries are all Mandelstam invariants~\cite{Gaiotto:2011dt}. This reflects the fact that Feynman integrals with massless propagators can only have branch points whenever a Mandelstam invariant is zero or infinite. The first entry condition was extended to integrals with massive propagators and thresholds in ref.~\cite{Abreu:2015zaa}. It was soon realized that the first entry condition can be generalized to a statement about the coaction on Feynman integrals: the coaction can always be written such that the branch points of each of the first entries are a subset of the branch points of the original integral~\cite{Chavez:2012kn,Drummond:2013nda,Drummond:2012bg,Dixon:2013eka,Dixon:2014voa,Dixon:2012yy}.

The first entry condition was given an even stronger formulation in ref.~\cite{Brown:2015fyf}, where it was shown that the first entries in the (motivic) coaction on (motivic) Feynman integrals are always themselves (motivic) Feynman integrals with fewer propagators. The discussion of ref.~\cite{Brown:2015fyf}, however, was restricted to finite integrals without the setting of dimensional regularization. Our diagrammatic coaction~\eqref{eq:Coaction_cut_integrals} is consistent with the findings of ref.~\cite{Brown:2015fyf}, at least in the case of one-loop integrals, and it extends it to integrals defined in dimensional regularization: the first entries in the coaction of a one-loop integral in dimensional regularization are themselves always one-loop integrals.

Our conjecture, however, goes beyond the usual statement of the first entry condition, because it 
 applies equally well to  cut and uncut integrals. We can therefore present a sharper version of the first entry condition, which treats  cut and uncut integrals in a uniform way (at least at one loop): 

\begin{center}
\emph{
In the coaction of a (cut) Feynman integral,
the first entries are 
 themselves Feynman integrals, with a subset of propagators but the same set of cut propagators. 
}
\end{center}

\noindent This statement is the strongest version of the first entry condition, and it contains all the previous versions as special cases.


\section{The diagrammatic coaction, differential equations and the symbol}\label{sec:diffEqAndAlphabet}

In this section we show the consistency of our conjecture (\ref{eq:graph_conjecture}) with the structure of differential equations satisfied by one-loop (cut) Feynman integrals. In particular, we identify the coefficient functions of the differential equations as derivatives of a small subset of cuts of the original integral, which are computed in Appendix~\ref{app:compact}. Using this information, we explicitly determine the differential equations satisfied by any general one-loop (cut) integral. Since the symbol of a pure function contains the same information as the differential equation it satisfies, we show how to iteratively construct the symbol of any 
one-loop Feynman integral. {We refer the reader to Appendix \ref{app:notation} for a summary of the notation used in this paper.}

\subsection{Differential equations for one-loop integrals}\label{sec:diffEq}
We first show that our conjectured coaction (\ref{eq:graph_conjecture}) is compatible with the 
fact that differential operators act only on the second entry,
 as stated in eq.~\eqref{eq:der_coproduct}. It is well known that Feynman integrals satisfy first-order linear differential equations~\cite{Kotikov:1990kg,Kotikov:1991hm,Kotikov:1991pm,Gehrmann:1999as,Henn:2013pwa}. In particular, the integrals $J_G$ form a basis, so the derivative of $J_G$ can be expressed as a linear combination of basis integrals with fewer propagators. Since in addition the basis integrals $J_G$ are all pure, the differential equation satisfied by these integrals can be cast in a particularly simple form~\cite{Henn:2013pwa}. More precisely, if $G$ is a one-loop graph and $X\subseteq E_G$, then we have
\beq\label{eq:diffeq}
dJ_{G_X} = \sum_{\emptyset\neq Y\subseteq X}\Omega_{X,Y}\,J_{G_Y}\,,\quad \Omega_{X,Y} = \Omega_{X,Y}^{(0)} + \eps\,\Omega_{X,Y}^{(1)}\,,
\eeq
where $\Omega_{X,Y}^{(i)}$ is a matrix whose entries are labeled by subsets of propagators and  given by logarithmic one-forms with algebraic arguments. 
Since cut integrals compute discontinuities, it was argued in ref.~\cite{Abreu:2017ptx} that one-loop cut integrals satisfy the same differential equations as their uncut analogues, in agreement with the method of reverse-unitarity~\cite{Anastasiou:2002yz,Anastasiou:2002wq,Anastasiou:2002qz,Anastasiou:2003yy,Anastasiou:2003ds} (see also the recent results in \cite{Frellesvig:2017aai,Zeng:2017ipr,Bosma:2017ens}),
\beq\label{eq:cutdiffeq}
d\cC_CJ_{G_X} = \sum_{C\subseteq Y\subseteq X}\Omega_{X,Y}\,\cC_CJ_{G_Y}\,,
\eeq
where the matrix $\Omega$ is identical to the one in eq.~\eqref{eq:diffeq}. 
{Therefore it suffices to analyze the case of uncut integrals.}

Let us now show that our conjecture is compatible with eq.~\eref{eq:der_coproduct}.
If we act with $\Delta{_{\textrm{MPL}}}$ on eq.~\eqref{eq:diffeq}, we find
\beq\bsp
\Delta{_{\textrm{MPL}}}(dJ_{G_X}) &\,= \sum_{\emptyset\neq Y\subseteq X}\Omega_{X,Y}\,\Delta{_{\textrm{MPL}}}(J_{G_Y})\\
&\,= \sum_{\substack{C\subseteq Y\subseteq X \\ C\neq\emptyset}}\Omega_{X,Y}\,\widehat{J}_{G_C}\otimes \cC_CJ_{G_Y}\\
&\,= \sum_{\emptyset\neq C\subseteq X }\widehat{J}_{G_C}\otimes d\cC_CJ_{G_X}\\
&\,=(\textrm{id}\otimes d)\Delta{_{\textrm{MPL}}}(J_{G_X})\,,
\esp\eeq
where in the last step we have used eq.~\eqref{eq:diffeq}. Hence, we see that our conjectured formula for the coaction is consistent with eq.~\eref{eq:der_coproduct}.

\subsection{Differential equations of one-loop integrals}
\label{sec:alph_diff_eq}
In previous sections we have mostly concentrated on providing evidence for the correctness of our conjecture~\eqref{eq:graph_conjecture}.
In this section we explore some implications of our conjecture for one-loop integrals. In particular, the previous discussion implies that it is possible to extract the system of differential equations satisfied by the one-loop basis integrals $J_G$ from our conjecture. In the remainder of this section we discuss this in more detail.

Consider a pure function $F_w$ of weight $w$. Then we can write
\beq
\Delta_{w-1,1}(F_w) = \sum_iF_{i,w-1}\otimes \log u_i\,,
\eeq
for some pure functions $F_{i,w-1}$ of weight $w-1$ and algebraic functions $u_i$. 
Equation~\eqref{eq:der_coproduct} then implies that
\beq\label{eq:pure_diffeq}
dF_w = \sum_iF_{i,w-1}\, d\log u_i\,,
\eeq
i.e., we can obtain a first order differential satisfied by $F_w$ if we know the $(w-1,1)$ component of its coaction.

Now consider the one-loop integral $J_G$. We can start from eq.~\eqref{Weight_contr} and isolate all the terms in the coaction that have a function of weight one in the second entry. We restrict our discussion to finite integrals, so that the only divergent terms in the coaction are $J_1$ and $\jhat_2$. The extension to divergent integrals is straightforward. Since our conjecture treats graphs with even and odd numbers of edges differently, we need to distinguish two separate cases:
\begin{itemize}
\item For $n_G$ odd, only the next-to-next-to-maximal, next-to-maximal and maximal cuts contribute to the relevant orders in the $\eps$ expansion: 
\beq\bsp
\label{eq:delta1wm1Odd}
	\Delta_{w_{G}^j-1,1}J^{(j)}_G&= J_G^{(j-1)}\otimes \cutRes^{(1)}_{E_G}J_G+
	\!\!\!\sum_{\substack{ X\subset E_G\\n_X=n_G-1}}\!\!\!\jhat^{(j)}_{G_X}\otimes\cutRes^{(0)}_XJ_G
\,+	\!\!\!\sum_{\substack{ X\subset E_G\\n_X=n_G-2}}\!\!\!J^{(j)}_{G_X}\otimes\cutRes^{(0)}_XJ_G\,.
\esp\eeq
Using \refE{eq:pure_diffeq} and summing the different orders in the $\eps$ expansion, we obtain
\beq\bsp\label{eq:diffeq_odd}
	dJ_G=&\, \eps\,J_G \,d\cutRes^{(1)}_{E_G}J_G+\!\!\!
	\sum_{\substack{ X\subset E_G\\n_X=n_G-1}}\!\!\!J_{G_X}\,d\cutRes^{(0)}_XJ_G\\
	+&\,\!\!\!	\sum_{\substack{ X\subset E_G\\n_X=n_G-2}}\!\!\!J_{G_X}\,\Big(d\cutRes^{(0)}_XJ_G+\frac{1}{2}\sum_{e\in E_G\setminus X}d\cutRes^{(0)}_{Xe}J_G\Big)\,.
\esp\eeq
\item For $n_G$ even, in principle there is one more contribution, from the next-to-next-to-next-to-maximal cut:
\beq\bsp\label{eq:delta1wm1Even}
	\Delta_{w_{G}^j-1,1}J^{(j)}_G&=\jhat_G^{(j-1)}\otimes \cutRes^{(1)}_{E_G}J_G
	+\!\!\!\sum_{\substack{ X\subset E_G\\n_X=n_G-1}}\!\!\!J^{(j-1)}_{G_X}\otimes\cutRes^{(1)}_XJ_G\\
&\,+\!\!\!	\sum_{\substack{ X\subset E_G\\n_X=n_G-2}}\!\!\!\jhat^{(j)}_{G_X}\otimes\cutRes_X^{(0)}J_G
+\!\!\!\sum_{\substack{ X\subset E_G\\n_X=n_G-3}}\!\!\!J^{(j)}_{G_X}\otimes\cutRes^{(0)}_XJ_G\,.
\esp\eeq
Using \refE{eq:pure_diffeq} and summing the different orders in the $\eps$ expansion, we obtain
\begin{align}\label{eq:diffeq_even}
\nonumber	dJ_G&=\eps\,J_G\, d\cutRes^{(1)}_{E_G}J_G
	+ \!\!\!\sum_{\substack{ X\subset E_G\\n_X=n_G-1}}\!\!\!\eps\,J_{G_X}\,\Big(d\cutRes^{(1)}_XJ_G+\frac{1}{2}d\cutRes^{(1)}_{E_G}J_G\Big)\\
&\,+	\sum_{\substack{ X\subset E_G\\n_X=n_G-2}}\!\!\!J_{G_X}\,d\cutRes_X^{(0)}J_G
+\sum_{\substack{ X\subset E_G\\n_X=n_G-3}}\!\!\!J_{G_X}\,\Big(d\cutRes^{(0)}_XJ_G+\frac{1}{2}\sum_{e\in E_G\setminus X}d\cutRes_{Xe}^{(0)}J_G\Big)\\
\nonumber&=\eps\,J_G\, d\cutRes^{(1)}_{E_G}J_G
	+ \!\!\!\sum_{\substack{ X\subset E_G\\n_X=n_G-1}}\!\!\!\!\eps\,J_{G_X}\,\Big(d\cutRes^{(1)}_XJ_G+\frac{1}{2}d\cutRes^{(1)}_{E_G}J_G\Big)+	\!\!\!\sum_{\substack{ X\subset E_G\\n_X=n_G-2}}\!\!\!J_{G_X}\,d\cutRes_X^{(0)}J_G\,,
\end{align}
where in the last step we use the fact that the terms with $n_X=n_G-3$  cancel due to eq.~\eqref{eq:DCI_cut_relations_app_anyN}.
\end{itemize}
To make these equations more concrete, consider the fully generic pentagon graph. According to \refE{eq:diffeq_odd}, the corresponding Feynman integral satisfies a differential equation which can be graphically represented as
\begin{align}\bsp
	\label{eq:pent}
	d\!\!&\left[\raisebox{-8mm}{\includegraphics[keepaspectratio=true, width=2cm]{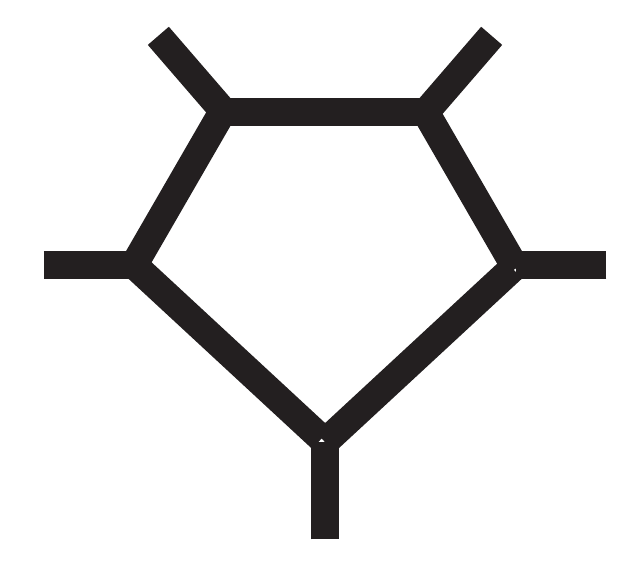}}\right]=
	\sum_{(ijk)}
	\raisebox{-6mm}{\includegraphics[keepaspectratio=true, width=1.6cm]{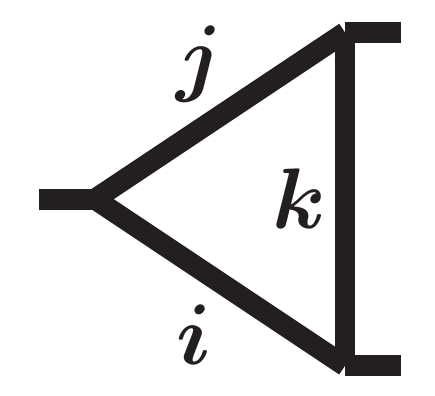}} \,\,\,d\!\!\left[\raisebox{-8mm}{\includegraphics[keepaspectratio=true, width=2cm]{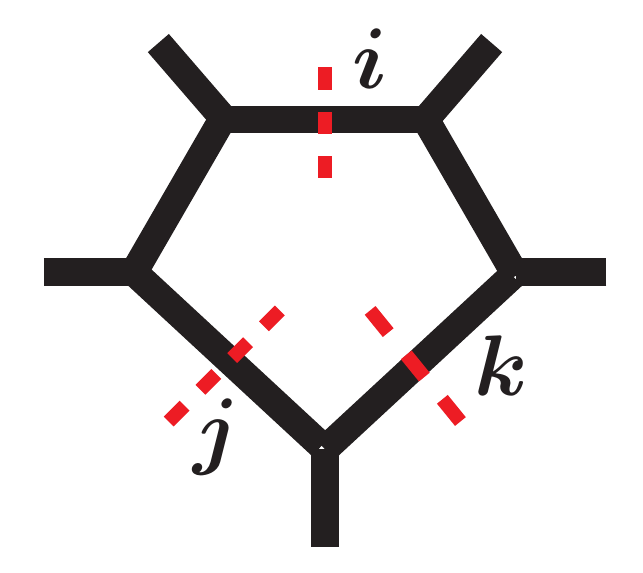}}\Bigg\vert_{\epsilon^0}+\frac{1}{2}\sum_{l}\raisebox{-8mm}{\includegraphics[keepaspectratio=true, width=2cm]{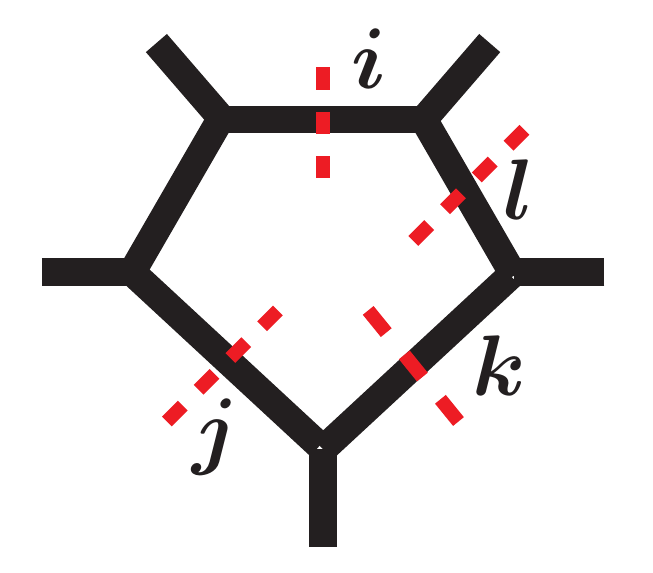}}\Bigg\vert_{\epsilon^0}\right]\\
	&+\sum_{(ijkl)}
	\raisebox{-6mm}{\includegraphics[keepaspectratio=true, width=2cm]{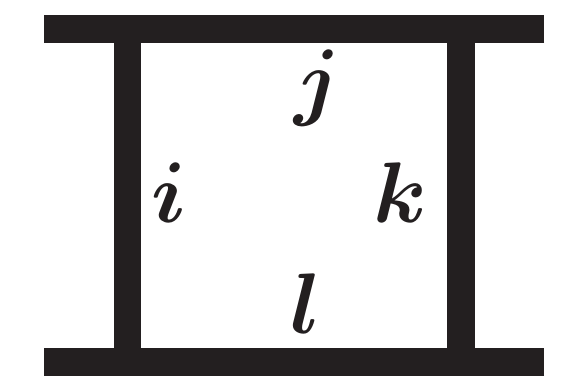}}
	\,\,\,d\!\!\left[\raisebox{-8mm}{\includegraphics[keepaspectratio=true, width=2cm]{./diagrams/pentagonTQuad2.pdf}}\Bigg\vert_{\epsilon^0}\right]+\eps\,\raisebox{-8mm}{\includegraphics[keepaspectratio=true, width=2cm]{./diagrams/pentagon2.pdf}}
\,\,\,d\!\!\left[	\raisebox{-8mm}{\includegraphics[keepaspectratio=true, width=2cm]{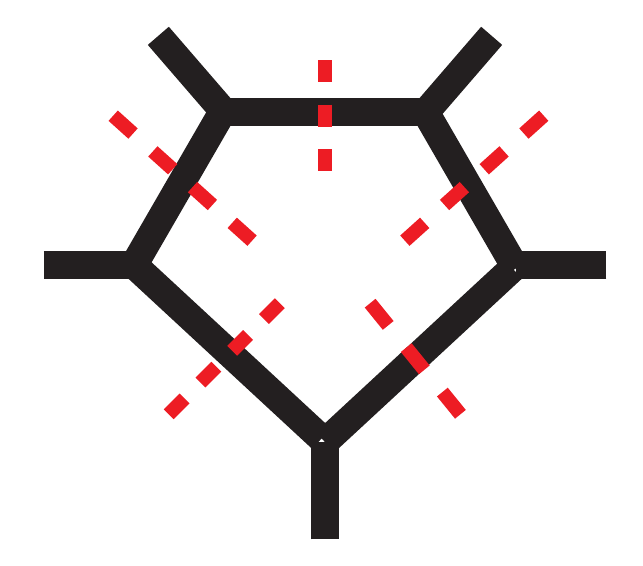}}\Bigg\vert_{\epsilon^1}\right]\,,
\esp\end{align}
where the labels on the edges of the diagrams denote the set of propagators being cut in the second entry.

From eqs.~\eqref{eq:diffeq_odd} and~\eqref{eq:diffeq_even} it follows that the differential equation for $J_G$ is determined by the cuts of $J_G$ with at most two uncut propagators. We compute the complete set of relevant cut integrals in Appendix~\ref{app:compact} and find:
\begin{itemize}
\item if $n_G$ is odd,
\beq\bsp\label{eq:cut_results_odd}
\cC^{(1)}_{E_G}J_G &\, = \ln\left(\frac{\Gram_{E_G}}{Y_{E_G}}\right),\\
\cC^{(0)}_{E_G\setminus e}J_G &\, = \ln\left(
\frac{\sqrt{Y_{E_G}\Gram_{E_G\setminus e}-\Gram_{E_G}Y_{E_G\setminus e}}-\sqrt{-\Gram_{E_G}Y_{E_G\setminus e}}}
{\sqrt{Y_{E_G}\Gram_{E_G\setminus e}-\Gram_{E_G}Y_{E_G\setminus e}}+\sqrt{-\Gram_{E_G}Y_{E_G\setminus e}}}
\right),\\
\cC^{(0)}_{E_G\setminus\{e,f\}}J_G &\, = \ln\left(\frac{a_1+a_2+a_3+a_4+a_5}{a_1+a_2+a_3+a_4-a_5}\right) ,
\esp\eeq
with the $a_i$ given in terms of determinants in \refE{eq:nnmaxOddComp}.

\item if $n_G$ is even,
\beq\bsp\label{eq:cut_results_even}
\cC^{(1)}_{E_G}J_G &\, = \ln\left(\frac{\Gram_{E_G}}{4\,Y_{E_G}}\right), \\
\cC^{(1)}_{E_G\setminus e}J_G &\, = -\ln\left(1+\sqrt{1-\frac{\Gram_{E_G}Y_{E_G\setminus e}}{Y_{E_G}\Gram_{E_G\setminus e}}}\right)-\frac{1}{2}\ln\left(\frac{\Gram_{E_G\setminus e}}{4\,Y_{E_G\setminus e}}\right),\\
\cC^{(0)}_{E_G\setminus\{e,f\}}J_G &\, = \frac{1}{2}\ln\left(
\frac{\sqrt{Y_{E_G\setminus e}Y_{E_G\setminus f}-Y_{E_G}Y_{E_G\setminus \{e,f\}}}+\sqrt{-Y_{E_G}Y_{E_G\setminus\{e,f\}}}}
{\sqrt{Y_{E_G\setminus e}Y_{E_G\setminus f}-Y_{E_G}Y_{E_G\setminus \{e,f\}}}-\sqrt{-Y_{E_G}Y_{E_G\setminus\{e,f\}}}}
\right).
\esp\eeq

\end{itemize}

Let us make some comments about this result. First we see that the differential equation of $J_G$ relates the integral only to integrals with at most two propagators less.\footnote{There is an exception to this rule coming from pentagon integrals with massless external legs. In this case the differential equation may involve triangle integrals that are reducible to bubble integrals.} We find it remarkable that for $n_G$ even the contribution from integrals with three propagators less cancels from the final result, although these terms would be allowed from weight considerations. For example, the differential of a hexagon integral in $D=6-2\eps$ (a weight three function) does not involve triangle integrals in $D=4-2\eps$ dimension (which are functions of weight two). 

Second, we stress that the result for the differential equations was obtained only using our conjecture for the coaction and the results for the cut integrals in eqs.~\eqref{eq:cut_results_odd} and~\eqref{eq:cut_results_even}. While it is in principle straightforward to obtain the same result for a specific integral $J_G$ using IBP relations, in practice the IBP reduction of the integrals may be computationally very heavy  for integrals depending on many scales. Our approach completely bypasses the problem of IBP reduction for integrals depending on many scales.

Finally, let us show what our differential equation~\eqref{eq:diffeq_even} becomes in the special case of DCI integrals. Using the notation of Section~\ref{sec:dci} (and restricting the discussion to convergent integrals) we find that, in the limit $\eps\to0$, eq.~\eqref{eq:diffeq_even} reduces to
\begin{align}\label{eq:diffeq_dci}
	dK_G&=	\sum_{\substack{ X\subset E_G\\n_X=n_G-2}}\!\!\!K_{G_X}\,d\cutRes_XK_G\,,\quad n_G\textrm{ even}\,.
\end{align}
We see that the differential of a DCI one-loop integral only involves DCI integrals with two propagators less, and in this way we reproduce a result of ref.~\cite{Spradlin:2011wp}. Starting from eq.~\eqref{eq:diffeq_odd}, we can obtain a similar differential equation for all (finite) one-loop integrals in integer dimensions with an odd number of propagators,
\beq\bsp\label{eq:diffeq_odd_finite}
	dK_G=&\,
	\sum_{\substack{ X\subset E_G\\n_X=n_G-1}}\!\!\!K_{G_X}\,d\cutRes_XK_G+\!\!\!	\sum_{\substack{ X\subset E_G\\n_X=n_G-2}}\!\!\!K_{G_X}\,\Big(d\cutRes_XK_G+\frac{1}{2}\sum_{e\in E_G\setminus X}d\cutRes_{Xe}K_G\Big)\,.
\esp\eeq
We have verified that this differential equation can also be obtained starting from eq.~\eqref{eq:diffeq_dci}: using the relation between cuts in eq.~\eqref{eq:DCI_cut_relations_app_anyN}, one rewrites eq.~\eqref{eq:diffeq_dci} in a form where a specific edge $i\in E_G$ is not cut. Then, upon sending the associated point to infinity, one recovers \refE{eq:diffeq_odd_finite}. This is done explicitly in Appendix \ref{app:pointToInf}.

By comparing eq.~\eqref{eq:diffeq} to eqs.~\eqref{eq:diffeq_odd} and~\eqref{eq:diffeq_even}, we can easily read off the explicit form of the matrix $\Omega_{XY}$ for one-loop integrals. All nonzero entries in the matrix are logarithmic one-forms, which reflects the fact that the basis integrals $J_G$ are pure functions.  Moreover, since the differential equation couples at most integrals with two propagators less, the matrix $\Omega$ has an interesting block-triangular structure. The arguments of the $d\log$-forms in the matrix are commonly referred to as the \emph{alphabet} of the functions and the elements of the alphabets are referred to as \emph{letters,} which we denote by $R^{(i)}_{X,Y}$. Using eqs.~\eqref{eq:diffeq_odd} and~\eqref{eq:diffeq_even}, we can easily write the letters in terms of linear combinations of specific terms in the Laurent expansion of cut integrals with at most two uncut propagators. Again, we distinguish the even and odd cases to find:
\begin{itemize}
\item If $n_X$ is odd and $e,f\in X$, we have
\begin{subequations}\label{eq:Omega_odd}
\begin{align}
\Omega_{X,X}^{(1)} &\,\equiv d\log R_{X,X}^{(1)}= d\cC_{X}^{(1)}J_{G_X}\,,\label{eq:Omega_odd_m0}\\
\Omega_{X,X\setminus e}^{(0)} &\,\equiv d\log R_{X,X\setminus e}^{(0)}= d\cC_{X\setminus e}^{(0)}J_{G_X}\,,\label{eq:Omega_odd_m1}\\
\!\!\Omega_{X,X\setminus \{e,f\}}^{(0)} &\,\equiv d\log R_{X,X\setminus \{e,f\}}^{(0)}= d\cC_{X\setminus \{e,f\}}^{(0)}J_{G_X}+\frac{1}{2}d\cC_{X\setminus e}^{(0)}J_{G_X}+\frac{1}{2}d\cC_{X\setminus f}^{(0)}J_{G_X}.\label{eq:Omega_odd_m2}
\end{align}
\end{subequations}
\item If $n_X$ is even and $e,f\in X$, we have
\begin{subequations}\label{eq:Omega_even}
\begin{align}
\Omega_{X,X}^{(1)} &\,\equiv d\log R_{X,X}^{(1)}= d\cC_{X}^{(1)}J_{G_X}\,,\label{eq:Omega_even_m0}\\
\Omega_{X,X\setminus e}^{(1)} &\,\equiv d\log R_{X,X\setminus e}^{(1)}= d\cC_{X\setminus e}^{(1)}J_{G_X}+\frac{1}{2}\,d\cC_{X}^{(1)}J_{G_X}\,,\label{eq:Omega_even_m1}\\
\Omega_{X,X\setminus \{e,f\}}^{(0)} &\,\equiv d\log R_{X,X\setminus \{e,f\}}^{(0)}= d\cC_{X\setminus \{e,f\}}^{(0)}J_{G_X}\,.\label{eq:Omega_even_m2}
\end{align}
\end{subequations}
\end{itemize}
Explicit expressions for the $R^{(i)}_{X,Y}$ in terms of Gram and Cayley determinants can be obtained from the results in eqs.~\eqref{eq:cut_results_odd} and \eqref{eq:cut_results_even}. Note that the expressions are different depending on whether $n_X$ is odd or even.

 Consequently, eqs.~\eqref{eq:Omega_odd} and~\eqref{eq:Omega_even}  determine the complete alphabet of all one-loop integrals. In particular, our findings indicate that the alphabet of all one-loop integrals is precisely the set of all cut integrals with at most two uncut propagators.  If we denote the alphabet of $J_G$ by $\cA(J_G)$, then the alphabet is given by: 
 \begin{itemize}
 	\item For $n_G$ odd:
\beq\label{eq:alphabet_recursion_odd}
\cA(J_G) =\{R_{E_G,E_G}^{(1)},R_{E_G,E_G\setminus e}^{(0)},R_{E_G,E_G\setminus \{e,f\}}^{(0)}:e,f\in E_G\}\cup\!\!\! \bigcup_{\substack{X\subset E_G\\ n_G-2\le n_X<n_G}}\!\!\!\cA(J_{G_X})\,.
\eeq
	 
	 \item For $n_G$ even:
\beq\label{eq:alphabet_recursion_even}
\cA(J_G) =\{R_{E_G,E_G}^{(1)},R_{E_G,E_G\setminus e}^{(1)},R_{E_G,E_G\setminus \{e,f\}}^{(0)}:e,f\in E_G\}\cup\!\!\! \bigcup_{\substack{X\subset E_G\\ n_G-2\le n_X<n_G}}\!\!\!\cA(J_{G_X})\,.
\eeq
\end{itemize}
where the $R_{E_G,Y}^{(i)}$ are defined in eq.~\eqref{eq:Omega_odd} for $n_G$ odd and eq.~\eqref{eq:Omega_even} for $n_G$ even.

Equations (\ref{eq:alphabet_recursion_odd}) and (\ref{eq:alphabet_recursion_even}) recursively determine the alphabet of every one-loop integral to all orders in the $\eps$ expansion. Note that since the letters are related to the cuts of the integral, they are in direct correspondence with the solutions to the Landau conditions of the integral, in agreement with the observation of ref.~\cite{Dennen:2015bet}.

Once the initial conditions are fixed, one can readily solve the differential equations and obtain results for all one-loop integrals. 
Since all integrals $J_G$ with at least five external legs are finite, the initial conditions can be chosen as the limit in which one of the external invariants vanishes. We will leave an explicit solution of the differential equations for one-loop integrals for future work. 

\subsection{Symbols of one-loop integrals}
Knowing a system of first order homogeneous differential equations for a set of pure functions is equivalent to knowing its symbol~\cite{ChenSymbol,Goncharov:2009tja,Brown:2009qja,Goncharov:2010jf,Duhr:2011zq}. Indeed, if the differential of a pure function $F_w$ satisfies eq.~\eqref{eq:pure_diffeq}, then its symbol can be defined recursively by~\cite{Goncharov:2010jf}
\beq
\label{recursive_symbol}
\cS(F_w) = \sum_i\cS(F_{i,w-1})\otimes u_i\,.
\eeq
Note that here and below we conform with the convention that each entry in the symbol is the argument of a logarithm, and we do not write the $\log$ symbol explicitly.

We can apply  (\ref{recursive_symbol}) to write a recursive formula for the symbols of all one-loop integrals. Since the differential equations look different for integrals with an odd or even number of edges, we need to distinguish two cases. Using respectively eqs.~(\ref{eq:Omega_odd}) and~(\ref{eq:Omega_even}) we find:
\begin{itemize}
\item If $n_G$ is odd, using eq.~(\ref{eq:diffeq_odd}) we obtain:
\begin{equation}\label{eq:symbol_odd}
\cS(J_G) =\, \eps\,\cS(J_G)\otimes R_{E_G,E_G}^{(1)}+\!\!\!
	\sum_{\substack{ X\subset E_G\\n_X=n_G-1}}\!\!\!\cS(J_{G_X})\otimes R_{E_G,X}^{(0)}
	+\,\!\!\!\!\!	\sum_{\substack{ X\subset E_G\\n_X=n_G-2}}\!\!\!\cS(J_{G_X})\otimes
	R_{E_G,X}^{(0)}\,.
\end{equation}
 \item If $n_G$ is even, using eq.~(\ref{eq:diffeq_even}) we obtain:
\begin{equation}\label{eq:symbol_even}
\cS(J_G) = \eps\,\cS(J_G)\otimes R_{E_G,E_G}^{(1)}+{\eps}\!\!\!
	\sum_{\substack{ X\subset E_G\\n_X=n_G-1}}\!\!\!\cS(J_{G_X})\otimes 
	{R_{E_G,X}^{(1)}}
	+\,\!\!\!	\sum_{\substack{ X\subset E_G\\n_X=n_G-2}}\!\!\!\cS(J_{G_X})\otimes R_{E_G,X}^{(0)}\,.
\end{equation}
\end{itemize}
The $R_{E_G,X}^{(i)}$ are explicitly written as combinations of cut integrals in eqs.~(\ref{eq:Omega_odd}) and~(\ref{eq:Omega_even}), respectively for $n_G$ odd or even.

Let us analyze the recursions for the one-loop symbol in more detail. First, as expected, the entries in the symbol of $J_G$ are precisely given by the alphabet $\cA(J_G)$. Second, we see that the right-hand sides of eqs.~\eqref{eq:symbol_odd} and~\eqref{eq:symbol_even} contain the symbol of the integral $J_G$. This term, however, is multiplied by an additional factor of $\eps$, and so we can solve eqs.~\eqref{eq:symbol_odd} and~\eqref{eq:symbol_even} order by order in $\eps$ once the lower-order symbols are known. The starting point of the recursion is the finite terms of the tadpole and bubble integrals in $D=2$ dimensions. In the case of nonvanishing propagator masses they are given by
\beq
J_1^{(0)} = \log m^2 {\rm~~and~~}J_2^{(0)} = \frac{1}{2}\ln\frac{w(1-\wbar)}{\wbar(1-w)}\,,
\eeq
where $w$ and $\bar{w}$ are defined in eq.~\eqref{eq:defWWbar}. Note that since the homogeneous term in the recursion is suppressed by a power of $\eps$, the letter $R_{G,E_G}$ corresponding to the maximal cut does not contribute to $\ord(\eps^0)$, and appears for the first time in the linear term in the $\eps$ expansion. Conversely, we see from the recursion for the symbol that beyond $\ord(\eps^1)$ no new letters will appear. 

Since the differential equations simplify in the limit $\eps\to0$, the recursion for the symbol must simplify in a similar manner. If $n_G$ is even, eq.~\eqref{eq:symbol_even} reduces to
\beq\label{eq:DCI_symbol}
\cS(K_G) =	\sum_{\substack{ X\subset E_G\\n_X=n_G-2}}\!\!\!\cS(K_{G_X})\otimes R^{(0)}_{E_G,X}\,,\quad n_G\textrm{ even}\,,
\eeq
where the $R^{(0)}_{E_G,X}$ for $n_X=n_G-2$ are written as cut integrals in \refE{eq:Omega_even_m2}.
This agrees with the recursion for the symbol of a DCI integral of ref.~\cite{Spradlin:2011wp}. The recursion~\eqref{eq:DCI_symbol} reveals a hierarchical structure in the symbol of $K_G$ which is absent if higher orders in $\eps$ are included: The $k$-th entry in the symbol of $K_G$ is an integral with $2k$ propagators of which $2(k-1)$ are cut, and these cut propagators are precisely the propagators (cut or uncut) of the integral in the $(k-1)$-th entry. As an example, the symbols of the box and hexagon integrals in $D=4$ and $D=6$ dimensions can be written in the form (for simplicity we only consider massive propagators):
\begin{align}\bsp\label{eq:symbol_even_graph}
	\cS\left(\raisebox{-6mm}{\includegraphics[keepaspectratio=true, width=2cm]{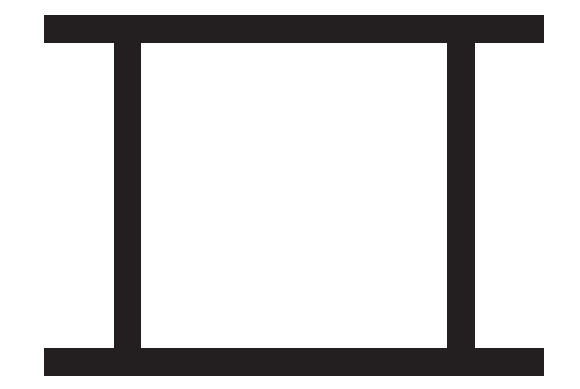}}\right) &\,=
	\sum_{(ij)}
	\raisebox{-5mm}{\includegraphics[keepaspectratio=true, width=2cm]{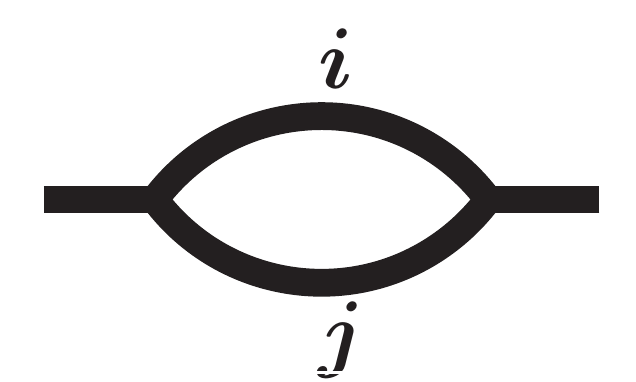}}
	\otimes
	\raisebox{-6mm}{\includegraphics[keepaspectratio=true, width=2cm]{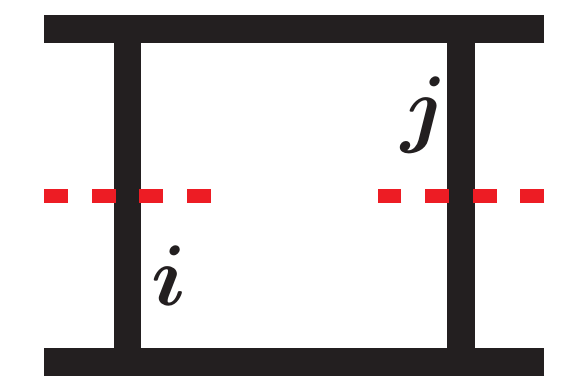}}\,,\\
	\cS\left(\raisebox{-7.5mm}{\includegraphics[keepaspectratio=true, width=2.1cm]{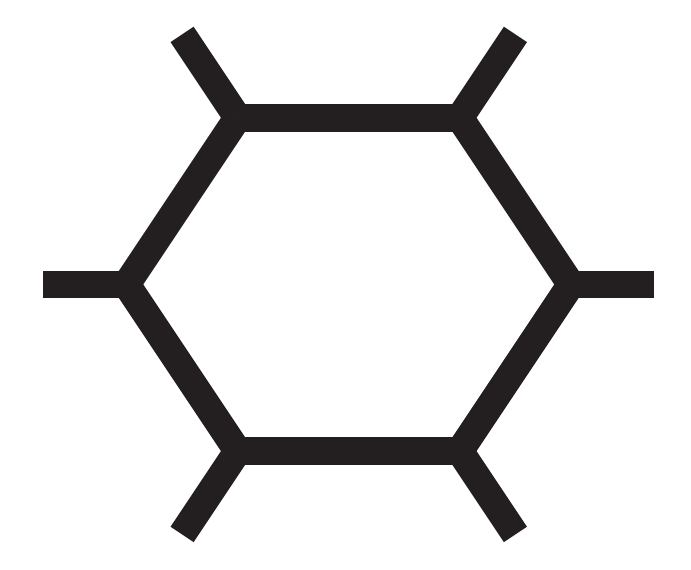}}\right)&\,
	=\sum_{(ij;kl)}
	\raisebox{-5mm}{\includegraphics[keepaspectratio=true, width=2cm]{./diagrams/bubij.pdf}}
	\otimes
	\raisebox{-6mm}{\includegraphics[keepaspectratio=true, width=2cm]{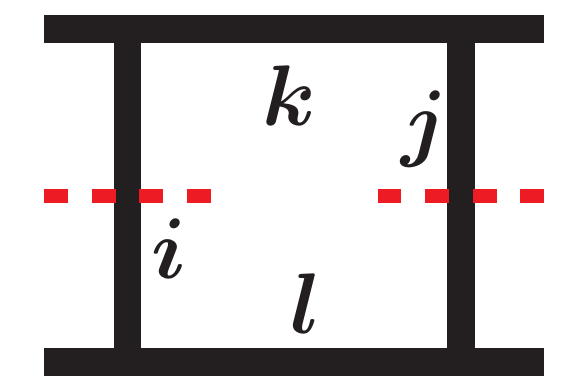}}
	\otimes
	\raisebox{-8mm}{\includegraphics[keepaspectratio=true, width=2.1cm]{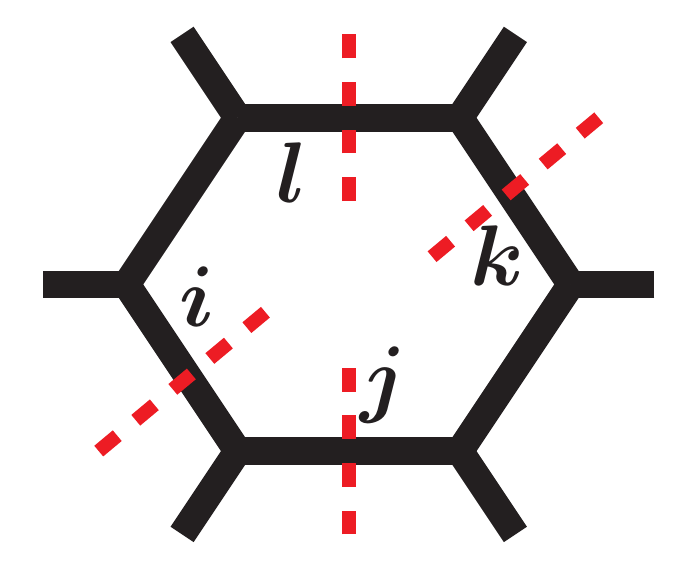}}\,,
\esp\end{align}
where the sums extend over all sequences of disjoint pairs of propagators. We stress that all the graphs in eq.~\eqref{eq:symbol_even_graph} are evaluated at $\eps=0$.

If $n_G$ is odd, then in the limit $\eps\to0$ the recursion in eq.~\eqref{eq:symbol_odd} reduces to
\beq\bsp
	\cS(K_G)=&\,
	\sum_{\substack{ X\subset E_G\\n_X=n_G-1}}\!\!\!\cS(K_{G_X})\otimes R^{(0)}_{E_G,X}
	\,+\!\!\!\sum_{\substack{ X\subset E_G\\n_X=n_G-2}}\!\!\!\cS(K_{G_X})\otimes R^{(0)}_{E_G,X}\,,\quad n_G\textrm{ odd}\,,
\esp\eeq
where the $R^{(0)}_{E_G,X}$ are written as cut integrals in \refE{eq:Omega_odd_m1} for $n_X=n_G-1$ and in \refE{eq:Omega_odd_m2}  for $n_X=n_G-2$.
Also in this case the symbol entries exhibit a hierarchy, with the difference that now  integrals with an odd number of edges can also appear. For example, for the triangle and pentagon integrals we find (restricting the discussion once again to finite integrals with massive propagators):
\begin{align}
\nonumber
	\cS&\!\left(\raisebox{-6mm}{\includegraphics[keepaspectratio=true, width=1.6cm]{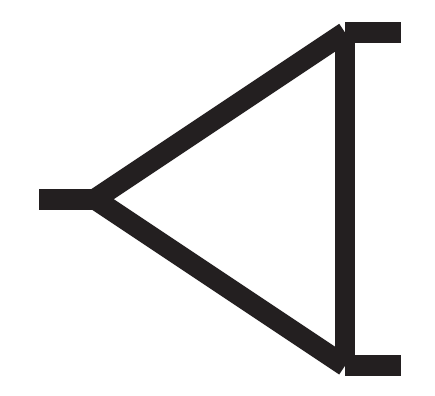}}\right)
	\!\!= \!\sum_{(ij)}
	\raisebox{-5mm}{\includegraphics[keepaspectratio=true, width=2cm]{./diagrams/bubij.pdf}}
	\otimes
	\raisebox{-6mm}{\includegraphics[keepaspectratio=true, width=1.6cm]{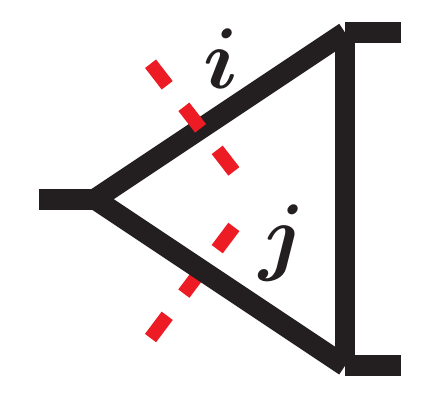}}\!\!+
	\!\sum_{(i)}
	\raisebox{-4mm}{\includegraphics[keepaspectratio=true, width=.8cm]{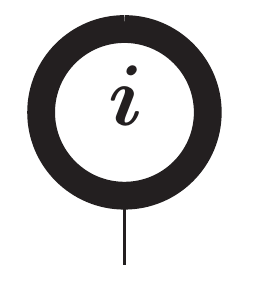}}
	\otimes\!\!
	\left(
	\raisebox{-6mm}{\includegraphics[keepaspectratio=true, width=1.6cm]{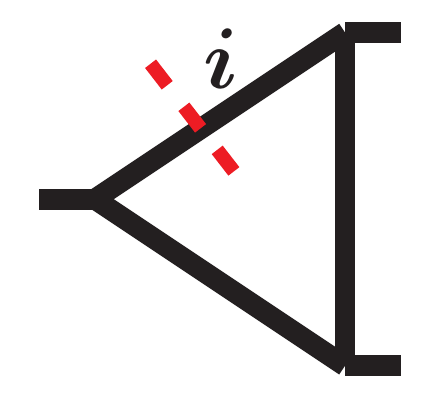}}
	+\frac{1}{2}\sum_{(j)}
	\raisebox{-6mm}{\includegraphics[keepaspectratio=true, width=1.6cm]{./diagrams/tCij.pdf}}
	\right)\!,\label{odd_Symbol_hierarchy}\\
	\cS&\left(\raisebox{-8mm}{\includegraphics[keepaspectratio=true, width=2cm]{./diagrams/pentagon2.pdf}}\right)
	=
	\sum_{(ij;kl)} \raisebox{-5mm}{\includegraphics[keepaspectratio=true, width=2cm]{./diagrams/bubij.pdf}}
	\otimes 
	\raisebox{-6mm}{\includegraphics[keepaspectratio=true, width=2cm]{./diagrams/bijklCij.pdf}}
	\otimes 
	\raisebox{-8mm}{\includegraphics[keepaspectratio=true, width=2cm]{./diagrams/pentagonTQuad2.pdf}}\\
\nonumber	&+\sum_{(ij;k)} 
	\raisebox{-5mm}{\includegraphics[keepaspectratio=true, width=2cm]{./diagrams/bubij.pdf}}
	\otimes
	\raisebox{-6mm}{\includegraphics[keepaspectratio=true, width=1.6cm]{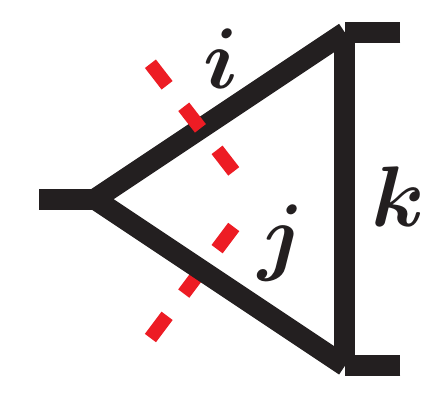}}
	\otimes
	\left(
	\raisebox{-8mm}{\includegraphics[keepaspectratio=true, width=2cm]{./diagrams/pentagonTriple2.pdf}}+
	\frac{1}{2}\sum_{(l)}
	\raisebox{-8mm}{\includegraphics[keepaspectratio=true, width=2cm]{./diagrams/pentagonTQuad2.pdf}}
	\right)\\
\nonumber	&+\sum_{(i;j;k)}
	\raisebox{-4mm}{\includegraphics[keepaspectratio=true, width=.8cm]{./diagrams/tadi.pdf}}
	\otimes
	\left(
	\raisebox{-6mm}{\includegraphics[keepaspectratio=true, width=1.6cm]{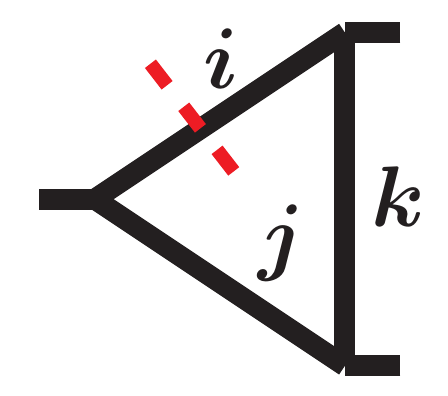}}
	+\frac{1}{2}\raisebox{-6mm}{\includegraphics[keepaspectratio=true, width=1.6cm]{./diagrams/tijkCij.pdf}}
	\right)
	\otimes
	\left(
	\raisebox{-8mm}{\includegraphics[keepaspectratio=true, width=2cm]{./diagrams/pentagonTriple2.pdf}}+
	\frac{1}{2}\sum_{(l)}
	\raisebox{-8mm}{\includegraphics[keepaspectratio=true, width=2cm]{./diagrams/pentagonTQuad2.pdf}}
	\right).
\end{align}


\section{Relation to other coactions}
\label{sec:master_formula}

In this section we discuss how the diagrammatic coaction on one-loop integrals is related to other coactions on integrals and graphs.
In particular, we show how the coaction on one-loop graphs can be derived from a more general conjecture formulated in ref.~\cite{Abreu:2017enx}. This conjecture  can be stated as follows: there is a coaction on certain classes of integrals given by
\beq\label{coaction}
\Delta\left(\int_{\gamma}\omega\right) = \sum_i \int_{\gamma}\omega_i\otimes \int_{\gamma_i}\omega\,.
\eeq
The sum runs over a basis of \emph{master integrands}, the $\omega_i$. Although the right-hand side of eq.~\eqref{coaction} depends on an explicit choice of basis, it is easy to check that the sum is independent of this choice. The integration contours $\gamma_i$ are the \emph{master contours}, dual to the master integrands in the following sense,
\beq
P_{ss}\left(\int_{\gamma_i}\omega_j\right) = \delta_{ij}\,,
\eeq
where $P_{ss}$ is the projection onto \emph{semi-simple} elements, i.e., elements on which the coaction acts via $\Delta(x) = x\otimes 1$. All algebraic numbers and functions are semi-simple~\cite{Brown:motivicperiods}. There are also semi-simple elements that are not algebraic. In particular, we see from eq.~\eqref{eq:Delta_ipi} that $i\pi$ is semi-simple. Functions such as classical logarithms and polylogarithms are not semi-simple, unless they happen to evaluate to powers of $2\pi i$. We consider semi-simple elements as a multiplicative group, and so $(i \pi)^n$ is semi-simple for every integer $n$. Before we discuss how one can derive the coaction on one-loop (cut) integrals in eq.~\eqref{eq:Coaction_cut_integrals} from eq.~\eqref{coaction}, we find it instructive to show how the coaction on MPLs follows from eq.~\eqref{coaction}.

\subsection{The coaction on MPLs}
For simplicity, we only discuss the generic case, i.e., we assume that the singularities $a_i$ in \refE{eq:Mult_PolyLog_def} are all nonzero and distinct from each other. We use the following notation,
\beq
G(\vec a;z) = \int_0^z\omega_{\vec a}\,.
\eeq
The set of master integrands for $G(\vec a;z)$ is $\omega_{\vec b}$ with $\vec b\subseteq\vec a$, and we set $\omega_{\emptyset} = dt$.

Let us now determine the master contours. We define ${\Gamma}_{\vec b}$ with $\vec b\neq \emptyset$ to be the path from $0$ to $z$ that encircles the poles in $\vec b$ in that order, and no other poles (cf. fig.~\ref{fig:paths}), and $\Gamma_\emptyset$ to be the straight line from $0$ to $z$. One can easily show that
\beq
\int_{{\Gamma}_{\vec b}}\omega_{\vec a} = \left\{\begin{array}{ll}
z\,,&\textrm{ if } \vec b=\vec a=\emptyset\,,\\
(2\pi i)^{|\vec b|}\,G_{\vec b}(\vec a;z)\,,&\textrm{ if }\vec b \subset \vec a\,,\\
(2\pi i)^{|\vec a|}\,,&\textrm{ if }\vec b = \vec a\neq \emptyset\,,\\
0\,,&\textrm{ otherwise}\,.
\end{array}\right.
\eeq
where $G_{\vec b}(\vec a;z)$ was defined following eq.~(\ref{eq:Delta_MPL}) and $|\vec a|$ is the length of the set $\vec{a}$.
We see that the contours ${\Gamma}_{\vec b}$ do not stand in one-to-one correspondence with the master integrands $\omega_{\vec a}$, because the integral does not vanish for $\vec b \subset \vec a$. However, since $G_{\vec b}(\vec a;z)$ is not semi-simple for $\vec b \subset \vec a$, this contribution is mapped to zero under the projection $P_{ss}$,
\beq
P_{ss}\left(\int_{{\Gamma}_{\vec b}}\omega_{\vec a}\right) = \left\{\begin{array}{ll}
z\,,&\textrm{ if } \vec b=\vec a=\emptyset\,,\\
(2\pi i)^{|\vec a|}\,,&\textrm{ if }\vec b = \vec a\neq \emptyset\,,\\
0\,,&\textrm{ otherwise}\,.
\end{array}\right.
\eeq
It thus follows that after the projection, there is a one-to-one correspondence between the master integrands $\omega_{\vec a}$ and the master contours ${\Gamma}_{\vec b}$, and for $\vec{b}=\vec{a}$ the two are dual to each other. Let us now define 
\beq
\gamma_{\vec b} = \mathcal{N}_{\vec b}\,{\Gamma}_{\vec b}\textrm{    ~~~~~ with ~~~~~    }
\mathcal{N}_{\vec b} = \left\{\begin{array}{ll}
z^{-1}\,,&\textrm{ if } \vec b=\emptyset\,,\\
(2\pi i)^{-|\vec b|}\,,&\textrm{ if } \vec b\neq\emptyset\,.
\end{array}\right.
\eeq
We then find
\beq
P_{ss}\left(\int_{{\gamma}_{\vec b}}\omega_{\vec a}\right) = \delta_{\vec b\vec a}\,.
\eeq
The previous equation implies that $\gamma_{\vec b}$ is the master contour associated to the master integrand $\omega_{\vec b}$. This contour is identical to the contour in Section~\ref{subsec:MPLs}
(see \refF{fig:paths}), and we can write
\beq
G_{\vec b}(\vec a;z) \equiv \int_{\gamma_{\vec b}}\omega_{\vec a}\,.
\eeq
Equation~\eqref{coaction} then implies
\beq\bsp
\Delta(G(\vec a;z)) &\,= \Delta\left(\int_0^z\omega_{\vec a}\right) \\
&\,= \sum_{\vec b\subseteq \vec a}\int_0^z\omega_{\vec b}\otimes \int_{\gamma_{\vec b}}\omega_{\vec a}\\
&\,=\int_0^z\omega_{\emptyset}\otimes \int_{\gamma_{\emptyset}}\omega_{\vec a} + \sum_{{\emptyset\neq\vec b\subseteq \vec a}}\int_0^z\omega_{\vec b}\otimes \int_{\gamma_{\vec b}}\omega_{\vec a}\\
&\,=1\otimes G({\vec a};z) + \sum_{\substack{\emptyset\neq\vec b\subseteq \vec a}}G({\vec b};z)\otimes G_{\vec b}({\vec a};z)\\
&\,=\Delta_{\textrm{MPL}}(G(\vec a;z))\,,
\esp\eeq
and we see that we can derive the formula for the coaction on MPLs, eq.~\eqref{eq:Delta_MPL}, from eq.~\eqref{coaction}.

\subsection{The coaction on one-loop integrals}
We now show that the coaction on one-loop cut integrals 
in eq.~\eqref{eq:Coaction_cut_integrals_psi}
can be derived from eq.~\eqref{coaction} in a similar way as the coaction on MPLs in the previous section. It is sufficient to discuss uncut integrals, for which $C=\emptyset$. 

Consider the one-loop integral $\tildeJ_G$, which we write in the form
\beq
\tildeJ_G \equiv \int \omega_G\,.
\eeq
We know from IBP identities that 
master integrals for $\tildeJ_G$ should not contain more propagators. In other words, the set of master integrands  is given by $\omega_X\equiv \omega_{G_X}$, with $\emptyset\neq X\subseteq E_G$.

Let us now determine a basis of master contours associated to this basis of master integrands. In ref.~\cite{Abreu:2017ptx}, it was shown that at one loop a basis of contours is given by $\Gamma_Y$, $Y\subseteq E_G$. This is a direct consequence of the homology relations in eq.~\eqref{eq:homo_even} and~\eqref{eq:homo_odd}, which allow us to write every contour associated to a Landau singularity of the second type as a linear combination of contours associated to singularities of the first type. 
This basis of contours, however, is not a basis of master contours for the master integrands $\omega_X$.
Instead, we need the following contours:
\beq
\gamma_Y = 
\mathcal{N}_Y\,\left(\Gamma_Y+(a-a_Y)\sum_{e\in E_G\setminus Y} \Gamma_{Ye}\right){\rm~~with~~} \mathcal{N}_Y = (2\pi i)^{-\lceil n_Y/2\rceil}\,j_Y^{-1}\,,
\eeq
where $a=1/2$, and $j_Y$ and $a_Y$ are defined in eqs.~\eqref{eq:LS_J_n} and~\eqref{eq:aXdef} respectively. 
We will now show that the contours $\gamma_Y$ are the master contours associated to the master integrands $\omega_X$, i.e., they satisfy the relation
\beq\label{eq:condition}
P_{ss}\left(\int_{\gamma_Y}\omega_X\right) = \delta_{YX}\,.
\eeq
 If $X=Y$, we immediately find that eq.~\eqref{eq:condition} holds,
\beq
P_{ss}\left(\int_{\gamma_X}\omega_X\right) = \mathcal{N}_X\,P_{ss}\left(\mathcal{C}_X\tildeJ_X\right) = 1\,.
\eeq
If $X\neq Y$, we start by noting that
\beq
\mathcal{C}_Y\tildeJ_{X} \equiv\int_{\Gamma_Y}\omega_X = 0{\rm~~if~~} Y\nsubseteq X\,,
\eeq
so we can restrict our attention to the cases where $Y \subset X$.
Then the weight of the cut integral $\mathcal{C}_Y\tildeJ_{X}$ is $\lceil n_X/2\rceil - \lceil n_Y/2\rceil$, and it is easy to check that $\mathcal{C}_Y\tildeJ_{X}$ is not semi-simple unless it evaluates to a function of weight zero. 
Since we are now studying the case where $n_Y<n_X$, we may obtain a nonzero semi-simple result only if $n_X$ is even and $n_Y=n_X-1$, i.e.~all propagators are cut except one. We now invoke a  relation between the maximal and next-to-maximal cuts of an integral with an even number of propagators~\cite{Abreu:2017ptx},
\beq
\mathcal{C}_Y\tildeJ_X = -\frac{1}{2} \mathcal{C}_X\tildeJ_X + \ord(\eps)\,,{\rm~~with~~} n_X=n_Y+1\, \textrm{ even}\,.
\eeq
Thus, if $Y=X\setminus e$ for some $e\in E_G\setminus Y$, we have
\beq
P_{ss}\left(\int_{\gamma_Y}\omega_X\right) = \mathcal{N}_Y\,\left[P_{ss}\left(\mathcal{C}_Y\tildeJ_X\right) + \frac{1}{2} P_{ss}\left(\mathcal{C}_X\tildeJ_X\right)\right] =0\,.
\eeq
We conclude  that whenever $X\neq Y$, eq.~\eqref{eq:condition} holds.

Having identified the right set of master contours, we can apply eq.~\eqref{coaction} and we find
\begin{align}
\Delta\left(\int\omega_G\right)&\, = \sum_{\substack{X\subseteq E_G\\ X\neq\emptyset}}\int\omega_X\otimes \int_{\gamma_X}\omega_G\\
\nonumber&\, = \sum_{\substack{X\subseteq E_G\\ X\neq\emptyset}}\int\frac{\omega_X}{j_X}\otimes (2\pi i)^{-\lceil n_X/2\rceil}\left(\int_{\Gamma_X}\omega_{G} + (a-{a}_C)
\sum_{e\in E_G\setminus X}\int_{\Gamma_{Xe}}\omega_{G}\right)\\
\nonumber&\,=\sum_{\substack{X\subseteq E_G\\ X\neq\emptyset}}J_X\otimes \left(\cC_{X}\widetilde{J}_G + (a-{a}_C)
\sum_{e\in E_G\setminus X}\cC_{Xe}\widetilde{J}_{G}\right)\,,
\end{align}
and we recover precisely eq.~\eqref{eq:Coaction_cut_integrals_psi} (with $C=\emptyset$). The analysis of the case with $C \neq \emptyset$ is similar.

{As a corollary, recall from Section~\ref{sec:dci} that DCI integrals in  even dimensions correspond to a class of integrals with singularities along a quadric studied by Goncharov in ref.~\cite{Goncharov:1996}, and their coactions agree as well.  The coaction on the DCI integrals is a special case of our conjecture~\eqref{eq:Coaction_cut_integrals_psi}.  It follows that the coaction of  ref.~\cite{Goncharov:1996} can be derived from eq.~\eqref{coaction}. }


\section{Conclusions}
\label{sec:summary}

In this paper we have defined a diagrammatic coaction acting on cut Feynman graphs. Our main result is the conjecture \eqref{eq:graph_conjecture}, stating that the coaction of eq.~(\ref{eq:phipsiaction}), defined purely in terms of graphs, exactly reproduces the combinatorics of the coaction on MPLs when applied 
to the Laurent coefficients in the $\eps$ expansion of (cut) Feynman integrals. The validity of the conjecture was tested explicitly through several orders in $\eps$ for a range of nontrivial one-loop integrals, finite or divergent, and with a variety of configurations of internal propagator masses and external scales. Furthermore, we have shown that in the case of finite integrals with arbitrarily large number of legs, the diagrammatic coaction is consistent with the extended dual conformal invariance. 

We have also seen that the diagrammatic coaction interacts with a discontinuity operator and with a differential operator as one would expect based on the coaction on MPLs, namely that a discontinuity operator acts on the first factor, while a differential operator acts on the second. 
It implies a stronger version of the first-entry condition: the first entry in the coaction of a Feynman integral is itself a Feynman integral. 
Furthermore, it provides a new way to derive differential equations for one-loop integrals, completely bypassing the need for IBP reduction, and it 
gives a novel perspective on the structure of these equations. Specifically,  
we have seen that the coefficients in these equations are fully determined by the first two orders in the $\eps$ expansion of the maximal, next-to-maximal and next-to-next-to-maximal cuts of Feynman integrals. Using a method developed in ref.~\cite{Abreu:2017ptx}, we computed these cuts explicitly and hence determined the system of differential equations for any one-loop integral. 
While solving the differential equations is left for future work, we explained here how to construct a recursive solution of these equations  at symbol level for any one-loop integral, and showed that these solutions display an interesting hierarchical structure, which is illustrated in eqs.~(\ref{eq:symbol_even_graph}) and (\ref{odd_Symbol_hierarchy}).

The search for a combinatorial coaction on Feynman graphs that agrees with the coaction on the functions and numbers 
resulting from their integral interpretation
has been an active area of research in pure mathematics (see e.g.~\cite{Bloch:2005bh}). 
Our diagrammatic coaction fits seamlessly into that line of research, but there are several significant differences from previous results. In particular, 
other studies in the pure mathematics literature of coproducts and coactions on Feynman graphs have focused only on graphs without cut edges \cite{Kreimer:1997dp,Connes:1998qv,Connes:1999yr,Kreimer:2009jt}. In our construction, cut Feynman integrals play a crucial role, and we find that it is not possible to define a diagrammatic coaction that reproduces the combinatorics of the coaction on MPLs in terms of (uncut) Feynman graphs alone. 
It would be interesting to investigate in detail how our results are related to the aforementioned studies, including a more rigorous mathematical formulation of our results and a formal mathematical proof of our conjectured coaction on one-loop Feynman graphs and integrals. Indeed, the goal of this paper is to provide compelling evidence for the existence of a coaction
on one-loop (cut) graphs which agrees with the coaction on MPLs when the graphs are replaced by the analytic expressions that 
they represent. A rigorous mathematical treatment of this topic goes beyond the scope of this paper, and we content ourselves here to highlight some of the aspects for which input from the pure mathematics community  may be desirable.\footnote{We thank the referee for some suggestions in this direction, which are summarized below.}

First, it would be interesting to put the notion of `cut Feynman graph' on solid mathematical footing. While Feynman graphs are by now  familiar and well-defined objects  in pure mathematics, their cut analogs have not yet received the same level of attention. Having a formal definition of cut Feynman graphs, including all the data that is needed to define them (e.g., external momenta, propagator masses, etc.), would allow one to look for a rigorous definition of their infinite-dimensional Hopf algebra for which our paper gives compelling evidence. For example, in a first step one could consider only finitely-generated Hopf algebras of some sort (e.g., generated only by cut graphs that share the same set of propagators -- a `topolology' in physics parlance), and in a second step then define the infinite-dimensional Hopf algebra of all cut graphs by taking a certain limit. A second item which definitely requires a more formal setting is the notion of working `mod $i\pi$.' While the intuitive meaning of this operation is clear from the physics perspective (because one-loop integrals are always real in the Euclidean region, and the structure of the imaginary part away from the Euclidean region is understood from the point of view of unitarity), a rigorous mathematical treatment of this concept would be useful to develop.
Finally, our construction relies heavily on the use of dimensional regularization. Indeed, we have seen that eq.~\eqref{eq:PCI}, which explicitly involves the dimensional regulator $\eps$, is crucial for divergences to cancel and to reproduce the correct trivial coaction components. 
Moreover, dimensional regularization makes it possible to consider any massless limit of the diagrammatic coaction before expansion in the regulator, retaining fully consistent results despite the fact that such limits do not commute with the expansion.
We do not know if these features are specific to dimensional regularization, and it would be interesting to explore extensions to other regularizations. Either way, given the importance of regularization in our construction  -- dimensional or other -- it will be crucial to understand how a notion of regularization can be embedded into a rigorous mathematical framework for one-loop cut graphs and their Hopf algebra structure.

%
%

Finally, throughout this paper, our discussion was restricted to one-loop Feynman integrals. However, in Section \ref{sec:master_formula} we have shown that the diagrammatic coaction on one-loop Feynman integrals can be viewed as a particular case of the much more general formula in \refE{coaction}, where the coaction is defined by pairing master integrands and master contours~\cite{Abreu:2017enx}, a construction that applies for example to the case of hypergeometric functions. We believe that this paves the way to defining a coaction on Feynman integrals beyond one-loop, certainly in the case where Feynman integrals evaluate to MPLs, but also when they do not. 
While we expect the diagrammatic interpretation to also extend beyond one-loop, the form that such a generalization would take remains to be determined.

\acknowledgments

We are grateful to Lance Dixon, Falko Dulat, Herbert Gangl,  David Broadhurst, Roman Lee, Bernhard Mistlberger, Erik Panzer and Volodya Smirnov for discussions and communications.

SA acknowledges the hospitality of Trinity College Dublin and the CERN Theoretical Physics Department at various stages of this work. CD acknowledges the hospitality of the Higgs Centre of the University of Edinburgh and of Trinity College Dublin at various stages of this work. EG acknowledges the hospitality of Trinity College Dublin.  

This work is supported by the Juniorprofessor Program of Ministry of Science, Research and the Arts of the state of Baden-W\"urttemberg, Germany, the European Research Council (ERC) under the Horizon 2020 Research and Innovation Programme  through grants 647356 and 637019, the National Science Foundation under Grant No.~NSF PHY11-25915, and by the STFC Consolidated Grant ``Particle Physics at the Higgs Centre.''

We would also like to thank the ESI institute in Vienna and the organizers of the program ``Challenges and Concepts for Field Theory and Applications in the Era of LHC Run-2,'' Nordita in Stockholm and the organizers of ``Aspects of Amplitudes'' over the summer of 2016, the MITP in Mainz and the organizers of the program AMPDEV2017, and the KITP in Santa Barbara and the organizers of SCAMP17, where certain ideas presented in this paper were consolidated.




\appendix

\section{Notation}\label{app:notation}

We summarize the different notations used throughout this paper as an easy reference point for the reader.

\paragraph{Graphs:} 
\begin{itemize}
	\item $G$: Feynman graph.
	\item $E_G$: set of the internal edges (propagators) of $G$. Each edge $e\in E_G$ carries information about its mass $m_e^2$ and the momentum $k-q_e$ flowing through it.
	\item $(G,C)$: cut Feynman graph, with the edges in $C\subseteq E_G$ cut.
	\item $G_X$, with $X\subseteq E_G$: graph obtained by contracting all edges not in $X$.
	\item $G \setminus e$: graph $G$ with edge $e$ contracted.
	\item $n_G$ is the number of edges in $G$, and $n_X$ is the number of edges in $X$.
\end{itemize}

\paragraph{Feynman integrals corresponding to graph G with implicit arguments:}
\begin{itemize}
	\item $I_G^D$: one-loop integral in $D$ dimensions.
	\item $\jtilde_G$: $I_G^{D}$ in $D=2\left\lceil\frac{n_G}{2}\right\rceil-2\epsilon$ dimensions.
	\item $ J_G$: $\jtilde_G$ normalized to its maximal cut $j_{n_G}$,
	\beq
		j_{n} = \left\{\begin{array}{ll}
\displaystyle 2^{1-n/2}\,i^{n/2}\,Y_{[n]}^{-1/2}\,, & \textrm{ for } n \textrm{ even}\,,\\
\displaystyle 2^{(1-n)/2}\,i^{(n-1)/2}\Gram_{[n]}^{-1/2}\,, & \textrm{ for } n \textrm{ odd}\,.
\end{array}\right.
	\eeq
	\item {A set of edges $X$ is enough to specify a one-loop Feynman graph, so we sometimes write $J_X$ for the integral  with these edges}.

\item
In our diagrammatic coaction, a specific linear combination of Feynman graphs appears in the first entry. We introduce the following notation for the corresponding function:
\beq
\label{eq:defjhat}
{\widehat J}_{G} \equiv  {J}_G + a_{E_G} \sum_{e\in E_G} {J}_{G\setminus e}\, ,
\eeq
where ${G\setminus e}$ is the graph $G$ with the edge $e$ pinched and
\begin{equation}
    a_{E_G} \,= \left\{\begin{array}{ll}\displaystyle
    a\,, & \textrm{ if } n_G \textrm{ even}\,,\\
    0 \,, & \textrm{ if } n_G \textrm{ odd}\,.
    \end{array}\right.
\end{equation}
In particular, for $n_G$ odd, $\jhat_G=J_G$.

\end{itemize}

\paragraph{Notation for cut integrals:}
\begin{itemize}
	\item $\cutRes_C I_n^D$, $\cutRes_C J_n$, $\cutRes_C \widetilde J_n$: cut Feynman integrals, where a subset $C$ of the edges of the Feynman diagram corresponding to $I^D_n$, $J_n$ or $\widetilde J_n$ are cut.
	\item $\cutRes_{e_i\ldots e_k}I_n^D$, $\cutRes_{e_i\ldots e_k}J_n$, $\cutRes_{e_i\ldots e_k}\widetilde J_n$: cut Feynman integrals where the edges $e_i\ldots e_k$ of the Feynman diagram are cut.
	\item We define $\cutRes_\emptyset I_n^D =I_n^D$.
\end{itemize}

\paragraph{Modified Cayley and Gram determinants:\\}

\begin{equation}
	Y_C=\det\left(\frac{1}{2}\left(-x_{ij}^2+m_i^2+m_j^2\right)\right)_{i,j\in C}\,,
\end{equation}
\begin{equation}
	\Gram_C=\det\left(x_{ie}\cdot x_{je}\right)_{i,j\in C\setminus e}\,,
\end{equation}
where $e$ is an arbitrary element of $C$ ($\Gram_C$ is independent of this choice). 

\paragraph{Standard notation for some important sets:\\}
Some sets appear repeatedly in the paper, and as far as possible we try to keep the following standard notation:
\begin{itemize}
	\item $C$ denotes a set of cut propagators, with size $n_C=\abs{C}$.
	\item In the diagrammatic coaction, $X$ denotes a set of cut propagators in the right entry, and of contracted propagators in the left entry.
	\item $[n]$ denotes the set of all integers from 1 to $n$ inclusive.
\end{itemize}

\section{Explicit expression for some Feynman integrals and their cuts}\label{sec:results}

\subsection{Tadpole}
\begin{equation}
	\widetilde J_1\left(m^2\right)=-\frac{e^{\gamma_E\epsilon}\Gamma(1+\epsilon)\left(m^2\right)^{-\epsilon}}{\epsilon}\,.
\end{equation}
\begin{equation}\label{eq:cut_tadpole}
	\cutRes_e \widetilde J_1\left(m^2\right)=\frac{e^{\gamma_E\epsilon}\left(-m^2\right)^{-\epsilon}}{\Gamma(
	1-\epsilon)}\,,
\end{equation}
\begin{equation}
	 j_1(m^2)=1\,.
\end{equation}

\subsection{Two-point functions with massive external legs}

\paragraph{Zero massive propagators.}\label{sec:resBubZeroMass}
\begin{equation}
	\widetilde J_2\left(p^2\right)=-\frac{2c_\Gamma}{\epsilon}\left(-p^2\right)^{-1-\epsilon},
\end{equation}
\begin{equation}
	\cutRes_{e_1,e_2}\widetilde  J_2\left(p^2\right)=-2\frac{e^{\gamma_E\epsilon}\Gamma(1-\epsilon)}{\Gamma(1-2\epsilon)}
	\left(p^2\right)^{-1-\epsilon}\,,
\end{equation}
\begin{equation}
	 j_2(p^2)=-\frac{2}{p^2}\,.
\end{equation}

\paragraph{One massive propagator.}\label{sec:resBubOneMass}
\begin{equation}\label{bubPM}
	\widetilde  J_2(p^2;m_1^2)=-\frac{e^{\gamma_E  \epsilon } \Gamma (1+\epsilon) }{\epsilon }\left(m_1^2-p^2\right)^{-1-\epsilon}
	\, _2F_1\left(-\epsilon ,1+\epsilon ;1-\epsilon
   ;\frac{p^2}{p^2-m_1^2}\right)\, ,
\end{equation}
\begin{equation}\label{bubPMCutM}
	\cutRes_{e_1}\widetilde  J_2(p^2;m_1^2)=\frac{e^{\gamma_E  \epsilon }}{\Gamma (1-\epsilon )}
	\frac{\left(-m_1^2\right)^{-\epsilon }}{p^2 }\, _2F_1\left(1,1+\epsilon ;1-\epsilon ;\frac{m_1^2}{p^2}\right),
\end{equation}
\begin{equation}\label{bubPMCutP}
	\cutRes_{e_1,e_2}\widetilde  J_2(p^2;m_1^2)=-2\frac{ e^{\gamma_E  \epsilon } \Gamma (1-\epsilon ) }{\Gamma (1-2 \epsilon )}
	\left(p^2\right)^{\epsilon } \left(p^2-m_1^2\right)^{-2 \epsilon -1}\,,
\end{equation}
\begin{equation}
	 j_2(p^2;m_1^2)=-\frac{2}{p^2-m_1^2}\,.
\end{equation}

\paragraph{Two massive propagators.}\label{sec:resBubTwoMass}
We use the variables $w, \wbar$ defined in \refE{eq:defWWbar}.
\begin{align}\bsp\label{eq:bubM1M2}
	\widetilde  J_2\left(p^2;m_1^2,m_2^2\right)=
	&-\frac{e^{\gamma_E\epsilon}\Gamma(1+\epsilon)}{\epsilon}\frac{\left(-p^2\right)^{-1-\epsilon}}{(w-\wbar)^{1+\epsilon}}
	\left[w ^{-\epsilon}\hypgeo{-\epsilon}{1+\epsilon}{1-\epsilon}{\frac{w }{w -\wbar }}\right.\\
	&\left.-(w -1)^{-\epsilon}\hypgeo{-\epsilon}{1+\epsilon}{1-\epsilon}{\frac{w -1}{w -\wbar }}\right]\\
	=&\frac{1}{p^2(w-\wbar)}\ln\left(\frac{\wbar(1-w)}{w(1-\wbar)}\right)+\mathcal{O}(\epsilon)\,,
\esp\end{align}
\begin{align}\bsp\label{bubM1M2CutM1}
	\cutRes_{e_1}\widetilde  J_2(p^2;m_1^2,m_2^2)=&-\frac{e^{\gamma_E\epsilon}}{\Gamma(1-\epsilon)}
	\frac{\left(-p_1^2\right)^{-1-\epsilon}}{w(w\wbar)^{\epsilon}}\hypgeo{1}{1+\epsilon}{1- \epsilon}{\frac{\wbar}{w}}\\
	=&\frac{1}{p^2(w-\wbar)}+\mathcal{O}(\epsilon)\,,
\esp\end{align}
\begin{align}\label{bubM1M2CutM2}
	\cutRes_{e_2}\widetilde  J_2(p^2;m_1^2,m_2^2)=&-\frac{e^{\gamma_E\epsilon}}{\Gamma(1-\epsilon)}
	\frac{\left(-p_1^2\right)^{-1-\epsilon}(1-w)^{-\epsilon}}{(w-\wbar)^{1+\epsilon}}
	\hypgeo{-\epsilon}{1+\epsilon}{1- \epsilon}{\frac{w-1}{w-\wbar}}\nonumber\\
	=&\frac{1}{p^2(w-\wbar)}+\mathcal{O}(\epsilon)\,,
\end{align}
\begin{align}\bsp\label{bubM1M2CutP}
	\cutRes_{e_1,e_2}\widetilde  J_2(p^2;m_1^2,m_2^2)=&-2\frac{e^{\gamma_E\epsilon}\Gamma(1-\epsilon)}{\Gamma(1-2\epsilon)}
	\left(p^2\right)^{-1-\epsilon}(w-\wbar)^{-1-2\epsilon}\\
	=&-\frac{2}{p^2(w-\wbar)}+\mathcal{O}(\epsilon).
\esp\end{align}
\begin{equation}\label{bubM1M2LS}
	j_2(p^2;m_1^2,m_2^2)=-\frac{2}{p^2(w-\wbar)}.
\end{equation}

\subsection{Three-point functions\label{sec:resTriangles}}

\begin{figure}[]
\includegraphics[width=3.5cm]{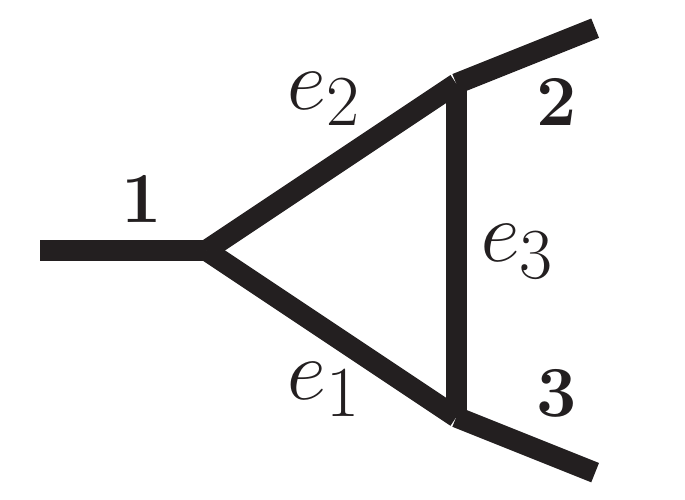}
\centering
\caption{Notation for external and internal edges of three-point functions. Propagator $e_i$ has mass $m_i^2$, and external leg $\bm{i}$ has mass $p_i^2$.}
\label{fig:triNotation}
\end{figure}

\paragraph{One external massive leg.}\label{sec:tp1Res}
We keep one external leg massive ($p^2\neq0$) and set all other scales to zero and find
\begin{equation}
	\widetilde J_3\left(p^2\right)=-\frac{c_\Gamma}{\epsilon^2}\left(-p^2\right)^{-1-\epsilon},
\end{equation}
\begin{equation}
	\cutRes_{e_1,e_2} \widetilde J_3\left(p^2\right)=-\frac{e^{\gamma_E\epsilon}\,\Gamma(1-\epsilon)}{\epsilon\,\Gamma(1-2\epsilon)}
	\left(p^2\right)^{-1-\epsilon}\,,
\end{equation}
\begin{equation}
	 j_3(p^2)=-\frac{1}{p^2}\,.
\end{equation}

\paragraph{Two external massive legs.}\label{sec:tp2p3Res}
We set $p_3^2=m_2^2=m_3^2=0$ --- see \refF{fig:triNotation} --- and find
\begin{equation}\label{eq:tp2p3}
	\widetilde J_3\left(p_1^2,p_2^2\right)=\frac{c_\Gamma}{\epsilon^2}\frac{\left(-p_1^2\right)^{-\epsilon}-\left(-p_2^2\right)^{-\epsilon}}{p_1^2-p_2^2},
\end{equation}
\begin{equation}
	\cutRes_{e_1,e_2} \widetilde J_3\left(p_1^2,p_2^2\right)=-\frac{e^{\gamma_E\epsilon}\Gamma(1-\epsilon)}{\epsilon\Gamma(1-2\epsilon)}
	\frac{\left(p_1^2\right)^{-\epsilon}}{p_1^2-p_2^2}\,.
\end{equation}
The result for $\cutRes_{e_2,e_3} \widetilde J_3\left(p_1^2,p_2^2\right)$ is obtained by symmetry.
\begin{equation}\label{eq:lstp2p3}
	 j_3\left(p_1^2,p_2^2\right)=-\frac{1}{p_1^2-p_2^2}\,.
\end{equation}

\paragraph{One external massive leg with adjacent massive propagator.}\label{sec:tp1m12Res}
We set $p_2^2=p_3^2=m_2^2=m_3^2=0$ --- see \refF{fig:triNotation} ---
 and find
\begin{align}
	&\widetilde J_3(p_1^2;m_{1}^2)=\frac{e^{\gamma_E\epsilon}\Gamma(1+\epsilon)}{\epsilon(1-\epsilon)}(m_1^2)^{-1-\epsilon}\hypgeo{1}{1+\epsilon}{2-\epsilon}{\frac{p_1^2}{m_1^2}}=\\
	&=\frac{1}{p_1^2}\left[\frac{\ln\left(\frac{m_1^2}{m_1^2-p_1^2}\right)}{\epsilon}+\text{Li}_2\left(\frac{p_1^2}{m_1^2}\right)+\log ^2\left(1-\frac{p_1^2}{m_1^2}\right)+\log \left(m_1^2\right) \log\left(1-\frac{p_1^2}{m_1^2}\right)\right]+\mathcal{O}\left(\epsilon\right)\,, \nonumber
\end{align}
\begin{align}\bsp
	\cutRes_{e_1}\widetilde J_3(p_1^2;m_{1}^2)&=
	\frac{ e^{\gamma_E\epsilon}}{\epsilon\Gamma(1-\epsilon)}
	\frac{(-m_1^2)^{-\epsilon}}{p_1^2}\hypgeo{1}{\epsilon}{1-\epsilon}{\frac{m_1^2}{p_1^2}}\\
	&=\frac{1}{p_1^2\epsilon}-\frac{1}{p_1^2}\left(\ln\left(m_1^2-p_1^2\right)
	+\ln\left(\frac{m_1^2}{p_1^2}\right)\right)+\mathcal{O}\left(\epsilon\right)\,,
\esp\end{align}
\begin{align}\bsp
	\cutRes_{e_1,e_2}\widetilde J_3(p_1^2;m_{1}^2)&=
	-\frac{e^{\gamma_E\epsilon}\Gamma(1-\epsilon)}{\epsilon\Gamma(1-2\epsilon)}
	\frac{(p_1^2-m_1^2)^{-2\epsilon}}{(p_1^2)^{1-\epsilon}}\\
	&=-\frac{1}{p_1^2\epsilon}-\frac{1}{p_1^2}\left(\ln\left(p_1^2\right)
	-2\ln\left(p_1^2-m_1^2\right)\right)+\mathcal{O}\left(\epsilon\right)\,,
\esp\end{align}
\begin{equation}
	 j_3(p_1^2;m_{1}^2)=-\frac{1}{p_1^2}\,.
\end{equation}

\paragraph{One external massive leg with opposite massive propagator.}\label{sec:tp1m23Res}
We set $p_2^2=p_2^2=m_1^2=m_2^2=0$ --- 
see \refF{fig:triNotation} --- and find
\begin{align}\bsp\label{eq:tp1m23}
	\widetilde J_3(p_1^2;m_{3}^2)
	=&\frac{e^{\gamma_E\epsilon}}{\epsilon}\left[\left(m_3^2\right)^{-1-\epsilon}\frac{\Gamma(1+\epsilon)\Gamma(1-\epsilon)}{\Gamma(2-\epsilon)}\hypgeo{1}{1}{2-\epsilon}{-\frac{p_1^2}{m_3^2}}\right.\\
	&\left.-\frac{(-p_1^2)^{-\epsilon}}{m_3^2}\frac{\Gamma(1+\epsilon)\Gamma^2(1-\epsilon)}{\Gamma(2-2\epsilon)}\hypgeo{1}{1-\epsilon}{2-2\epsilon}{-\frac{p_1^2}{m_3^2}}\right]\\
	=&-\frac{1}{p_1^2}\left(\text{Li}_2\left(\frac{m_3^2+p_1^2}{m_3^2}\right)-\frac{\pi ^2}{6}\right)+\mathcal{O}(\epsilon)\, ,
\esp\end{align}

\begin{align}\bsp\label{eq:tp1m23Cutm23}
	\cutRes_{e_3}\widetilde J_3(p_1^2;m_{3}^2)&=
	-\frac{e^{\gamma_E\epsilon}}{\Gamma(2-\epsilon)}\frac{(-m_3^2)^{-\epsilon}}{p_1^2+m_3^2}\hypgeo{1}{1-\epsilon}{2-\epsilon}{\frac{p_1^2}{p_1^2+m_3^2}}\\
	&=\frac{1}{p_1^2}\log\left(\frac{m_3^2}{m_3^2+p_1^2}\right)+\mathcal{O}(\epsilon)\, ,
\esp\end{align}
\begin{align}\bsp
	\cutRes_{e_1,e_2}\widetilde J_3(p_1^2;m_{3}^2)
	&= \frac{e^{\gamma_E\epsilon}\Gamma(1-\epsilon)}{\Gamma(2-2\epsilon)}\frac{(p_1^2)^{-\epsilon}}{p_1^2+m_3^2}\hypgeo{1}{1-\epsilon}{2-2\epsilon}{\frac{p_1^2}{p_1^2+m_3^2}}\\
	&=\frac{1}{p_1^2}\ln\left(\frac{p_1^2+m_3^2}{m_3^2}\right)+\mathcal{O}(\epsilon)\, ,
	\label{eq:tp1m23CutP1}
\esp\end{align}
\begin{align}\bsp
	\cutRes_{e_1,e_2,e_3}\widetilde J_3(p_1^2;m_{3}^2)
	=&-\frac{e^{\gamma_E\epsilon}}{\Gamma(1-\epsilon)}\frac{(p_1^2)^{-1+\epsilon}(m_3^2)^{-\epsilon}}{(p_1^2+m_3^2)^{\epsilon}}\\
	=&-\frac{1}{p_1^2}+\mathcal{O}(\epsilon)\, ,
	\label{eq:tp1m23CutP1M23}
\esp\end{align}
\begin{equation}
	 j_3(p_1^2;m_{3}^2)=-\frac{1}{p_1^2}\,.
\end{equation}

The coaction of the $\epsilon^0$-coefficient of this integral normalized to $j_3(p_1^2;m_{3}^2)$ is
\begin{align}\bsp\label{eq:tp1m23Coprod}
	\Delta \left[J^{(0)}_3\left(p_1^2;m_3^2\right)\right]=&
	1\otimes\text{Li}_2\left(\frac{m_3^2+p_1^2}{m_3^2}\right)
	+\ln(m^2_3)\otimes\ln\left(\frac{m_3^2+p_1^2}{m_3^2}\right)\\
	&+\ln(-p_1^2)\otimes\ln\left(\frac{m_3^2}{m_3^2+p_1^2}\right)
	+\left(\text{Li}_2\left(\frac{m_3^2+p_1^2}{m_3^2}\right)-\frac{\pi^2}{6}\right)\otimes 1\,.
\esp\end{align}

\paragraph{Three external massive legs.}
\label{sec:tp1p2p3Res}
We set $m_1^2=m_2^2=m_3^2=0$ --- see \refF{fig:triNotation} --- and use the variables in \refE{eq:ZZbarDef} to find
\begin{equation}
	\widetilde J_3\left(p_1^2,p_2^2,p_3^2;\epsilon\right)=
	\frac{(-p_1^2)^{-1-\epsilon}}{(\zz-\zbar)}\left[\mathcal{T}(\zz,\zbar)
	+\mathcal{O}(\epsilon)\right]\,,
\end{equation}
\begin{equation}\label{tp1p2p3Ep0}
	\mathcal{T}(\zz,\zbar)=-2\Li_2(\zz)+2\Li_2(\zbar)-\ln(\zz\zbar)
	\ln\left(\frac{1-\zz}{1-\zbar}\right)\,,
\end{equation}
\begin{align}\bsp
	\label{eq:tp1p2p3CutP1}
	\cutRes_{e_1,e_2}\widetilde J_3\left(p_1^2,p_2^2,p_3^2\right)=&
	\frac{e^{\gamma_E  \epsilon} \Gamma (1-\epsilon)}{\Gamma (2-2 \epsilon)}
	\frac{\left(p_1^2\right)^{-1-\epsilon}}{(1-z) \zbar }\,
	_2F_1\left(1,1-\epsilon;2-2 \epsilon;\frac{z-\zbar}{(z-1)\zbar}\right)\\
	=&\frac{1}{p_1^2(z-\zbar)}\log \left(\frac{z (1-\zbar)}{\zbar(1-z) }\right)+\mathcal{O}(\epsilon)\, ,
\esp\end{align}
\begin{align}\bsp
	\label{eq:tp1p2p3CutP2}
	\cutRes_{e_2,e_3}\widetilde J_3\left(p_1^2,p_2^2,p_3^2\right)&=\frac{e^{\gamma_E  \epsilon} \Gamma (1-\epsilon)}{\Gamma (2-2 \epsilon)}\frac{\left(-p_1^2\right)^{-1-\epsilon}}{(1-z) (-z\zbar)^{\epsilon} }
	\,_2F_1\left(1,1-\epsilon;2-2 \epsilon;\frac{z-\zbar}{z-1}\right)\\
	&=\frac{1}{p_1^2(z-\zbar)}\log \left(\frac{1-z }{ 1-\zbar}\right)+\mathcal{O}(\epsilon)\, ,
\esp\end{align}
\begin{align}\bsp
	\label{eq:tp1p2p3CutP3}
	\cutRes_{e_1,e_3}\widetilde J_3\left(p_1^2,p_2^2,p_3^2\right)
	=&\frac{ e^{\gamma_E  \epsilon} \Gamma (1-\epsilon)}{\Gamma (2-2 \epsilon)}
	\frac{\left(-p_1^2\right)^{-1-\epsilon}}{\zbar((z-1)(1-\zbar))^{\epsilon}  }\,
	_2F_1\left(1,1-\epsilon;2-2 \epsilon;\frac{\zbar-z}{\zbar}\right)\\
	=&\frac{1}{p_1^2(z-\zbar)}\log \left(\frac{\zbar }{ z}\right)+\mathcal{O}(\epsilon)\, ,
\esp\end{align}
\begin{align}\bsp
	\label{eq:tp1p2p3CutMax}
	\cutRes_{e_1,e_2,e_3}\widetilde J_3\left(p_1^2,p_2^2,p_3^2\right)
	&=-\frac{e^{\gamma_E  \epsilon}}{\Gamma (1-\epsilon)}
	\frac{(p_1^2)^{-1-\epsilon}}{(z-\zbar)^{1-2\epsilon}}(z\zbar(1-z)(1-\zbar))^{-\epsilon}\\
	&=-\frac{1}{p_1^2(z-\zbar)}+\mathcal{O}(\epsilon)\,,
\esp\end{align}
\begin{equation}
	j_3\left(p_1^2,p_2^2,p_3^2\right)=-\frac{1}{p_1^2(z-\zbar)}\,.
\end{equation}

\subsection{Four-point functions}

\begin{figure}[]
\includegraphics[width=4cm]{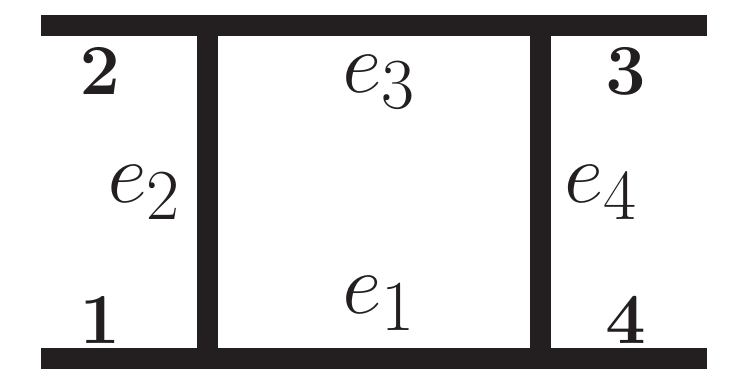}
\centering
\caption{Notation for external and internal edges of four-point functions. Propagator $e_i$ has mass $m_i^2$, and external leg $\bm{i}$ has mass $p_i^2$.}
\label{fig:boxNotation}
\end{figure}

See Figure~\ref{fig:boxNotation} for our labeling conventions.

\paragraph{Massless external and internal edges.}\label{sec:bst}
We set $p_i^2=m_i^2=0$ for all $i$, and recall $s=(p_1+p_2)^2$ and $t=(p_2+p_3)^2$:
\begin{align}\label{bst}
	\widetilde J_4(s,t)=&\,\frac{2\,c_\Gamma}{\epsilon}\left\{\frac{\Gamma^2(1-\epsilon)\Gamma^2(1+\epsilon)}{\epsilon\, \Gamma(1+2\epsilon)\Gamma(1-2\epsilon)}\frac{2}{st}\left(-\frac{s+t}{st}\right)^{\epsilon}-\frac{(-s)^{-2-\epsilon}}{1+\epsilon}\hypgeo{1}{1}{2+\epsilon}{-\frac{t}{s}}\right.\nonumber\\
	&\left.-\frac{(-t)^{-2-\epsilon}}{1+\epsilon}\hypgeo{1}{1}{2+\epsilon}{-\frac{s}{t}}\right\}\\
	=&\frac{2}{st}\left(\frac{2}{\epsilon^2}-\frac{\ln(s t)}{\epsilon}+\log(-s)\log(-t)-\frac{2\pi^2}{3}\right)+\mathcal{O}(\epsilon),\nonumber
\end{align}
\begin{align}\bsp
	\cutRes_{e_1,e_3}\widetilde J_4(s,t)=&
	2\frac{e^{\gamma_E\epsilon}\Gamma(1-\epsilon)}{\epsilon\Gamma(1-2\epsilon)}
	s^{-2-\epsilon}\hypgeo{1}{1}{1-\epsilon}{1+\frac{t}{s}}\\
	=&-\frac{2}{st}\left(\frac{1}{\epsilon}-\ln(-t)\right)+\mathcal{O}(\epsilon).
\esp\end{align}
By symmetry, $\cutRes_{e_2,e_4}\widetilde J_4(s,t)$ is obtained by swapping $s\leftrightarrow t$.
\begin{align}\bsp\label{eq:bstCutQuad}
	\cutRes_{e_1,e_2,e_3,e_4}\widetilde J_4(s,t)=&
	2\,\frac{e^{\gamma_E\epsilon}\Gamma(1-\epsilon)}{\Gamma(1-2\epsilon)}
	\frac{(st)^{-1-\epsilon}}{(s+t)^{-\epsilon}},
\esp\end{align}
\begin{equation}
	j_4\left(s,t\right)=\frac{2}{st}.
\end{equation}

\paragraph{`Two-mass-easy' box.}
We set $p_2^2=p_4^2=0$, $m_i^2=0$ for all $i$ and $s=(p_1+p_2)^2$ and $t=(p_2+p_3)^2$. We recall the result given in ref.~\cite{Brandhuber:2005kd},
\begin{align}\bsp\label{bstp1p3}
	\widetilde J_4&(s,t,p_1^2,p_3^2)=\frac{2\,c_\Gamma}{\epsilon^2(st-p_1^2p_3^2)}
	\Bigg\{(-s)^{-\epsilon}+(-t)^{-\epsilon}-(-p_1^2)^{-\epsilon}-(-p_3^2)^{-\epsilon}\\
	&+\sum_{j=0}^3(-1)^{j}\left(\frac{s+t-p_1^2-p_3^2}{\alpha_j}\right)^{\epsilon}\hypgeo{\epsilon}{\epsilon}{1+\epsilon}{\frac{st-p_1^2p_3^2}{\alpha_j}}\Bigg\},
\esp\end{align}
with
\begin{align*}\bsp
	\alpha_0=&\left(p_1^2-s\right)\left(p_1^2-t\right)\qquad\qquad \alpha_1=\left(p_1^2-s\right)\left(s-p_3^2\right)\\
	\alpha_2=&\left(p_3^2-s\right)\left(p_3^2-t\right)\qquad\qquad \alpha_3=\left(p_1^2-t\right)\left(t-p_3^2\right).
\esp\end{align*}
\begin{align}\bsp
	\cutRes_{e_1,e_3}&\widetilde J_4(s,t,p_1^2,p_3^2)=-\frac{2}{\left(st-p_1^2 p_3^2\right)}\left[\frac{1}{\epsilon}-\log \left(\frac{s(p_1^2 p_3^2-s t)}{\left(s-p_1^2\right)
	\left(s-p_3^2\right)}\right)\right]+\mathcal{O}(\epsilon),\\
\esp\end{align}
$\cutRes_{e_2,e_4}\widetilde J_4(s,t,p_1^2p_3^2)$ is obtained by symmetry.
\begin{align}\bsp
	\cutRes_{e_1,e_2}&\widetilde J_4(s,t,p_1^2,p_3^2)=\frac{2}{\left(st-p_1^2 p_3^2\right)}\left[\frac{1}{\epsilon}-\ln\left(\frac{p_1^2(s t-p_1^2p_3^2)}{(p_1^2-s)(p_1^2-t)}\right)\right]+\mathcal{O}(\epsilon),\\
\esp\end{align}
$\cutRes_{e_3,e_4}\widetilde J_4(s,t,p_1^2p_3^2)$ is obtained by symmetry.
\begin{align}\bsp
	\cutRes_{e_1,e_2,e_3,e_4}\widetilde J_4(s,t,p_1^2,p_3^2)=2\frac{ e^{\gamma_E  \epsilon } \Gamma (1-\epsilon ) }{\Gamma (1-2 \epsilon )}
	\frac{\left(s+t-p_1^2-p_3^2\right){}^{\epsilon }}{\left(s t-p_1^2 p_3^2\right){}^{1+\epsilon}}\,.
\esp\end{align}
\begin{equation}
	j_4\left(s,t,p_1^2,p_3^2\right)=\frac{2}{\left(st-p_1^2 p_3^2\right)}\,.
\end{equation}

\paragraph{`Two-mass-hard' box.}
\label{sec:2mh-box}
We set $p_3^2=p_4^2=0$, $m_i^2=0$ for all $i$, $s=(p_1+p_2)^2$ and $t=(p_2+p_3)^2\equiv sx$. Given that the triangle with three-external masses $\widetilde J_3(s,p_1^2,p_2^2)$ appears in the diagrammatic coaction of this diagram, it is convenient to introduce $y$ and $\bar{y}$ such that
\begin{equation}\label{eq:ydef}
	y\bar y=\frac{p_1^2}{s},\qquad (1-y)(1-\bar y)=\frac{p_2^2}{s}.
\end{equation}
Up to order $\epsilon^0$, the expression for the two-mass-hard box can be found e.g.~in ref.~\cite{Bern:1993kr}:
\begin{align}\bsp\label{eq:bstp1p2}
	\widetilde J_4(s,t,p_1^2,p_2^2)&=(-s)^{-2-\epsilon}c_\Gamma\frac{(y\bar y)^{\epsilon}x^{-1-2\epsilon}}{((1-y)(1-\bar y))^{-\epsilon}}\left[\frac{1}{\epsilon^2}+2\Li_2\left(1-\frac{x}{y\bar y}\right)\right.\\
	&\left.+2\Li_2\left(1-\frac{x}{(1-y)(1-\bar y)}\right)-\frac{\pi^2}{12}\right]+\mathcal{O}(\epsilon).
\esp\end{align}
The cuts are found to be
\begin{align}\bsp
	\cutRes_{e_1,e_3}\widetilde J_4(s,t,p_1^2,p_2^2)=-\frac{s^{-2-\epsilon}}{x^{1+2\epsilon}}
	\left[\frac{1}{\epsilon}+\ln(y\bar y)+\ln((1-y)(1-\bar y))\right]+\mathcal{O}(\epsilon)\,,
\esp\end{align}
\begin{align}\bsp
	\cutRes_{e_2,e_4}\widetilde J_4(s,t,p_1^2,p_2^2)=-2\frac{s^{-2-\epsilon}}{x^{1+2\epsilon}}
	\left[\frac{1}{\epsilon}+\ln(\left(x-y\bar y\right)\left(x-(1-y)(1-\bar y)\right))\right]+\mathcal{O}(\epsilon)\,,
\esp\end{align}
\begin{align}\bsp
	\cutRes_{e_1,e_2}\widetilde J_4(s,t,p_1^2,p_2^2)=\frac{s^{-2-\epsilon}}{x^{1+2\epsilon}}
	\left[\frac{1}{\epsilon}+\ln\left(\frac{(1-y)(1-\bar y)}{y\bar y}\right)+2\ln(x-y\bar y)\right]+\mathcal{O}(\epsilon)\,,
\esp\end{align}
\begin{align}
	\cutRes_{e_2,e_3}\widetilde J_4(s,t,p_1^2,p_2^2)=\frac{s^{-2-\epsilon}}{x^{1+2\epsilon}}
	&\left[\frac{1}{\epsilon}-\ln\left(\frac{(1-y)(1-\bar y)}{y\bar y}\right)+2\ln(x-(1-y)(1-\bar y))\right]\nonumber\\
	&+\mathcal{O}(\epsilon)\,,
\end{align}
\begin{align}\bsp
	\cutRes_{e_1,e_2,e_3}\widetilde J_4(s,t,p_1^2,p_2^2)=-\frac{1}{s^2 x}+\mathcal{O}(\epsilon)\,,
\esp\end{align}
\begin{align}\bsp
	\cutRes_{e_1,e_2,e_3,e_4}\widetilde J_4(s,t,p_1^2,p_2^2)=2
	\frac{ e^{\gamma_E  \epsilon } \Gamma (1-\epsilon ) }{\Gamma (1-2 \epsilon )}
	\frac{\left(x-\bar y(1-y)\right)^{\epsilon}\left(x- y(1-\bar y)\right)^{\epsilon}}{s^{2+\epsilon}x^{1+2\epsilon}}\,,
\esp\end{align}
\begin{equation}
	j_4\left(s,t,p_1^2,p_2^2\right)=\frac{2}{s^2 x}\,.
\end{equation}

\paragraph{Box with a single massive propagator.}\label{sec:bstm1}
We set $p_i^2$ for all $i$, $m_{2}^2=m_{3}^2=m_{4}^2=0$, $s=(p_1+p_2)^2$ and $t=(p_2+p_3)^2$:
\begin{align}\bsp
	&\widetilde J_4(s,t,m_1^2)=\frac{(-t)^{-\epsilon}}{t\left(s-m_1^2\right)}\left[\frac{1}{\epsilon^2}+\frac{2}{\epsilon}\ln\left(\frac{m_1^2}{m_1^2-s}\right)-\frac{1}{2}\ln^2\left(-\frac{m_1^2}{t}\right)\right.\\
	&+G\left(1,0,-\frac{m_1^2}{t}\right)
	+4 \ln\left(\frac{s}{t}\right) G\left(-\frac{s}{t},-\frac{m_1^2}{t}\right)-4 G\left(-\frac{s}{t},0,-\frac{m_1^2}{t}\right)\\
	&\left.+4 G\left(-\frac{s}{t},-\frac{s}{t},-\frac{m_1^2}{t}\right)-\frac{13 \pi ^2}{12}\right]+\mathcal{O}(\epsilon)\,,
\esp\end{align}
\begin{align}\bsp
	\cutRes_{e_1}\widetilde J_4(s,t,m_1^2)=\frac{1}{t\left(m_1^2-s\right)}
	&\left[-\frac{2}{\epsilon}+2\ln(-t)+\ln\left(\frac{m_1^2}{m_1^2+t}\right)+4\ln\left(\frac{m_1^2-s}{s}\right)\right]+\mathcal{O}(\epsilon)\,,
\esp\end{align}
\begin{align}\bsp
	\cutRes_{e_1,e_3}\widetilde J_4(s,t,m_1^2)=\frac{2}{t\left(m_1^2-s\right)}
	\left[\frac{1}{\epsilon}+2\ln\left(\frac{s}{s-m_1^2}\right)-\ln(-t)\right]
	+\mathcal{O}(\epsilon)\,,
\esp\end{align}
\begin{align}\bsp
	\cutRes_{e_2,e_4}\widetilde J_4(s,t,m_1^2)=\frac{1}{t\left(m_1^2-s\right)}
	&\left[\frac{1}{\epsilon}+\ln\left(\frac{m_1^2(m_1^2+t)}{t}\right)+2\ln\left(m_1^2-s\right)\right]+\mathcal{O}(\epsilon)\,,
\esp\end{align}
\begin{align}\bsp
	\cutRes_{e_1,e_2,e_3}\widetilde J_4(s,t,m_1^2)=\frac{1}{t\left(m_1^2-s\right)}
	+\mathcal{O}(\epsilon)\,,
\esp\end{align}
\begin{align}\bsp
	\cutRes_{e_1,e_2,e_3,e_4}\widetilde J_4(s,t,m_1^2)=2
	\frac{e^{\gamma_E  \epsilon } \Gamma (1-\epsilon ) }{\Gamma (1-2 \epsilon )}
	\frac{s^{\epsilon}(s+t)^{\epsilon}}{t^{1+\epsilon}\left(s-m_1^2\right)^{1+2\epsilon}}\,,
\esp\end{align}
\begin{equation}
	j_4\left(s,t;m_1^2\right)=\frac{2}{t\left(s-m_1^2\right)}\,.
\end{equation}


\section{The Hopf algebras of graphs and MPLs}
\label{app:hopf}

In this appendix we present a brief overview and references on Hopf algebras and comodules. We start by a general introduction and then focus on the special instances of Hopf algebras and comodules on cut graphs encountered in this paper. All vector spaces and algebras will always be assumed to be defined over $\mathbb{Q}$.

\subsection{Bialgebras, Hopf algebras and comodules}

A \emph{bialgebra} is an algebra $H$ together with two maps, the \emph{coproduct} $\Delta:H\to H\otimes H$ and the \emph{counit} $\varepsilon: H\to\mathbb{Q}$, satisfying the following properties:
\begin{enumerate}
\item Coassociativity: 
\beq
(\Delta\otimes \textrm{id})\Delta = (\textrm{id}\otimes \Delta)\Delta\,.
\eeq
\item $\Delta$ and $\varepsilon$ are algebra homomorphisms:
\beq
\Delta(a\cdot b) = \Delta(a)\cdot \Delta(b) {\rm~~and~~}
\varepsilon(a\cdot b) = \varepsilon(a)\cdot \varepsilon(b)\,,\quad \forall a,b\in H\,.
\eeq
\item The counit and the coproduct are related by
\beq
(\varepsilon\otimes \textrm{id})\Delta = (\textrm{id}\otimes \varepsilon)\Delta = \textrm{id}\,.
\eeq
\end{enumerate}
An element $a\in H$ is called \emph{group-like} if $\Delta(a)=a\otimes a$.

A \emph{Hopf algebra} is a bialgebra $H$ together with an involution $S:H\to H$ that satisfies
\beq
m(\textrm{id}\otimes S)\Delta = m(S\otimes\textrm{id})\Delta = \varepsilon{\rm~~and~~} S(a\cdot b) = S(b)\cdot S(a)\quad \forall a,b\in H \,,
\eeq
where $m$ denotes the multiplication in $H$.
If $H$ is a Hopf algebra, then a $H$ (right-)\emph{comodule} is a vector space $A$ together with a map $\rho:A\to A\otimes H$ such that
\beq
(\rho\otimes \textrm{id})\rho = (\textrm{id}\otimes \Delta)\rho {\rm~~and~~} (\textrm{id}\otimes \varepsilon)\rho = \textrm{id}\,.
\eeq

\subsection{Incidence Hopf algebras}
In this section we discuss a special incarnation of a Hopf algebra, which plays an important role in this paper. 
More precisely, we present the incidence algebra, which is a simple combinatorial construction
on partially ordered sets.  
The incidence algebra is defined as follows.  Let $P$ be a \emph{partially ordered set}, i.e., a set together with a partial order $\le$ on its elements. If $A\le B$, we define the \emph{interval} $[A,B] = \{X\in P| A\le X\le B\}$, and we denote by $P_2$ the algebra generated by all intervals $[A,B]$. Then $P_2$ can be turned into a bialgebra, known as the \emph{incidence  bialgebra}~\cite{JoniRota,Schmitt1}, with the coproduct
\beq \label{eq:math-incidence-coproduct}
\Delta([A,B]) =\sum_{A\le X\le B}[A,X]\otimes [X,B]\,,
\eeq
and the counit
\beq
\varepsilon([A,B]) = \left\{\begin{array}{ll}
1\,,&\textrm{ if } A=B\,,\\
0\,,&\textrm{ otherwise}\,.
\end{array}\right.
\eeq
It is easy to see that elements of the form $[A,A]$ are grouplike, $\Delta([A,A])=[A,A]\otimes[A,A]$.

 An incidence bialgebra can be turned into a Hopf algebra by the construction of an antipode, if and only if the elements $[A,A]$ are invertible~\cite{Schmitt2}. Let us define $\widetilde{P}_2$ as the bialgebra obtained by adding to $P_2$ new grouplike generators $[A,A]^{-1}$, 
 $A\in P$, which are multiplicative inverses of $[A,A]$. 
Then $\widetilde{P}_2$ is a Hopf algebra, and the antipode is unique and involutive (i.e.~$S^2=1$). It is given by~\cite{Schmitt2}
\beq
S([A,B]) = \left\{\begin{array}{ll}
\displaystyle[A,A]^{-1}\,,& \textrm{ if } A=B\,,\\
\displaystyle\sum_{\substack{A=X_0\le X_1\le\ldots\le  X_{k}=B\\ X_i\neq X_j}}\frac{(-1)^k}{[A,A]}\,\prod_{i=1}^{k}\frac{[X_{i-1},X_{i}]}{[X_{i},X_{i}]}\,,&\textrm{ otherwise}\,.
\end{array}\right.
\eeq
Since $S$ is involutive, this implies $S([A,A]^{-1}) = S^2([A,A]) = [A,A]$.

\subsection{{Examples of incidence Hopf algebras}}

\paragraph{The Hopf algebra of MPLs.}  A first example of an incidence Hopf algebra is the Hopf algebra $\cH$ of MPLs (for generic values of the arguments). It is easy to see that the coproduct on MPLs in eq.~\eqref{eq:Delta_MPL} can be cast in the form of eq.~\eqref{eq:math-incidence-coproduct} if we identify the set $P$ with the set of all words $\vec a$ formed from the alphabet $\{a_1,\ldots,a_n\}$ (including the empty word). The partial order on $P$ is the natural ordering induced by the inclusion of words. We can then identify an interval of words $[\vec b,\vec a]$ with the function $G_{\vec b}(\vec a;z)$, and the interval $[\emptyset,\vec a]$ is identified with the MPL $G(\vec a;z)$.

\paragraph{The incidence Hopf algebra of pairs of subsets.}
A second example of an incidence Hopf algebra is provided by the set $\cP(E)$ of all subsets of a set $E$. The subsets are ordered by inclusion, and the set of intervals $\cP_2(E)$ is the set of all pairs $[A,B]$ with $A\subseteq B\subseteq E$. The incidence coproduct on $\cP_2(E)$ takes the form
\beq
\Delta([A,B]) =\sum_{A\subseteq X\subseteq B}[A,X]\otimes [X,B]\,,
\eeq

There are two natural families of Hopf subalgebras in $\widetilde{\cP}_2(E)$:
\begin{enumerate}
\item 
Nested subsets containing a given subset $C$. 
Define $\widetilde{\cP}_2(E,C)
$
 as the subalgebra generated by elements $[A,B]$ and $[A,A]^{-1}$ with $C\subseteq A\subseteq B\subseteq E$. It is easy to check that $\widetilde{\cP}_2(E,C)$ is a Hopf subalgebra of $\widetilde{\cP}_2(E)$.
\item 
Nested subsets of a minimum cardinality $k$.
For $k$ a positive integer, we define $\widetilde{\cP}_{2,k}(E)
$ as the subalgebra generated by elements $[A,B]$ and $[A,A]^{-1}$ with $|A|\ge k$. Here too, $\widetilde{\cP}_{2,k}(E)$ is a Hopf subalgebra of $\widetilde{\cP}_2(E)$.
\end{enumerate}

Further, we can use these Hopf subalgebras to define certain comodules. Suppose that we consider pairs in which the smaller subset is fixed. To be precise,
for $C\subseteq E$, we define $\widetilde{\mathcal{Q}}_2(E,C)$ as the subalgebra of $\widetilde{\cP}_2(E)$ generated by elements $[C,B]$ with $C\subseteq B\subseteq E$. 
However, $\widetilde{\mathcal{Q}}_2(E,C)$ is not a Hopf subalgebra. Instead, we can see it as a $\widetilde{\cP}_{2,|C|+1}(E)$ comodule. The coaction is given by the map $\rho: \widetilde{\mathcal{Q}}_{2}(E,C) \to \widetilde{\mathcal{Q}}_{2}(E,C)\otimes \widetilde{\cP}_{2,|C|+1}(E)$ defined by
\beq\label{eq:incidence_coaction_2}
\rho([C,B]) = \sum_{\substack{C\subset X\subseteq B\\ X\neq C}}[C,X]\otimes [X,B]\,.
\eeq

\paragraph{The incidence Hopf algebra of cut graphs.}
It is readily apparent that the coaction on cut graphs defined in \refE{eq:incidence_coaction} takes the form of an incidence algebra.
The incidence algebra for the family of cut graphs obtained by pinches of a cut graph $(G,E_G)$ is then $\widetilde{\cP}_2(E_G)$.  For a cut graph $(G,C)$, it is $\widetilde{\cP}_2(E_G,C)$.
However, we introduce several modifications:
\begin{itemize}
\item The ordering of pairs is reversed compared to the mathematical literature such as \refE{eq:math-incidence-coproduct}, so that in the body of this paper, the subsets appear on the right.
\item  As a matter of notation, we write the elements in the form $(G,C)$ rather than $[E_G,C]$, to reinforce the notion that we are acting on graphs with physical interpretations.
\item  
We specifically exclude the empty set from the sum.  (Compare the indexing in \refE{eq:incidence_coaction} with \refE{eq:math-incidence-coproduct}, where there is no such restriction, and no special treatment for the case $A = \emptyset$.)  In this respect, our operation differs from the mathematical literature on incidence algebras.  In fact, this modification means {that we are working with the comodule $\widetilde{\mathcal{Q}}_2(E_G,C)$ on which the Hopf algebra $\widetilde{\cP}_{2,|C|+1}(E_G)$ {\em coacts} according to \refE{eq:incidence_coaction_2}. In other words, we can identify the coaction $\Delta_{\textrm{Inc}}$ on cut graphs defined in eq.~\eqref{eq:incidence_coaction} with the coaction $\rho$ of \refE{eq:incidence_coaction_2}.}
\end{itemize}

This incidence algebra does not involve any information about the graphical structure other than the list of edges. We note that a similar construction was suggested in \cite{Goncharov:2005sla} but, importantly, it did not consider cut graphs.


\section{Compactification of cut and uncut one-loop integrals}\label{app:compact}

In ref.~\cite{Abreu:2017ptx} it was shown that in order to study cuts of one-loop Feynman integrals, it is convenient to write them
as integrals over a compact quadric in the complex projective space $\mathbb{CP}^{D+1}$~\cite{PhamCompact,SimmonsDuffin:2012uy,Caron-Huot:2014lda}. In this appendix we briefly summarize the main results presented there, and we compute the cut integrals necessary to completely determine the differential equations satisfied by a generic one-loop integral as discussed in Section \ref{sec:alph_diff_eq}. Throughout this appendix, we will work in Euclidean kinematics to match ref.~\cite{Abreu:2017ptx}.

We equip $\mathbb{CP}^{D+1}$ with the bilinear form
\beq\label{eq:scalarProd}
	(Z_1Z_2) = z_1^\mu\,z_{2\mu} - \frac{1}{2}\,Z_1^+Z_2^- - \frac{1}{2}\,Z_1^-Z_2^+\,,
\eeq
where $z_1^\mu\,z_{2\mu}$ denotes the usual Euclidean scalar product.
If we work in the coordinate patch $Z^+=1$, then to each propagator $D_i = (k^E-q^E_i)^2+m_i^2$ we associate  the point $X_i\in \mathbb{CP}^{D+1}$ defined by
\beq
X_i = \left[\begin{array}{c}(q^E_i)^\mu\\ (q_i^E)^2+m_i^2\\ 1\end{array}\right]\,,\qquad 1\le i\le n\,.
\eeq
A one-loop integral is then written as
\begin{equation}\label{eq:uncutApp}
	I_n^D = \frac{(-1)^n\,e^{\gamma_E\eps}}{\pi^{D/2}}\int_\Sigma  \frac{d^{D+2}Y\,\delta((YY))}{\textrm{Vol}(GL(1))}\frac{[-2(X_{\infty}Y)]^{n-D}}{[-2(X_1Y)]\ldots[-2(X_{n}Y)]}\,,
\end{equation}
where ${\Sigma}$ is the real quadric defined by $(YY)=0$ (take e.g.~$Y=[(k^E)^\mu,(k^E)^2,1]^T$), and we have  introduced the lightlike `point at infinity'
\beq\label{eq:infinityPoint}
X_{\infty}= \left[\begin{array}{c}0^\mu\\ 1\\ 0\end{array}\right]\,.
\eeq

In ref.~\cite{Abreu:2017ptx}, it was shown that all one-loop integrals,
cut or uncut, can be written in terms of the same class of functions
$Q_{n}^D$ defined as
\begin{equation}
\label{eq:Q_n_def}
Q_{n}^D(X_1,\ldots,X_n,X_{0}) = \frac{(-1)^{n}\,e^{\gamma_E\eps}}{\pi^{D/2}}\int_{\Sigma}\frac{d^{D+2}Y\,\delta((YY))}{\textrm{Vol}(GL(1))}\,\frac{[-2(X_{0}Y)]^{n-D}}{[-2(X_1Y)]\ldots [-2(X_nY)]}\,,
\end{equation}
where the integration runs over the real quadric $(YY)=0$.
In the special case where $X_0=X_\infty$, \refE{eq:Q_n_def} reduces to an ordinary one-loop integral,
\begin{equation}\label{eq:IComp}
	I_n^D=Q_{n}^D(X_1,\ldots,X_n,X_{\infty})\,.
\end{equation}

The cut integral $\cutRes_CI_n^D$, where $C$ is a subset of the propagators in
\refE{eq:uncutApp} has a simple interpretation in $\mathbb{CP}^{D+1}$. Let $P_i=\{X\in\mathbb{CP}^{D+1}:(YX_i)=0\}$ and $P_C=\bigcap_{i\in C}P_i$. Then the cutting operation is, loosely speaking, a projection onto the hyperplane $P_C$: the integration contour is given by $\Sigma\cap P_C$, which is still a quadric, and the points $Y$ and $X_j$ for $j\notin C$ are transformed into $Y_\bot$ and $X'_{C,j}$, defined as
\begin{equation}
	Y_\bot=Y-\sum_{i\in C}a_iX_i\,,\qquad\quad
	X'_{C,j}=X_j-\sum_{i\in C}a_{ij}X_i\,,
\end{equation}
and such that $(Y_\bot X_i)=(X'_{C,j} X_i)=0$ for $i\in C$. In particular, while $X_\infty$ is lightlike, $X'_{C,\infty}$ might not be. It is then easy to see that, taking
$C=\{1,\ldots,c\}$,
\begin{align}
\nonumber
\cC_CI_n^D &= \frac{(-1)^{n}\,(2\pi i)^{\lfloor c/2\rfloor}\,e^{\gamma_E\eps}}{(-2)^c\,\pi^{D/2}\sqrt{(-1)^cY_C}}\int_{\tilde{S}_C}\!\frac{d^{D-c+2}Y_{\bot}\,\delta((Y_\bot Y_\bot))}{\textrm{Vol}(GL(1))}
\,\frac{[-2(X'_{C,\infty}Y_{\bot})]^{(n-c)-(D-c)}}{[-2(X'_{C,c+1}Y_{\bot})]\ldots [-2(X'_{C,n}Y_{\bot})]}\\
\label{eq:cut-loop-duality}
&\,= \frac{2^{-c}\,(2\pi i)^{\lfloor c/2\rfloor}}{\pi^{c/2}\sqrt{(-1)^cY_C}}\,Q_{n-c}^{D-c}(X'_{C,c+1},\ldots,X'_{C,n},X'_{C,\infty})\,.
\end{align}
Here, $Y_C$ is the modified Cayley determinant. Both $Y_C$ and the Gram determinant $\Gram_C$ have simple expressions in terms of the $X_i$:
\beq\label{eq:compact_Gram}
\det(X_iX_j)_{i,j\in C} = \left\{\begin{array}{ll}
(-1)^c\,Y_C\,,&\textrm{ if } \infty \notin C\,,\\
\frac{(-1)^{c-1}}{4}\,\Gram_{C\setminus \{\infty\}}\,,&\textrm{ if } \infty \in C\,.
\end{array}\right.
\eeq
It is often useful to write the scalar products $(X'_{C,i}X'_{C,j})$ in terms of the $X_i$:
\beq\label{eq:proj_scalar_product}
(X'_{C,i}X'_{C,j}) = \frac{1}{(-1)^c\, Y_C}\,\det\left(\begin{array}{cccc}
(X_1X_1) & \ldots &(X_cX_1)& (X_iX_1)\\
\vdots && \vdots&\vdots \\
(X_1X_c) & \ldots &(X_cX_c)& (X_iX_c)\\
(X_1X_j) & \ldots &(X_cX_j)& (X_iX_j)
\end{array}\right)\,.
\eeq

The functions  $Q_{n}^D$ can easily be written as parametric integrals after combining denominators with Feynman parameters \cite{SimmonsDuffin:2012uy}. In particular, in the case where none of the $X_i$ are lightlike, which is the case relevant for cuts of a generic one-loop integral, we find
\begin{align}\bsp\label{eq:Q_Feynman}
	&Q_n^D(X_1,\ldots,X_n,X_0)=\\
	&=\frac{(-1)^ne^{\gamma_E\eps}\Gamma(D/2)}{\Gamma(D-n)}\!
	\int [db]\left(\frac{b_0}{\sqrt{(X_0X_0)}}\right)^{D-n-1}\!
	\Bigg(-\sum_{i=0}^n b_i^2-2\!\!\sum_{\substack{i, j=0 \\ i<j}}^n b_i b_j u_{ij}\Bigg)^{-D/2}\!\!,
\esp\end{align}
where
\begin{equation}
	u_{ij}\equiv\frac{(X_iX_j)}{\sqrt{(X_iX_i)(X_jX_j)}}\,,\qquad\quad
	\int [db]\equiv\left(\prod_{i=0}^n\int_0^\infty \frac{db_i}{\sqrt{(X_iX_i)}}\right)
	\,\delta\left(1-h(b)\right)\,,
\end{equation}
with $h(b)=\sum_{i=0}^n h_i b_i$ such that the $h_i\geq 0$ are not all zero (see e.g.~ref.~\cite{Panzer:2015ida}). The $Y$-integration in \refE{eq:Q_n_def} is performed after Feynman parametrization using
\begin{equation}\label{eq:confInt}
	I(X)=\int_{\Sigma}\frac{d^{D+2}Y\,\delta((Y Y))}{\textrm{Vol}(GL(1))}
	[-2(Y X)]^{-D}=\frac{\pi^{D/2}\Gamma(D/2)}{\Gamma(D)}
	[-(X X)]^{-D/2}\,.
\end{equation}

Note that eq.~\eqref{eq:Q_Feynman} is only valid for $D\neq n$, because otherwise the integrand has a pole at $b_0=0$ {and a problematic factor of $\Gamma(D-n)$}. The case $D=n$, however, is important, because it covers precisely the DCI integrals considered in Section~\ref{sec:dci}. {In this case, we see
from \refE{eq:Q_n_def} that $Q^n_n$ does not depend on $X_0$ and} it
is straightforward to repeat the derivation of eq.~\eqref{eq:Q_Feynman} for $D=n$. We find
\begin{align}\bsp\label{eq:Q_Feynman_DCI}
	&Q_n^n(X_1,\ldots,X_n)=(-1)^ne^{\gamma_E\eps}\,\Gamma(n/2)
	\int [db]
	\Bigg(-\sum_{i=1}^n b_i^2-2\!\!\sum_{\substack{i, j=1 \\ i<j}}^n b_i b_j u_{ij}\Bigg)^{-n/2}\!\!,
\esp\end{align}
where the integration measure $[db]$ does not involve the variable $b_0$. Equation~\eqref{eq:Q_Feynman_DCI} describes an integral over a projective simplex of an integrand with singularities along a quadric. Integrals of this type have been considered by Goncharov in ref.~\cite{Goncharov:1996}.


\subsection{Explicit results for some cut integrals}

In this appendix, we present results for the cut integrals needed to obtain
the differential equations of one-loop integrals. The maximal cuts
are trivial and given by \refE{eq:confInt}, so we will simply quote the result
below. For the non-maximal cuts, we will need the functions $Q_n^D$ for specific values of $D$ and $n$.
For $\cutRes_C\jtilde_G$ with $n_G$ even, $D=n_G-2\eps$:
\begin{itemize}
	\item for $n_{C}=n_G-1$:
	\begin{equation}\label{eq:forNMaxEven}
		Q_1^{1-2\eps}(X_1,X_0)=-\frac{\sqrt{\pi}\,i^{2\eps-1}}{\sqrt{(X_1X_1)}}
		\left[1+2\eps\ln(1+u_{10})+\eps\ln(X_0X_0)\right]
		+\mathcal{O}(\eps^2)\,.
	\end{equation}
	\item for $n_C=n_G-2$:
	\begin{equation}
		Q_2^{2-2\eps}(X_1,X_2,X_0)=\frac{i^{2\eps-2}}{2\sqrt{(X_1X_1)(X_2X_2)}}
		\frac{1}{\sqrt{u^2_{12}-1}}
		\ln\left(\frac{u_{12}+\sqrt{u_{12}^2-1}}{u_{12}-\sqrt{u_{12}^2-1}}\right)
		+\mathcal{O}(\eps)\,.
	\end{equation}
\end{itemize}
For $\cutRes_CJ_G$ with $n_G$ odd, $D=n_G+1-2\eps$:
\begin{itemize}
	\item for $n_C=n_G-1$:
	\begin{equation}
		Q_1^{2-2\eps}(X_1,X_0)=\frac{i^{2\eps-2}}{2\sqrt{(X_0X_0)(X_1X_1)}}
		\frac{1}{\sqrt{u^2_{10}-1}}
		\ln\left(\frac{u_{10}+\sqrt{u_{10}^2-1}}{u_{10}-\sqrt{u_{10}^2-1}}\right)
		+\mathcal{O}(\eps)\,.
	\end{equation}
	\item for $n_C=n_G-2$:
	\begin{align}\bsp\label{eq:forNNMaxOdd}
		Q_2^{3-2\eps}(X_1,X_2,X_0)
		=&\,
		\frac{\sqrt{\pi}\,i^{2\eps-3}}{2\sqrt{(X_0X_0)(X_1X_1)(X_2X_2)}}
		\frac{1}{\Delta(u_{10},u_{20},u_{12})}\\
		&\ln\left(\frac{1+u_{10}+u_{20}+u_{12}+\Delta(u_{10},u_{20},u_{12})}
		{1+u_{10}+u_{20}+u_{12}-\Delta(u_{10},u_{20},u_{12})}\right)
		+\mathcal{O}(\eps)\,,
	\esp\end{align}
	where we have defined the function
	$\Delta(a,b,c)=\sqrt{a^2+b^2+c^2-2abc-1}$.
\end{itemize}

These expressions appear in \refE{eq:cut-loop-duality} evaluated at specific points $X'_{C,i}$, in combinations that can be written in terms of Gram and Cayley determinants. We will explain how this is done through an example, the next-to-maximal cut of an integral with an even number of propagators. The remaining cases can be treated in an analogous way. For our example, we set  $C=[n-1]$ without loss of generality. We must evaluate \refE{eq:forNMaxEven} at $X_1=X'_{C,n}$ and
$X_0=X'_{C,\infty}$.
From eqs.~\eqref{eq:compact_Gram} and \eqref{eq:proj_scalar_product}, it is easy to find that
\begin{equation}
	(X'_{C,n}X'_{C,n})=-\frac{Y_{[n]}}{Y_{[n-1]}}\,,\qquad\quad
	(X'_{C,\infty}X'_{C,\infty})=\frac{\Gram_{[n-1]}}{4\,Y_{[n-1]}}\,.
\end{equation}
To evaluate $u_{n\infty}(C)$, i.e.~$u_{10}$ for $X_1=X'_{C,n}$ and
$X_0=X'_{C,\infty}$, we first note that
\beq
\det\left(\begin{array}{cc}
(X'_{C,n}X'_{C,n})& (X'_{C,n}X'_{C,\infty})\\
(X'_{C,n}X'_{C,\infty})& (X'_{C,\infty}X'_{C,\infty})
\end{array}\right)
=\frac{\det(X_i X_j)_{i,j\in Ce\infty}}{(-1)^c\,Y_C}\,.
\eeq
This is just eq.~(7.15) of ref.~\cite{Abreu:2017ptx}, whose proof was presented there.
The determinant on the left-hand side is easily written in terms of $u_{n\infty}(C)$,
$(X'_{C,n}X'_{C,n})$ and $(X'_{C,\infty}X'_{C,\infty})$ to finally get
\begin{equation}
u^2_{n\infty}(C)
=1-\frac{\Gram_{[n]}Y_{[n-1]}}{Y_{[n-1]}\Gram_{[n-1]}}\,.
\end{equation}
The same procedure allows one to determine all the required $u_{ij}$. The function $\Delta$ appearing in \refE{eq:forNNMaxOdd} is obtained from the determinant of the three-dimensional matrix involving the scalar products of the points $X'_{[n-2],n-1}$, $X'_{[n-2],n}$ and $X'_{[n-2],\infty}$, assuming propagators $n-1$ and $n$ are uncut.

We now combine all the results listed above to give the explicit expressions for the cuts required for full determination of the differential equations of one-loop integrals, as discussed in Section \ref{sec:alph_diff_eq}. For diagrams with an even number of propagators, we find the following results.
\begin{itemize}
	\item Maximal cut:
	\begin{equation}\label{eq:maxEvenComp}
		\cutRes_{E_G}\jtilde_G=
		\frac{2^{1-n/2}i^{n/2}}{\sqrt{Y_{E_G}}}\left[1+\eps\ln\left(\frac{\Gram_{E_G}}{4\,Y_{E_G}}\right)\right]
		+\mathcal{O}\left(\eps^2\right)\,.
	\end{equation}
	\item Next-to-maximal cut:
	\begin{align}\label{eq:nmaxEvenComp}
		\cutRes_{E_G\setminus e}\jtilde_G=&
		-\frac{2^{-n/2}\,i^{n/2}}{\sqrt{Y_{E_G}}}
		\left[
		1+2\eps\ln\left(1+\sqrt{\frac{Y_{E_G}\Gram_{E_G\setminus e}-\Gram_{E_G}Y_{E_G\setminus e}}{Y_{E_G}\Gram_{E_G\setminus e}}}\right)+\right.\nonumber\\
		&\left.\eps\ln\left(\frac{\Gram_{E_G\setminus e}}{4\,Y_{E_G\setminus e}}\right)
		\right]+\mathcal{O}\left(\eps^2\right)\,.
	\end{align}
	\item Next-to-next-to-maximal cut:
	\begin{align}\label{eq:nnmaxEvenComp}
		&\cutRes_{E_G\setminus \{e,f\}}\jtilde_G=\\
		&=\frac{2^{-n/2}\,i^{n/2}}{\sqrt{Y_{E_G}}}
		\ln\left(
		\frac{\sqrt{Y_{E_G\setminus e}Y_{E_G\setminus f}-Y_{E_G}Y_{E_G\setminus \{e,f\}}}+\sqrt{-Y_{E_G}Y_{E_G\setminus\{e,f\}}}}
		{\sqrt{Y_{E_G\setminus e}Y_{E_G\setminus f}-Y_{E_G}Y_{E_G\setminus \{e,f\}}}-\sqrt{-Y_{E_G}Y_{E_G\setminus\{e,f\}}}}
		\right)
		+\mathcal{O}\left(\eps\right)\,.\nonumber
	\end{align}
\end{itemize}

For diagrams with an odd number of propagators, we find the following results.
\begin{itemize}
	\item Maximal cut:
	\begin{equation}\label{eq:maxOddComp}
		\cutRes_{E_G}\jtilde_G=
		\frac{2^{(1-n)/2}\,i^{(n-1)/2}}{\sqrt{\Gram_{E_G}}}\left[1+\eps\ln\left(\frac{\Gram_{E_G}}{Y_{E_G}}\right)\right]
		+\mathcal{O}\left(\eps^2\right)\,.
	\end{equation}
	\item Next-to-maximal cut: 
	\begin{align}\label{eq:nmaxOddComp}
		\cutRes_{E_G\setminus e}\jtilde_G=&\frac{2^{(1-n)/2}\,i^{(n-1)/2}}{\sqrt{\Gram_{E_G}}}\\
		&\ln\left(
		\frac{\sqrt{Y_{E_G}\Gram_{E_G\setminus e}-\Gram_{E_G}Y_{E_G\setminus e}}-\sqrt{-\Gram_{E_G}Y_{E_G\setminus e}}}
		{\sqrt{Y_{E_G}\Gram_{E_G\setminus e}-\Gram_{E_G}Y_{E_G\setminus e}}+\sqrt{-\Gram_{E_G}Y_{E_G\setminus e}}}
		\right)
		+\mathcal{O}\left(\eps\right)\,. \nonumber
	\end{align}
	\item Next-to-next-to-maximal cut:
	\begin{align}
		\cutRes_{E_G\setminus \{e,f\}}\jtilde_G=&
		\frac{2^{(1-n)/2}\,i^{(n-1)/2}}{\sqrt{\Gram_{E_G}}}
		\ln\left(\frac{a_1+a_2+a_3+a_4+a_5}{a_1+a_2+a_3+a_4-a_5}\right)\,.
	\end{align}
	where 
	\begin{align}\bsp\label{eq:nnmaxOddComp}
	a_1&=\sqrt{\Gram_{E_G\setminus \{e,f\}}Y_{E_G\setminus e}Y_{E_G\setminus f}}\,,\\
	a_2&=\sqrt{\Gram_{E_G\setminus \{e,f\}}\left(Y_{E_G\setminus e}Y_{E_G\setminus f}-Y_{E_G}Y_{E_G\setminus \{e,f\}}\right)}\,,\\
	a_3&=\sqrt{Y_{E_G\setminus f}\left(Y_{E_G\setminus e}\Gram_{E_G\setminus \{e,f\}}-\Gram_{E_G\setminus e}Y_{E_G\setminus \{e,f\}}\right)}\,,\\
	a_4&=\sqrt{Y_{E_G\setminus e}\left(Y_{E_G\setminus f}\Gram_{E_G\setminus \{e,f\}}-\Gram_{E_G\setminus f}Y_{E_G\setminus \{e,f\}}\right)}\,,\\
	a_5&=\sqrt{-\Gram_{E_G}Y^2_{E_G\setminus \{e,f\}}}\,.
\esp\end{align}
\end{itemize}

Let us make some comments on the explicit results listed above.
First, we stress that they only apply when the cuts are finite. Divergent cuts can be computed
through a similar procedure, but they are less interesting for us in the context of Section \ref{sec:alph_diff_eq}, as they correspond to cuts of known integrals. Second, we recall that the cuts defined in ref.~\cite{Abreu:2017ptx} are only determined up to an overall sign and modulo $i\pi$. Here, we have fixed the overall signs such that relations like eqs.~\eqref{eq:homo_even} and \eqref{eq:homo_odd} coming from the study of the homology groups of one-loop integrals are satisfied \cite{PhamInTeplitz,HwaTeplitz}. In particular, we note that the maximal and next-to-maximal cuts of a diagram with an even number of propagators, eqs.~\eqref{eq:maxEvenComp} and \eqref{eq:nmaxEvenComp}, are related by a factor of -1/2, and that eqs.~\eqref{eq:nnmaxEvenComp} and \eqref{eq:nmaxOddComp} are related by sending one point to infinity (e.g., the one associated with the edge $f$ in \eqref{eq:nnmaxEvenComp}).


\subsection{Sending a point to infinity}\label{app:pointToInf}

We now make precise the operation of `sending a point to infinity' 
\cite{Broadhurst:1993ib,Drummond:2006rz}, which is particularly simple to explain in the
compactified picture discussed above. We define $K_G$ as
\begin{equation}
	\widetilde K_G \equiv  \lim_{\eps\to0} \jtilde_G\,,
\end{equation}
and assume completely generic kinematics so that the limit exists (the only divergent integral
is the tadpole, but this is a trivial case that can be dealt with separately).
Our discussion extends to integrals with some scales set to zero,
as long as they are finite in the limit $\eps\to0$.
In the following, we assume that $G$ is a graph with an even number of internal edges, and $G'$
is the same graph after contracting one edge $e$,
  so that $E_G=E_{G'}\cup \{e\}$ and 
$n_{G}=n_{G'}+1$.
The operation of sending a point to infinity relates $\widetilde K_G$ and $\widetilde K_{G'}$.

From \refE{eq:IComp}, it follows that
\begin{equation}
	\widetilde K_G=\frac{1}{\pi^{n_G/2}}
	\int_\Sigma  \frac{d^{n_G+2}Y\,\delta((YY))}{\textrm{Vol}(GL(1))}
	\frac{1}{[-2(X_1Y)]\ldots[-2(X_{n_G}Y)]}\,.
\end{equation}
For $G'$, we have
\begin{equation}
	\widetilde K_{G'}=-\frac{1}{\pi^{n_G/2}}
	\int_\Sigma  \frac{d^{n_G+2}Y\,\delta((YY))}{\textrm{Vol}(GL(1))}
	\frac{1}{[-2(X_1Y)]\ldots[-2(X_{n_G'-1}Y)][-2(X_{\infty}Y)]}\,.
\end{equation}
We thus see that, up to a sign, $\widetilde K_{G'}$ is obtained from 
$\widetilde K_G$ by taking the point
$X_e$ associated with an edge $e$ of $G$ to be $X_\infty$. Without loss of generality, we
can choose this point to be $X_{n_G}$, so that
\begin{equation}\label{eq:toInfinityJT}
	\widetilde K_{G'}(X_1,\ldots,X_{n_G'})=
	-\widetilde K_{G}(X_1,\ldots,X_{n_G-1},X_{n_G}=X_\infty)\,.
\end{equation}

In this paper, we find it convenient to work with integrals normalized to the leading term in the
$\eps$-expansion of their maximal cuts, as defined in \refE{eq:LS_J_n}. Because of this
normalization, the relation \eqref{eq:toInfinityJT} is modified. For $G$, the normalized integral
is:
\begin{equation}
	K_G=\frac{\sqrt{Y_{E_G}}}{2^{1-n_G/2}\,i^{n_G/2}}\widetilde K_G\,.
\end{equation}
For $G'$, it is:
\begin{equation}
	K_{G'}=\frac{\sqrt{\Gram_{E_{G'}}}}{2^{1-n_G/2}\,i^{n_G/2-1}}\widetilde K_{G'}\,.
\end{equation}
To obtain the equivalent of \refE{eq:toInfinityJT} for the normalized integrals, we must
determine how the normalization factor in $K_G$ transforms as we take
the point $X_{n_G}$ to infinity. We use \refE{eq:compact_Gram} to write
the modified Cayley determinant as a determinant of the points $X_i$ and then send
$X_{n_G}$ to infinity:
\begin{equation}
	\frac{\sqrt{\det(X_iX_j)_{i,j\in E_G}}}{2^{1-n_G/2}\,i^{n_G/2}}
	\xrightarrow[X_{n_G}\to X_\infty]{}
	\frac{\sqrt{\det(X_iX_j)_{i,j\in E_G'\cup\infty}}}{2^{1-n_G/2}\,i^{n_G/2}}
	=\frac{\sqrt{\Gram_{E_{G'}}}}{2^{2-n_G/2}\,i^{n_G/2-1}}\,,
\end{equation}
where in the last equality we again used \refE{eq:compact_Gram} to relate
determinants involving the point at infinity with the usual Gram determinant. We thus find
that the relation between the normalized integrals is
\begin{equation}\label{eq:toInfinityJ}
	 K_{G'}(X_1,\ldots,X_{n_G'})=
	-2  K_{G}(X_1,\ldots,X_{n_G-1},X_{n_G}=X_\infty)\,.
\end{equation}
This relation extends trivially to cut integrals, with the restriction that the point that is
taken to infinity cannot correspond to a cut edge.

We note that the factor of 2 in the above relation is related to the deformation parameter
$1/2$ appearing
in the definition of $\jhat_G$, in terms of which we write our diagrammatic coaction,
see  eqs.~\eqref{eq:Coaction_cut_integrals} and \eqref{hatG_J}.
As an illustration of this observation, we recover the differential equation
for a finite integral with an odd number of propagators at $\eps=0$
by sending a point to infinity in \refE{eq:diffeq_dci}, i.e., from the
differential equation of a finite integral with an even number of propagators.
The same differential equation was obtained directly from the diagrammatic coaction
in \refE{eq:diffeq_odd_finite}. We start
from \refE{eq:diffeq_dci} and single out the terms containing edge {$e$}
in the sum over subsets of edges:
\begin{equation}\label{eq:temp211153}
	dK_{G}=\sum_{\substack{ X\subset E_{G'}\\n_X=n_{G'}-1}}\!\!\!K_{G_X}\,d\,\cutRes_XK_G +
	\sum_{\substack{ X\subset E_{G'}\\n_X=n_{G'}-2}}\!\!\!K_{G_{X {e}}}
	\,d\,\cutRes_{X {e}}K_G\,,
\end{equation}
where  $E_G=E_{G'}\cup {\{e\}}$ and   ${X {e}} \equiv X\cup \{e\}$. The cuts in the second term on the right-hand side
involve the edge ${e}$, but using \refE{eq:DCI_cut_relations_app_anyN} they
can be rewritten in terms of cuts not involving ${e}$,
\begin{equation}
	\cutRes_{X {e}}K_G=
	-2\,\cutRes_XK_G-\sum_{e'\in E_{G'}\setminus X}\cutRes_{Xe'}K_G\,.
\end{equation}
We can now rewrite \refE{eq:temp211153} in such a way that the edge ${e}$ is never cut
in any of the contributions in the right-hand side:
\begin{equation}
	dK_{G}=\!\!\!\sum_{\substack{ X\subset E_{G'}\\n_X=n_{G'}-1}}\!\!\!K_{G_X}\,d\,\cutRes_XK_G +
	\sum_{\substack{ X\subset E_{G'}\\n_X=n_{G'}-2}}\!\!\!K_{G_{X {e}}}
	\,\left(-2\,d\cutRes_XK_G-\sum_{e'\in E_{G'}\setminus X}d\cutRes_{Xe'}K_G\right).
\end{equation}
We can now use \refE{eq:toInfinityJ} and take the point corresponding to edge ${e}$ to infinity
to obtain
\begin{equation}
	dK_{G'}=\sum_{\substack{ X\subset E_{G'}\\n_X=n_{G'}-1}}\!\!\!K_{G'_X}\,d\,\cutRes_XK_{G'}+
	\sum_{\substack{ X\subset E_{G'}\\n_X=n_{G'}-2}}\!\!\!K_{G'_{X}}
	\,\left(d\cutRes_XK_{G'}+\frac{1}{2}\sum_{e\in E_{G'}\setminus X}d\cutRes_{Xe}K_{G'}\right),
\end{equation}
which reproduces \refE{eq:diffeq_odd_finite}. While in \refE{eq:diffeq_odd_finite} the factor
of 1/2 appeared directly from the diagrammatic coaction, in the derivation above it appeared
because of the relation~\eqref{eq:toInfinityJ}. For the operation of sending a point to infinity
to be consistent with the diagrammatic coaction, the deformation parameter must thus be
consistent with the coefficient appearing in \refE{eq:toInfinityJ}.

\bibliographystyle{JHEP}
\bibliography{bibMain.bib}

\end{document}